\documentclass[prb,twocolumn,showpacs,floatfix,amsmath,amssymb,superscriptaddress]{revtex4-1}
\usepackage{latexsym}
\usepackage{graphicx}
\usepackage{times}
\usepackage{amsmath}
\usepackage{dcolumn}
\usepackage{latexsym,amsmath,amssymb,bm,euscript}
\begin{document}

\title{New directions in the pursuit of Majorana fermions in solid state systems}

\author{Jason Alicea}
\affiliation{Department of Physics and Astronomy, University of California, Irvine, California 92697}

\date{\today}

\begin{abstract}

The 1937 theoretical discovery of Majorana fermions---whose defining property is that they are their own anti-particles---has since impacted diverse problems ranging from neutrino physics and dark matter searches to the fractional quantum Hall effect and superconductivity.  Despite this long history the unambiguous observation of Majorana fermions nevertheless remains an outstanding goal.  This review article highlights recent advances in the condensed matter search for Majorana that have led many in the field to believe that this quest may soon bear fruit.  We begin by introducing in some detail exotic `topological' one- and two-dimensional superconductors that support Majorana fermions at their boundaries and at vortices.  We then turn to one of the key insights that arose during the past few years; namely, that it is possible to `engineer' such exotic superconductors in the laboratory by forming appropriate heterostructures with \emph{ordinary} $s$-wave superconductors.  Numerous proposals of this type are discussed, based on diverse materials such as topological insulators, conventional semiconductors, ferromagnetic metals, and many others.  The all-important question of how one experimentally detects Majorana fermions in these setups is then addressed.  We focus on three classes of measurements that provide smoking-gun Majorana signatures: tunneling, Josephson effects, and interferometry.  Finally, we discuss the most remarkable properties of condensed matter Majorana fermions---the non-Abelian exchange statistics that they generate and their associated potential for quantum computation.

\end{abstract}

\maketitle


\section{Introduction}

Three quarters of a century ago Ettore Majorana introduced into theoretical physics what are now known as `Majorana fermions': particles that, unlike electrons and positrons, constitute their own antiparticles.\cite{MajoranaInvention}  The monumental significance of this development required many intervening decades to fully appreciate, and despite being an `old' idea Majorana fermions remain central to diverse problems across modern physics.  In the high-energy context, Ettore's original suggestion that neutrinos may in fact be Majorana fermions endures as a serious proposition even today.\cite{NeutrinoReview}  Supersymmetric theories further postulate that bosonic particles such as photons have a corresponding Majorana `superpartner' that may provide one of the keys to the dark matter puzzle.\cite{WilczekPerspective}  Experiments at the large hadron collider are well-positioned to critically test these predictions in the near future.  Condensed matter physicists, too, are fervently chasing Majorana's vision in a wide variety of solid state systems, motivated both by the pursuit of exotic fundamental physics and quantum computing applications.  While a definitive sighting of Majorana fermions has yet to be reported in any setting, there is palpable optimism in the condensed matter community that this may soon change.\cite{WilczekPerspective,SternPerspective,FranzPerspective,Service:2011,HosurViewpoint}  

Unlike the Majorana fermions sought by high-energy physicists, those pursued in solid state systems are not fundamental particles---the constituents of condensed matter are, inescapably, ordinary electrons and ions.  This fact severely constrains the likely avenues of success in this search.  In conventional metals, for example, electron and hole excitations can annihilate, but since they carry opposite charge are certainly not Majorana fermions.  In operator language this is reflected by the fact that if $c_\sigma^\dagger$ adds an electron with spin $\sigma$, then its Hermitian conjugate $c_\sigma$ is a physically distinct operator that creates a hole.  If Majorana is to surface in the solid state it must therefore be in the form of nontrivial \emph{emergent} excitations.  

Superconductors (and other systems where fermions pair and condense) provide a natural hunting ground for such excitations.  Indeed, because Cooper pair condensation spontaneously violates charge conservation, quasiparticles in a superconductor involve \emph{superpositions} of electrons and holes.  Unfortunately, however, this is not a sufficient condition for the appearance of Majorana fermions.  With only exceedingly rare exceptions superconductivity arises from $s$-wave-paired electrons carrying opposite spins; quasiparticle operators then (schematically) take the form $d = u c_\uparrow^\dagger + v c_\downarrow$, which is still physically distinct from $d^\dagger = v^* c_\downarrow^\dagger + u^* c_\uparrow$.  Thus whereas charge prevents Majorana from emerging in a metal, spin is the culprit in conventional $s$-wave superconductors.

As the preceding discussion suggests, `spinless' superconductors---\emph{i.e.}, paired systems with only one active fermionic species rather than two---provide ideal platforms for Majorana fermions.  By Pauli exclusion, Cooper pairing in a `spinless' metal must occur with odd parity, resulting in $p$-wave superconductivity in one dimension (1D) and, in the most relevant case for our purposes, $p+ip$ superconductivity in two dimensions (2D).  These superconductors are quite special: as Sec.\ \ref{ToyModels} describes in detail, they realize \emph{topological phases} that support exotic excitations at their boundaries and at topological defects.\cite{VolovikBook,1DwiresKitaev,ReadGreen}  Most importantly, zero-energy modes localize at the ends of a 1D topological $p$-wave superconductor\cite{1DwiresKitaev}, and bind to superconducting vortices in the 2D $p+ip$ case\cite{Kopnin}.  These zero-modes are precisely the condensed matter realization of Majorana fermions\cite{1DwiresKitaev,ReadGreen} that are now being vigorously pursued.  

Let $\gamma$ denote the operator corresponding to one of these modes (the specific realization is unimportant for now).  This object is its own `anti-particle' in the sense that $\gamma = \gamma^\dagger$ and $\gamma^2 = 1$.  We caution, however, that labeling $\gamma$ as a particle---emergent or otherwise---is a misnomer because unlike an ordinary electronic state in a metal there is no meaning to $\gamma$ being occupied or unoccupied.  Rather, $\gamma$ should more appropriately be viewed as a  \emph{fractionalized} zero-mode comprising `half' of a regular fermion.  More precisely, a \emph{pair} of Majorana zero-modes, say $\gamma_1$ and $\gamma_2$, must be combined via $f = (\gamma_1 + i\gamma_2)/2$ to obtain a fermionic state with a well-defined occupation number.  While this new operator represents a conventional fermion in that it satisfies $f \neq f^\dagger$ and obeys the usual anticommutation relations, $f$ remains nontrivial in two critical respects.  First, $\gamma_1$ and $\gamma_2$ may localize arbitrarily far apart from one another; consequently $f$ encodes highly non-local entanglement.  Second, one can empty or fill the non-local state described by $f$ with no energy cost, resulting in a ground-state degeneracy.  These two properties underpin by far the most interesting consequence of Majorana fermions---the emergence of non-Abelian statistics.

A brief digression is in order to put this remarkable phenomenon in proper perspective.  Exchange statistics characterizes the manner in which many-particle wavefunctions transform under interchange of indistinguishable particles, and is one of the cornerstones of quantum theory.  There indeed exists a rather direct path from particle statistics to the existence of metals, superfluids, superconductors, and many other quantum phases, not to mention the periodic table as we know it.\cite{TQCreview,SternReview}  It has long been appreciated that for topological reasons 2D systems allow for particles whose statistics is neither fermionic nor bosonic.\cite{Leinaas,WilczekAnyons}  Such \emph{anyons} come in two flavors: Abelian and non-Abelian.  Upon exchanging Abelian anyons---which arise in most fractional quantum Hall states\cite{HalperinAnyons,ArovasAnyons,TQCreview,SternReview}---the wavefunction acquires a statistical phase $e^{i\theta}$ that is intermediate between $-1$ and $1$.  Non-Abelian anyons are far more exotic (and elusive); under their exchange the wavefunction does not simply acquire a phase factor, but rather can \emph{change to a fundamentally different quantum state}.  As a result subsequent exchanges do not generally commute, hence the term `non-Abelian'.  An important step toward finding experimental realizations of the second flavor came in 1991 when Moore and Read introduced a set of `Pfaffian' trial wavefunctions for fractional quantum Hall states that support non-Abelian anyons.\cite{MooreRead,NayakWilczekBraiding,ReadRezayiBraiding,ReadBraiding,BondersonNonAbelianStatistics}  Several theoretical and experimental works\cite{TQCreview,FiveHalvesCharge,FiveHalvesTunneling,Willett1,Willett2,FiveHalvesNeutralModes,FiveHalvesKang,FiveHalvesSpinPolarization} indicate that the observed quantum Hall state at filling factor\cite{FiveHalvesDiscovery} $\nu = 5/2$ may provide the first realization of such a non-Abelian phase.  Read and Green\cite{ReadGreen} later provided a key breakthrough that in many ways served as a stepping stone for the new directions reviewed here.  In particular, these authors established an intimate connection between the superficially very different Moore-Read Pfaffian states and a topological spinless 2D $p+ip$ superconductor---deducing that universal properties of the former such as non-Abelian statistics must also be shared by the latter (which crucially can arise in \emph{weakly} interacting systems). 

With this backdrop let us now describe how non-Abelian statistics arises in a 2D spinless $p+ip$ superconductor.  Consider a setup with $2N$ vortices binding Majorana zero-modes $\gamma_{1,\ldots,2N}$.  One can (arbitrarily) combine pairs of Majoranas to define $N$ fermion operators $f_j = (\gamma_{2j-1} +i\gamma_{2j})/2$ corresponding to zero-energy states that can be either filled or empty.  Thus the vortices generate $2^N$ degenerate ground states\footnote{Some authors quote a $2^{N-1}$-fold ground state degeneracy, which is the actual number of states available if fermion parity is fixed.  In this review we will ignore parity restrictions when evaluating ground-state degeneracies.} that can be labeled in terms of occupation numbers $n_j = f_j^\dagger f_j$ by 
\begin{equation}
  |n_1,n_2,\ldots,n_N\rangle.  
  \label{DegenerateGroundStates}
\end{equation}
Suppose that one prepares the system into an arbitrary ground state and then adiabatically exchanges a pair of vortices.  Because this process swaps the positions of two Majorana modes, each being `half' of a fermion, the system generally ends up in a different ground state from which it began.  More formally the exchange unitarily rotates the wavefunction inside of the ground-state manifold in a non-commutative fashion.  The vortices---because of the Majorana zero-modes that they bind---therefore exhibit non-Abelian statistics.\cite{ReadGreen,Ivanov,SternBraiding,StoneBraiding}

One might naively conclude that in this regard the Majorana zero-modes bound to the ends of a 1D topological $p$-wave superconductor are substantially less interesting than those arising in 2D.  After all, exchange statistics of any type is ill-defined in 1D because particles inevitably `collide' during the course of an exchange.\cite{TQCreview}  This is the root, for instance, of the equivalence between hard-core bosons and fermions in 1D.  Fortunately this obstacle can be very simply surmounted by fabricating \emph{networks} of 1D superconductors; envision, say, an array of wires forming junctions, with topological $p$-wave superconductors binding Majorana zero-modes interspersed at various locations.  Such networks allow the positions of Majorana zero-modes to be meaningfully exchanged,\cite{AliceaBraiding} which remarkably still gives rise to non-Abelian statistics despite the absence of vortices.\cite{AliceaBraiding,ClarkeBraiding,SauBraiding,HalperinBraiding}  Thus 1D and 2D topological superconductors can both be appropriately described as non-Abelian phases of matter.  [As an interesting aside, Teo and Kane first showed that non-Abelian statistics can even appear in three dimensions, where exchange has long been assumed to be trivial.\cite{TeoKane,Freedman3D,Freedman3Db,HalperinBraiding}]

The observation of Majorana fermions in condensed matter would certainly constitute a landmark achievement from a fundamental physics standpoint, both because it could mean the first realization of Ettore Majorana's theoretical discovery and, far more importantly, because of the non-Abelian statistics that they harbor.  Moreover, success in this search might ultimately prove essential to overcoming one of the grand challenges in the field---the synthesis of a scalable quantum computer.\cite{Freedman98,kitaev,Freedman03,TopologicalQubits,MeasurementOnlyTQC,TQCreview}  The basic idea is that the occupation numbers $n_j = 0,1$ specifying the degenerate ground states of Eq.\ (\ref{DegenerateGroundStates}) can be used to encode `topological qubits'.\cite{TopologicalQubits}  Crucially, this quantum information is stored highly non-locally due to the arbitrary spatial separation between pairs of Majorana modes corresponding to a given $n_j$.  Suppose now that temperature is low compared to the bulk gap; if manipulations are carried out adiabatically the system then essentially remains confined to the ground-state manifold.  The user can controllably manipulate the state of the qubit by adiabatically exchanging the positions of Majorana modes, owing to the existence of non-Abelian statistics.  In principle the environment can also induce (unwanted) exchanges, thereby corrupting the qubit, but this happens with extraordinarily low probability due to the non-locality of such processes.  This is the basis of fault-tolerant topological quantum computation schemes that elegantly beat decoherence at the hardware level.\cite{Freedman98,kitaev,Freedman03,TopologicalQubits,MeasurementOnlyTQC,TQCreview}  While braiding of Majorana fermions alone permits somewhat limited topological quantum information processing,\cite{TQCreview} the additional unprotected operations needed for universal quantum computation come with unusually high error thresholds.\cite{BravyiKitaev,Bravyi}  The search for Majorana fermions is thus fueled also by the potential for revolutionary technological applications down the road.

In the beginning of this introduction we noted that researchers are optimistic that this search may soon come to fruition.  One might reasonably wonder why, given that we live in three dimensions, electrons carry spin, and $p$-wave pairing is scarce in nature.  To a large extent this optimism stems from the recent revelation that one can engineer low-dimensional topological superconductors by judiciously forming heterostructures with \emph{conventional} bulk $s$-wave superconductors.  This new line of attack could eventually lead to `designer topological phases' persisting up to relatively high temperatures, perhaps measuring in the 10K range or beyond.  The conceptual breakthrough here originated with the seminal work of Fu and Kane in the context of topological insulators,\cite{FuKane,MajoranaQSHedge} which paved the way for many subsequent proposals of a similar spirit.  We devote a large fraction of this review---Secs.\ \ref{1D_toy_model_realizations} and \ref{2D_toy_model_realizations}---to discussing these new routes to Majorana fermions.  `Classic' settings such as the $\nu = 5/2$ fractional quantum Hall state and Sr$_2$RuO$_4$ (which of course remain highly relevant to the field) will also be discussed, but only briefly.  An omission that we regret is a discussion of Helium-3, where seminal work related to this subject was carried out early on by Volovik and others; see the excellent book in Ref.\ \onlinecite{VolovikBook}.  Section \ref{Detection} explores the key question of how one experimentally identifies Majorana modes once a suitable topological phase is fabricated.  The long-term objectives of observing non-Abelian statistics and realizing quantum computation are taken up in Sec.\ \ref{NonAbelianStatistics}.  Finally, we offer some closing thoughts in Sec.\ \ref{Outlook}.  For additional perspectives on this fascinating problem we would like to refer the reader to several other reviews and popular articles: Refs.\ \onlinecite{WilczekPerspective,SternPerspective,FranzPerspective,Service:2011,HosurViewpoint,TQCreview,SternReview,KaneReview,QiReview,BeenakkerReview}.

\section{Toy Models for Topological Superconductors Supporting Majorana modes}
\label{ToyModels}

This section introduces toy models for topological 1D and 2D superconductors that support Majorana fermions.  We will explore the anatomy of the phases realized in these exotic superconductors and elucidate how they give rise to Majorana modes in some detail.  Later parts of this review rely heavily on the material discussed here.  Indeed, our perspective is that all of the recent experimental proposals highlighted in Secs.\ \ref{1D_toy_model_realizations} and \ref{2D_toy_model_realizations} are, in essence, practical realizations of these toy models.  The ideas developed here will also prove indispensable when we discuss experimental detection schemes in Sec.\ \ref{Detection} and non-Abelian statistics in Sec.\ \ref{NonAbelianStatistics}.

\subsection{1D spinless $p$-wave superconductor}
\label{1D_toy_model}

We begin by reviewing Kitaev's toy lattice model\cite{1DwiresKitaev}, introduced nearly a decade ago, for a 1D spinless $p$-wave superconductor.  This model's many virtues include the fact that in this setting Majorana zero-modes appear in an extremely simple and intuitive fashion.  Following Kitaev, we introduce operators $c_x$ describing spinless fermions that hop on an $N$-site chain and exhibit long-range-ordered $p$-wave superconductivity.  The minimal Hamiltonian describing this setup reads
\begin{equation}
  H = -\mu \sum_{x} c_x^\dagger c_x - \frac{1}{2}\sum_{x}(t c_x^\dagger c_{x+1} + \Delta e^{i\phi} c_x c_{x+1}+ h.c.),
  \label{Hkitaev}
\end{equation}
where $\mu$ is the chemical potential, $t\geq 0$ is the nearest-neighbor hopping strength, $\Delta \geq 0$ is the $p$-wave pairing amplitude and $\phi$ is the corresponding superconducting phase.  For simplicity we set the lattice constant to unity.  

\begin{figure}
\includegraphics[width = \linewidth]{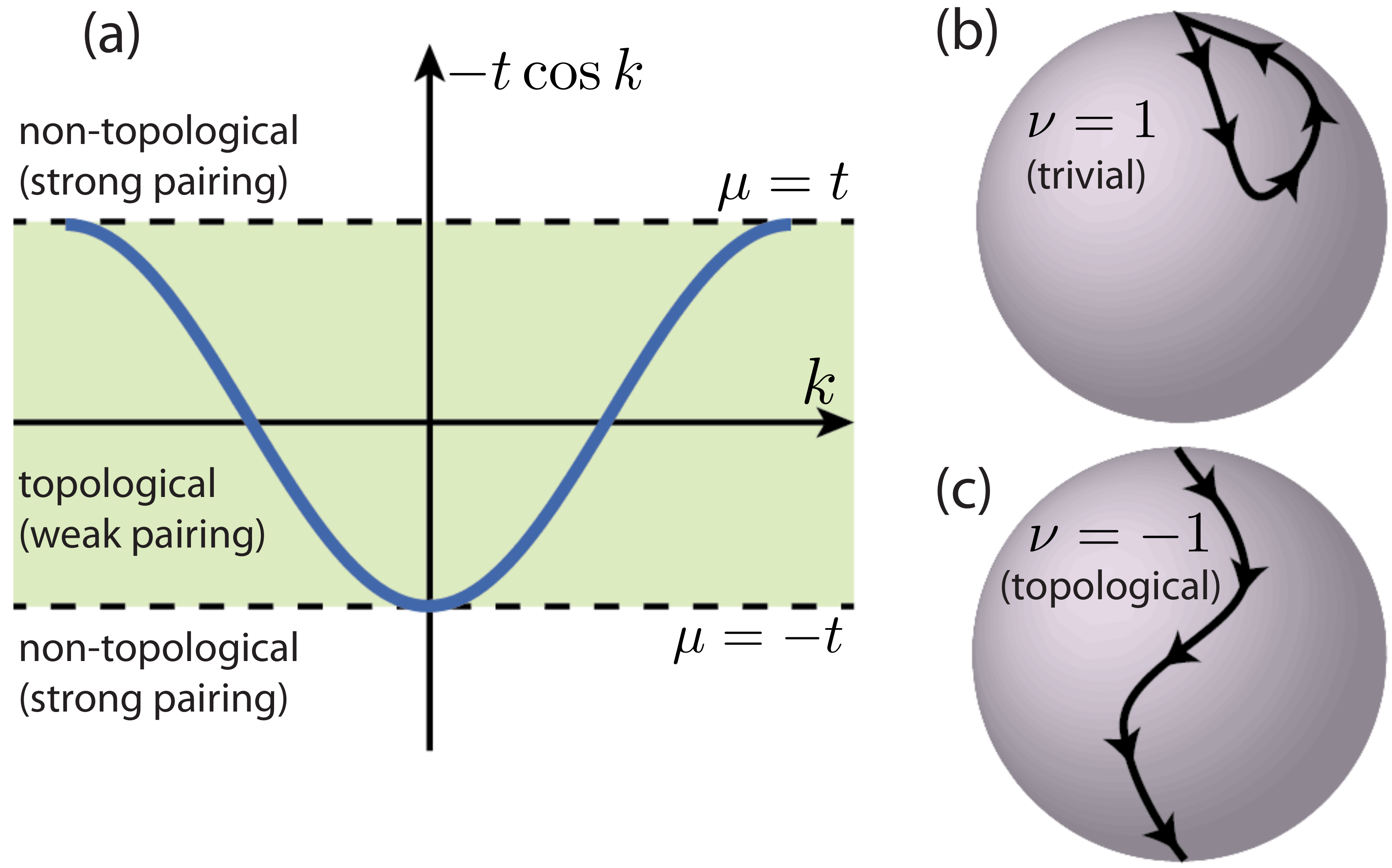}
\caption{(a) Kinetic energy in Kitaev's model for a 1D spinless $p$-wave superconductor.  The $p$-wave pairing opens a bulk gap except at the chemical potential values $\mu = \pm t$ displayed above.  For $|\mu|>t$ the system forms a non-topological strong pairing phase, while for $|\mu|<t$ a topological weak pairing phase emerges.  The topological invariant $\nu$ distinguishing these states can be visualized by considering the trajectory that ${\bf \hat{h}}(k)$ [derived from Eq.\ (\ref{hk_definition})] sweeps on the unit sphere as $k$ varies from 0 to $\pi$; (b) and (c) illustrate the two types of allowed trajectories.   }
\label{BandStructureFigKitaevModel}
\end{figure}

It is instructive to first understand the chain's bulk properties, which can be conveniently studied by imposing periodic boundary conditions on the system (thereby wrapping the chain into a loop and removing its ends).  Upon passing to momentum space and introducing a two-component operator $C^\dagger_k = [c_k^\dagger, c_{-k}]$, one can write $H$ in the standard Bogoliubov-de Gennes form:
\begin{eqnarray}
  H &=& \frac{1}{2} \sum_{k \in BZ}C^\dagger_k \mathcal{H}_k C_k,
   ~~~~~ \mathcal{H}_k = \left( \begin{array}{cc}
  \epsilon_k & \tilde \Delta_k^* \\
  \tilde \Delta_k & -\epsilon_k 
  \end{array}\right),
  \label {Hk}
\end{eqnarray}
with $\epsilon_k = -t\cos k-\mu$ the kinetic energy and $\tilde\Delta_k = -i \Delta e^{i \phi} \sin k$ the Fourier-transformed pairing potential. 
The Hamiltonian becomes simply
\begin{equation}
  H = \sum_{k \in BZ} E_{\rm bulk}(k)a_k^\dagger a_k
  \label{Hdiagonal}
\end{equation}
when expressed in terms of quasiparticle operators 
\begin{eqnarray}
  a_k &=& u_k c_k + v_k c_{-k}^\dagger
  \label{ak} \\
  u_k &=& \frac{\tilde\Delta}{|\tilde\Delta|}\frac{\sqrt{E_{\rm bulk}+\epsilon}}{\sqrt{2E_{\rm bulk}}}, ~~~ v_k = \left(\frac{E_{\rm bulk}-\epsilon}{\tilde\Delta}\right)u_k,
  \label{uv}
\end{eqnarray}
where the bulk excitation energies are given by
\begin{equation}
  E_{\rm bulk}(k) = \sqrt{\epsilon_k^2 + |\tilde\Delta_k|^2}.
  \label{Ebulk}
\end{equation}
Equation (\ref{Ebulk}) demonstrates that the chain admits gapless bulk excitations only when the chemical potential is fine-tuned to $\mu = t$ or $-t$, where the Fermi level respectively coincides with the top and bottom of the conduction band as shown in Fig.\ \ref{BandStructureFigKitaevModel}(a).  The gap closure at these isolated $\mu$ values reflects the $p$-wave nature of the pairing required by Pauli exclusion.  More precisely, since $\tilde\Delta_k$ is an odd function of $k$, Cooper pairing at $k = 0$ or $k = \pm\pi$ is prohibited, thereby leaving the system gapless at the Fermi level when $\mu = \pm t$.  Note that the phases that appear at $\mu < -t$ and $\mu >t$ are related by a particle-hole transformation; thus to streamline our discussion we will hereafter neglect the latter chemical potential range.

The physics of the chain is intuitively rather different in the gapped regimes with $\mu < -t$ and $|\mu| < t$---the former connects smoothly to the trivial vacuum (upon taking $\mu \rightarrow -\infty$) where no fermions are present, whereas in the latter a partially filled band acquires a gap due to $p$-wave pairing.  One can make this distinction more quantitative following Read and Green\cite{ReadGreen} by examining the form of the ground-state wavefunction in each regime.  Equation (\ref{Hdiagonal}) implies that the ground state $|{\rm g.s.}\rangle$ must satisfy $a_k|{\rm g.s.}\rangle = 0$ for all $k$ so that no quasiparticles are present.  Equations (\ref{ak}) and (\ref{uv}) allow one to explicitly write the ground state as follows,
\begin{eqnarray}
  |{\rm g.s.}\rangle &\propto& \prod_{0<k<\pi}[1+\varphi_{\rm C.p.}(k)c_{-k}^\dagger c_{k}^\dagger]|0\rangle
  \nonumber \\
  \varphi_{\rm C.p.}(k) &=& \frac{v_k}{u_k} = \left(\frac{E_{\rm bulk}-\epsilon}{\tilde\Delta}\right),
\end{eqnarray}
where $|0\rangle$ is a state with no $c_k$ fermions present with momenta in the interval $0<|k|<\pi$.  One can loosely interpret $\varphi_{\rm C.p.}(k)$ as the wavefunction for a Cooper pair formed by fermions with momenta $k$ and $-k$.  An important difference between the $\mu < -t$ and $|\mu|<t$ regimes is manifested in the real-space form $\varphi_{\rm C.p.}(x) = \int_k e^{i k x}\varphi_{\rm C.p.}(k)$ at large $x$:\cite{LesikUnpublished}
\begin{eqnarray}
  |\varphi_{\rm C.p.}(x)| \sim \left\{ \begin{array}{c}
 e^{-|x|/\zeta},~~~~\mu<-t~~~~ ({\rm strong ~pairing}) \\
 {\rm const}, ~~~~|\mu|<t~~~~({\rm weak ~pairing}).
       \end{array} \right.
\end{eqnarray}
It follows that $\mu <-t$ corresponds to a \emph{strong pairing} regime in which `molecule-like' Cooper pairs form from two fermions bound in real space over a length scale $\zeta$, whereas in the \emph{weak pairing} regime $|\mu|<t$ the Cooper pair size is infinite\cite{ReadGreen}.  
We emphasize that this distinction by itself does not guarantee that the weak and strong pairing regimes constitute distinct phases.  Indeed, similar physics occurs in the well-studied ``BCS-BEC crossover'' in $s$-wave paired systems where no sharp transition arises\cite{BCS_BEC_crossover,VictorLeo}.  The fact that the weak and strong pairing regimes are distinct phases separated by a phase transition at which the bulk gap closes is rooted in topology.  

There are several ways in which one can express the `topological invariant' (akin to an order parameter in the theory of conventional phase transitions) distinguishing the weak and strong pairing phases\cite{1DwiresKitaev}.  We will follow an approach that closely parallels the 2D case we address in Sec.\ \ref{2D_toy_model}.  Let us revisit the Hamiltonian in Eq.\ (\ref{Hk}), but now allow for additional perturbations that preserve translation symmetry.\footnote{The physics in no way relies on translation symmetry, which is merely a crutch that allows one to simply define a topological invariant.  In fact the interesting physical consequences arise only when translation symmetry is broken.}  The resulting $2\times 2$ matrix $\mathcal{H}_k$ can be expressed in terms of a vector of Pauli matrices ${\bm {\sigma}} = \sigma^x {\bf \hat{x}} + \sigma^y {\bf \hat{y}} + \sigma^z{\bf \hat{z}}$ as follows,
\begin{equation}
  \mathcal{H}_k = {\bf h}(k)\cdot {\bm \sigma}
  \label{hk_definition}
\end{equation}
for some vector ${\bf h}(k)$.  (A term proportional to the identity can also be added, but will not matter for our purposes.)  Although we are considering a rather general Hamiltonian here, the structure of ${\bf h}(k)$ is not entirely arbitrary.  In particular, since the two-component operator $C_k$ in Eq.\ (\ref{Hk}) satisfies $(C_{-k}^\dagger)^T = \sigma^x C_k$, the vector ${\bf h}(k)$ must obey the important relations
\begin{equation}
  h_{x,y}(k) = -h_{x,y}(-k), ~~~h_z(k) = h_z(-k).
  \label{hconstraint}
\end{equation}
Thus it suffices to specify ${\bf h}(k)$ only on the interval $0 \leq k \leq \pi$, since ${\bf h}(k)$ on the other half of the Brillouin zone follows from Eq.\ (\ref{hconstraint}).  

Suppose now that ${\bf h}(k)$ is non-zero throughout the Brillouin zone so that the chain is fully gapped.  One can then always define a unit vector ${\bf \hat{h}}(k) = {\bf h}(k)/|{\bf h}(k)|$ that provides a map from the Brillouin zone to the unit sphere.  The relations of Eq.\ (\ref{hconstraint}) strongly restrict this map at $k = 0$ and $\pi$ such that
\begin{equation}
  {\bf \hat{h}}(0) = s_0{\bf\hat{z}}, ~~~ {\bf \hat{h}}(\pi) = s_\pi {\bf \hat{z}},
\end{equation}
where $s_0$ and $s_\pi$ represent the sign of the kinetic energy (measured relative to the Fermi level) at $k = 0$ and $\pi$, respectively.  Thus as one sweeps $k$ from 0 to $\pi$, ${\bf \hat{h}}(k)$ begins at one pole of the unit sphere and either ends up at the same pole (if $s_0 = s_\pi$) or the opposite pole (if $s_0 = -s_\pi$).  These topologically distinct trajectories, illustrated schematically in Figs.\ \ref{BandStructureFigKitaevModel}(b) and (c), are distinguished by the $Z_2$ \emph{topological invariant} 
\begin{equation}
  \nu = s_0 s_\pi,
\end{equation}
which can only change sign when the chain's bulk gap closes [resulting in ${\bf \hat{h}}(k)$ being ill-defined somewhere in the Brillouin zone].\footnote{Note that individually, the signs $s_0$ and $s_\pi$ are not particularly meaningful; for example, a particle-hole transformation $C_k \rightarrow \sigma^x C_k$ changes the sign of both, but leaves their product $\nu$ invariant.}  Physically, $\nu = +1$ if at a given chemical potential there exists an even number of pairs of Fermi points, while $\nu = -1$ otherwise.   
From this perspective it is clear that $\nu = +1$ in the (topologically trivial) strong pairing phase while $\nu = -1$ in the (topologically nontrivial) weak pairing phase.  

The nontrivial topology inherent in the weak pairing phase leads to the appearance of Majorana modes in a chain with open boundary conditions, which we will now consider.   The new physics associated with the ends of the chain can be most simply accessed by decomposing the spinless fermion operators $c_x$ in the original Hamiltonian of Eq.\ (\ref{Hkitaev}) in terms of two Majorana fermions via
\begin{equation}
  c_x = \frac{e^{-i \phi/2}}{2}(\gamma_{B,x}+i \gamma_{A,x}).
\end{equation}
The operators on the right-hand side obey the canonical Majorana fermion relations 
\begin{equation}
  \gamma_{\alpha,x} = \gamma_{\alpha,x}^\dagger, ~~~~~~
  \{\gamma_{\alpha,x},\gamma_{\alpha',x'}\} = 2\delta_{\alpha\alpha'}\delta_{xx'}.  
\end{equation}  
In this basis $H$ becomes
\begin{eqnarray}
  H &=& -\frac{\mu}{2}\sum_{x = 1}^N(1+i\gamma_{B,x}\gamma_{A,x})
  \nonumber \\
  &-&\frac{i}{4}\sum_{x = 1}^{N-1}[(\Delta+t)\gamma_{B,x}\gamma_{A,x+1}+(\Delta-t)\gamma_{A,x}\gamma_{B,x+1}].
  \label{Hkitaev2}
\end{eqnarray}
Generally the parameters $\mu, t$, and $\Delta$ induce relatively complex couplings between these Majorana modes; however, the problem becomes trivial in two limiting cases\cite{1DwiresKitaev}.

\begin{figure}
\includegraphics[width = 8cm]{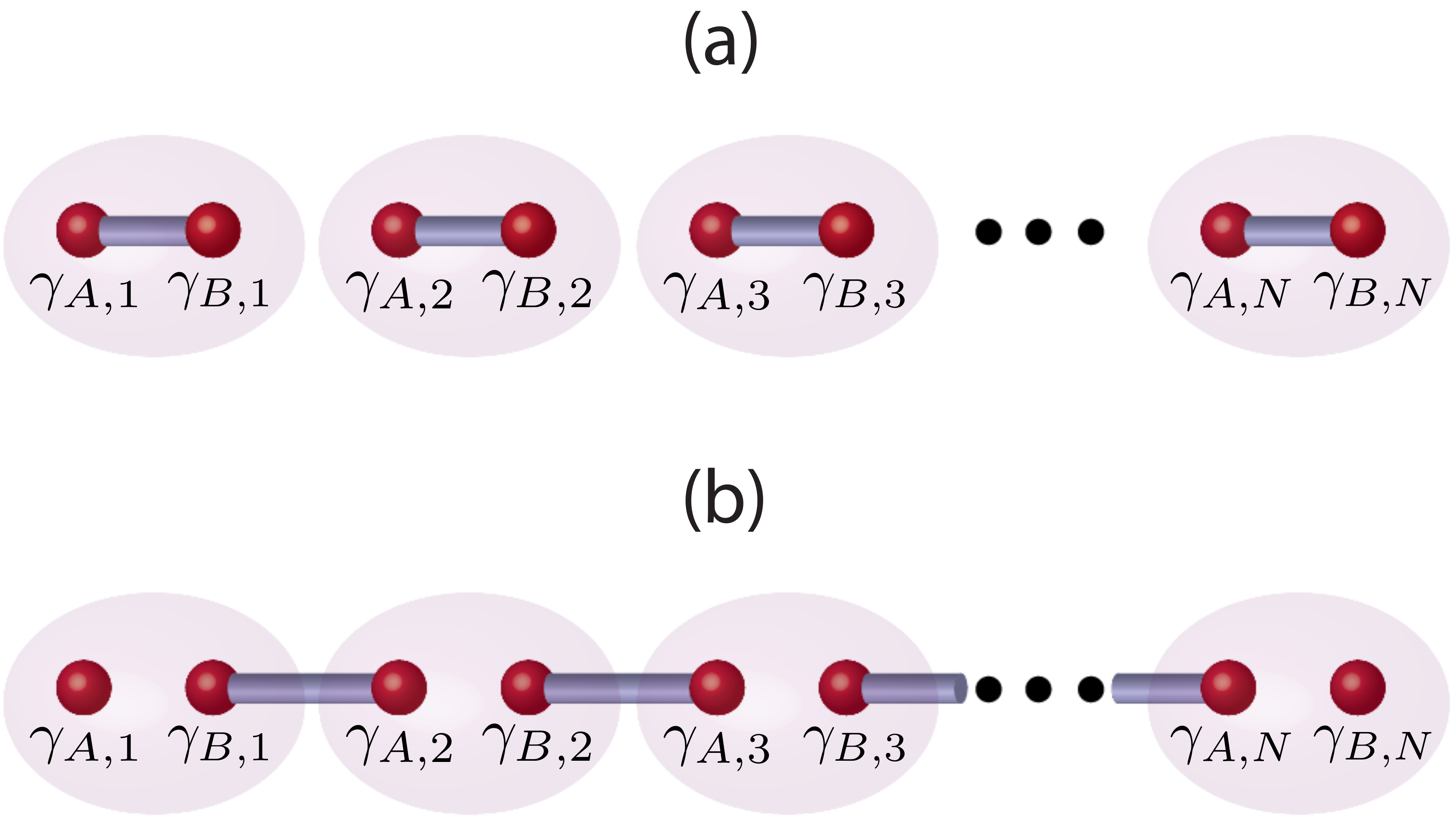}
\caption{Schematic illustration of the Hamiltonian in Eq.\ (\ref{Hkitaev2}) when (a) $\mu \neq 0$, $t = \Delta = 0$ and (b) $\mu = 0$, $t = \Delta \neq 0$.  In the former limit Majoranas `pair up' at the same lattice site, resulting in a unique ground state with a gap to all excited states.  In the latter, Majoranas couple at adjacent lattice sites, leaving two `unpaired' Majorana zero-modes $\gamma_{A,1}$ and $\gamma_{B,N}$ at the ends of the chain.  Although there remains a bulk energy gap in this case, these end-states give rise to a two-fold ground state degeneracy.  }
\label{KitaevModelFig}
\end{figure}

The first corresponds to $\mu < 0$ but $t = \Delta = 0$, where the chain resides in the topologically trivial phase.  Here the second line of Eq.\ (\ref{Hkitaev2}) vanishes, leaving a coupling only between Majorana modes $\gamma_{A,x}$ and $\gamma_{B,x}$ at the same lattice site as Fig.\ \ref{KitaevModelFig}(a) schematically illustrates.  In this case there is a unique ground state corresponding to the vacuum of $c_x$ fermions.  Clearly the spectrum is gapped since introducing a spinless fermion into the chain costs a finite energy $|\mu|$.  Note that this is entirely consistent with our treatment of the chain with periodic boundary conditions; in the trivial phase the ends of the chain have little effect.  We emphasize that these conclusions hold even away from this fine-tuned limit provided the gap persists so that the chain remains in the same trivial phase.    

The second simplifying limit corresponds to $\mu = 0$ and $t = \Delta \neq 0$, where the topological phase appears.  Here the Hamiltonian is instead given by
\begin{equation}
  H = -i\frac{t}{2}\sum_{x = 1}^{N-1}\gamma_{B,x}\gamma_{A,x+1},
  \label{H_gammas}
\end{equation}
which couples Majorana fermions only at \emph{adjacent} lattice sites as Fig.\ \ref{KitaevModelFig}(b) illustrates.  In terms of new ordinary fermion operators $d_x = \frac{1}{2}(\gamma_{A,x+1}+i \gamma_{B,x})$, the Hamiltonian can be written
\begin{equation}
  H = t\sum_{x = 1}^{N-1}\left(d_x^\dagger d_x - \frac{1}{2}\right).
\end{equation}
In this form it is apparent that a bulk gap remains here too---consistent with our results with periodic boundary conditions---since one must pay an energy $t$ to add a $d_x$ fermion.  However, as Fig.\ \ref{KitaevModelFig}(b) illustrates the ends of the chain now support `unpaired' \emph{zero-energy Majorana modes} $\gamma_1 \equiv \gamma_{A,1}$ and $\gamma_2 \equiv \gamma_{B,N}$ that are explicitly absent from the Hamiltonian in Eq.\ (\ref{H_gammas}).  These can be combined into an ordinary---though highly non-local---fermion,
\begin{equation}
  f = \frac{1}{2}(\gamma_{1}+i \gamma_{2}),
  \label{f}
\end{equation}
that costs zero energy and therefore produces a two-fold ground-state degeneracy.  In particular, if $|0\rangle$ is a ground state satisfying $f|0\rangle = 0$, then $|1\rangle \equiv f^\dagger |0\rangle$ is necessarily also a ground state (with opposite fermion parity).  Note the stark difference from conventional gapped superconductors, where typically there exists a unique ground state with even parity so that all electrons can form Cooper pairs.  

The appearance of localized zero-energy Majorana end-states and the associated ground-state degeneracy arise because the chain forms a topological phase while the vacuum bordering the chain is trivial.  (It may be helpful to imagine adding extra sites to the left and right of the chain, with $\mu < -t$ for those sites so that the strong pairing phase forms there.)  These phases cannot be smoothly connected, so the gap necessarily closes at the chain's boundaries.  Because this conclusion has a topological origin it is very general and does not rely on the particular fine-tuned limit considered above, with one caveat.  In the more general situation with $\mu \neq 0$ and $t \neq \Delta$ (but still in the topological phase) the Majorana zero-modes $\gamma_1$ and $\gamma_2$ are no longer simply given by $\gamma_{A,1}$ and $\gamma_{B,N}$; rather, their wavefunctions decay exponentially into the bulk of the chain.  The overlap of these wavefunctions results in a splitting of the degeneracy between $|0\rangle$ and $|1\rangle$ by an energy that scales like $e^{-L/\xi}$, where $L$ is the length of the chain and $\xi$ is the coherence length (which diverges at the transition to the trivial phase).  Provided $L \gg \xi$, however, this splitting can easily be negligible compared to all relevant energy scales in the problem; unless specified otherwise we will assume that this is the case and simply refer to the Majorana end-states as zero-energy modes despite this exponential splitting.  

Finally we comment on the importance of the fermions being spinless in Kitaev's toy model.  This property ensures that a single zero-energy Majorana mode resides at each end of the chain in its topological phase.  Suppose that instead spinful fermions---initially without spin-orbit interactions---formed a $p$-wave superconductor.  In this case spin merely doubles the degeneracy for every eigenstate of the Hamiltonian, so that when $|\mu|<t$ each end supports \emph{two} Majorana zero-modes, or equivalently one \emph{ordinary} fermionic zero-mode.  Unless special symmetries are present these ordinary fermionic states will move away from zero energy upon including perturbations such as spin-orbit coupling.  (Note that even for a spinless chain it is in principle possible for multiple nearby Majorana modes to coexist at zero energy if certain symmetries are present; see Refs.\ \onlinecite{FidkowskiKitaev1,FidkowskiKitaev2,Sri} for examples.  Time-reversal symmetry can also protect pairs of Majorana end-states in `class DIII' 1D superconductors with spin.\cite{RyuClassification,KitaevClassification,Wong})  

This by no means implies that it is impossible to experimentally realize Kitaev's toy model and the Majorana modes it supports with systems of electrons (which always carry spin).  Rather these considerations only imply that a prerequisite to observing isolated Majorana zero-modes is lifting Kramer's degeneracy such that the electron's spin degree of freedom becomes effectively `frozen out'.  We will discuss several ways of achieving this, as well as the requisite $p$-wave superconductivity, in Sec.\ \ref{1D_toy_model_realizations}.

\subsection{2D spinless $p+ip$ superconductor}
\label{2D_toy_model}

In two dimensions, the simplest system that realizes a topological phase supporting Majorana fermions is a spinless 2D electron gas exhibiting $p+ip$ superconductivity.  We will study the following model for such a system,
\begin{eqnarray}
  H &=& \int d^2{\bf r} \bigg{\{}\psi^\dagger \left(-\frac{\nabla^2}{2m}-\mu\right)\psi 
  \nonumber \\
  &+& \frac{\Delta}{2} \left[e^{i \phi}\psi(\partial_x +  i\partial_y)\psi + H.c.\right]\bigg{\}},
  \label{Hp+ip}
\end{eqnarray}
where $\psi^\dagger({\bf r})$ creates a spinless fermion with effective mass $m$, $\mu$ is the chemical potential, and $\Delta\geq 0$ determines the $p$-wave pairing amplitude while $\phi$ is the corresponding superconducting phase.  For the moment we take the superconducting order parameter to be uniform, though we relax this assumption later when discussing vortices.  To understand the physics of Eq.\ (\ref{Hp+ip}) we will adopt a similar strategy to that of the previous section---first identifying signatures of topological order encoded in bulk properties of the $p+ip$ superconductor, and then turning to consequences of the nontrivial topology for the boundaries of the system.  

\begin{figure}
\includegraphics[width = \linewidth]{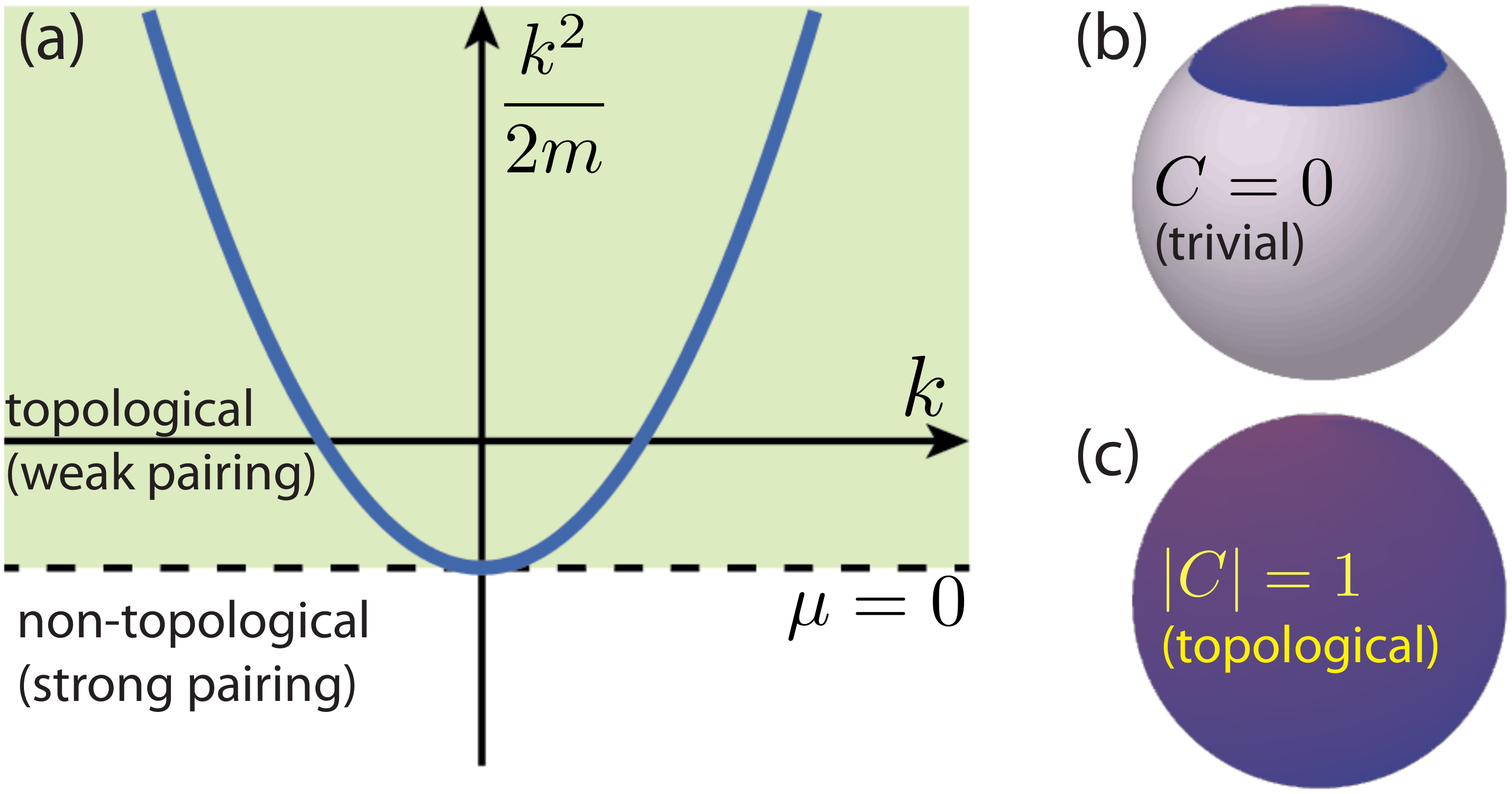}
\caption{(a) Kinetic energy for a spinless 2D electron gas exhibiting $p+ip$ superconductivity.  The pairing opens a bulk gap except when $\mu = 0$.  This gapless point marks the transition between a weak pairing topological phase at $\mu>0$ and a trivial strong pairing phase at $\mu < 0$.  These phases are distinguished by the Chern number $C$ which specifies how many times the map ${\bf \hat{h}}({\bf k})$ [derived from Eq.\ (\ref{hk_definition_2D})] covers the entire unit sphere as one sweeps over all momenta ${\bf k}$.  As $|{\bf k}|$ increases from zero, in the trivial phase ${\bf \hat{h}}({\bf k})$ covers the shaded area in (b) but then `uncovers' the same area, resulting in Chern number $C = 0$, whereas in the topological phase the map covers the entire unit sphere once as illustrated in (c) leading to $|C| = 1$.   }
\label{BandStructureFig2D}
\end{figure}

In a system with periodic boundary conditions along $x$ and $y$ (\emph{i.e.}, a superconductor on a torus with no edges) translation symmetry allows one to readily diagonalize Eq.\ (\ref{Hp+ip}) by going to momentum space.  Defining $\Psi({\bf k})^\dagger = [\psi^\dagger({\bf k}),~ \psi(-{\bf k})]$, one obtains
\begin{eqnarray}
  H &=& \frac{1}{2} \int \frac{d^2{\bf k}}{(2\pi)^2}\Psi^\dagger({\bf k}) \mathcal{H}({\bf k}) \Psi({\bf k}),
\nonumber \\
    \mathcal{H}({\bf k}) &=& \left( \begin{array}{cc}
  \epsilon(k) & \tilde \Delta({\bf k})^* \\
  \tilde \Delta({\bf k}) & -\epsilon(k) 
  \end{array}\right)
  \label {Hp+ip2} 
\end{eqnarray}
with $\epsilon(k) = \frac{k^2}{2m}-\mu$ and $\tilde\Delta({\bf k}) = i \Delta e^{i\phi}(k_x+ik_y)$.  
A canonical transformation of the form $a({\bf k}) = u({\bf k}) \psi({\bf k})+v({\bf k})\psi^\dagger(-{\bf k})$ diagonalizes the remaining $2\times 2$ matrix.  In terms of these quasiparticle operators the Hamiltonian reads 
\begin{equation}
  H = \int \frac{d^2{\bf k}}{(2\pi)^2}E_{\rm bulk}({\bf k}) a^\dagger({\bf k})a({\bf k}).
\end{equation}
The coherence factors $u({\bf k})$ and $v({\bf k})$ take the same form as in Eq.\ (\ref{uv}), and the bulk excitation energies are similarly given by
\begin{equation}
  E_{\rm bulk}({\bf k}) = \sqrt{\epsilon(k)^2+|\tilde\Delta({\bf k})|^2}.
\end{equation}
For any $\mu > 0$ the bulk is fully gapped since here the pairing field $\tilde\Delta({\bf k})$ is non-zero everywhere along the Fermi surface.  As one depletes the band the bulk gap decreases and eventually closes at $\mu = 0$, where the Fermi level resides precisely at the bottom of the band as shown in Fig.\ \ref{BandStructureFig2D}(a).  (The gap closure here arises because Pauli exclusion prohibits $p$-wave pairing at ${\bf k} = 0$.)  Further reducing $\mu$ reopens the gap, which remains finite for any $\mu < 0$.  

As in the 1D case the intuitively different $\mu > 0$ and $\mu < 0$ gapped regimes can be quantitatively distinguished by examining the ground-state wavefunction\cite{ReadGreen}, which can be written as
\begin{eqnarray}
  |{\rm g.s.}\rangle &\propto& \prod_{k_x \geq 0,k_y}[1+\varphi_{\rm C.p.}({\bf k})\psi(-{\bf k})^\dagger \psi({\bf k})^\dagger]|0\rangle
  \nonumber \\
  \varphi_{\rm C.p.}({\bf k}) &=& \frac{v({\bf k})}{u({\bf k})} = \left(\frac{E_{\rm bulk}-\epsilon}{\tilde\Delta}\right),
\end{eqnarray}
where $|0\rangle$ is a state with no $\psi({\bf k})$ fermions present with non-zero momentum.  The `Cooper pair wavefunction' $\varphi_{\rm C.p.}({\bf k})$ again encodes a key difference between the $\mu>0$ and $\mu < 0$ regimes.  In real space one finds the asymptotic forms\cite{ReadGreen}
\begin{eqnarray}
  |\varphi_{\rm C.p.}({\bf r})| \sim \left\{ \begin{array}{c}
 e^{-|{\bf r}|/\zeta},~~~~\mu<0~~~~ ({\rm strong ~pairing}) \\
 |{\bf r}|^{-1}, ~~~~\mu >0~~~~({\rm weak ~pairing}).
       \end{array} \right.
\end{eqnarray}
demonstrating that $\mu < 0$ corresponds to a `BEC-like' strong pairing regime, whereas with $\mu > 0$ a `BCS-like' weakly paired condensate forms from Cooper pairs loosely bound in space.

Also as in the 1D case, topology underlies the fact that the weak and strong pairing regimes constitute distinct phases that cannot be smoothly connected without closing the bulk gap.  To expose the topological invariant that distinguishes these phases, consider a 2D superconductor described by a Hamiltonian of the form\cite{VolovikChernNumber} 
\begin{equation}
  \mathcal{H}({\bf k}) = {\bf h}({\bf k})\cdot {\bm \sigma}
  \label{hk_definition_2D}
\end{equation}
with ${\bf h}({\bf k})$ a smooth function that is non-zero for all momenta so that the bulk is fully gapped.  One can then define a unit vector ${\bf \hat{h}}({\bf k})$ that maps 2D momentum space onto a unit sphere.  Assuming that ${\bf \hat{h}}({\bf k})$ tends to a unique vector as $|{\bf k}| \rightarrow \infty$ (independent of the direction of ${\bf k}$), the number of times this map covers the entire unit sphere defines an integer topological invariant given formally by the \emph{Chern number}
\begin{equation}
  C = \int \frac{d^2{\bf k}}{4\pi}[{\bf \hat{h}}\cdot(\partial_{k_x}{\bf \hat{h}}\times \partial_{k_y}{\bf \hat{h}})].
\end{equation}
The integrand above determines the solid angle (which can be positive or negative) that ${\bf \hat{h}}({\bf k})$ sweeps on the unit sphere over an infinitesimal patch of momentum space centered on ${\bf k}$.  Performing the integral over all ${\bf k}$ yields an integer that remains invariant under smooth deformations of ${\bf \hat{h}}({\bf k})$.  The Chern number can change only when the gap closes, making ${\bf \hat{h}}({\bf k})$ ill-defined at some momentum.  

Consider now the Hamiltonian in Eq.\ (\ref{Hp+ip2}) for which $h_x({\bf k}) = {\rm Re}[\tilde \Delta({\bf k})]$, $h_y({\bf k}) = {\rm Im}[\tilde \Delta({\bf k})]$, and $h_z({\bf k}) = \epsilon(k)$.  Notice that for momenta with fixed $|{\bf k}|$, $\hat{h}_x$ and $\hat{h}_y$ always sweep out a circle on the unit sphere at height $\hat{h}_z$.  As $|{\bf k}|$ increases from zero in the $\mu < 0$ strong pairing phase, $\hat{h}_z$ begins at the north pole, descends towards the equator, and then returns to the north pole as $|{\bf k}| \rightarrow \infty$.  Thus in the (topologically trivial) strong pairing phase ${\bf \hat{h}}({\bf k})$ initially sweeps out the shaded region in the northern hemisphere of Fig.\ \ref{BandStructureFig2D}(b) but then `unsweeps' the same area, resulting in a vanishing Chern number.  In contrast, for the (topologically nontrivial) $\mu > 0$ weak pairing phase $\hat{h}_z$ transitions from the south pole at ${\bf k} = 0$ to the north pole when $|{\bf k}| \rightarrow \infty$; the map ${\bf \hat{h}}({\bf k})$ therefore covers the entire unit the sphere exactly one time as shown schematically in Fig.\ \ref{BandStructureFig2D}(c), leading to a nontrivial Chern number $C = -1$.  [Note that other integer Chern numbers are also possible.  For instance, a $p-ip$ superconductor carries a Chern number $C = +1$ in the topological phase.  An $f$-wave superconductor with $\tilde \Delta({\bf k}) \propto (k_x + i k_y)^3$ provides a more nontrivial example.  In this case for momenta with fixed $|{\bf k}|$, $\hat{h}_x$ and $\hat{h}_y$ trace out a circle on the unit sphere \emph{three} times, yielding a Chern number $C = -3$ in the weak pairing phase (see, \emph{e.g.}, Ref.\ \onlinecite{HoleDopedMajoranas}).]

\begin{figure}
\includegraphics[width = 8cm]{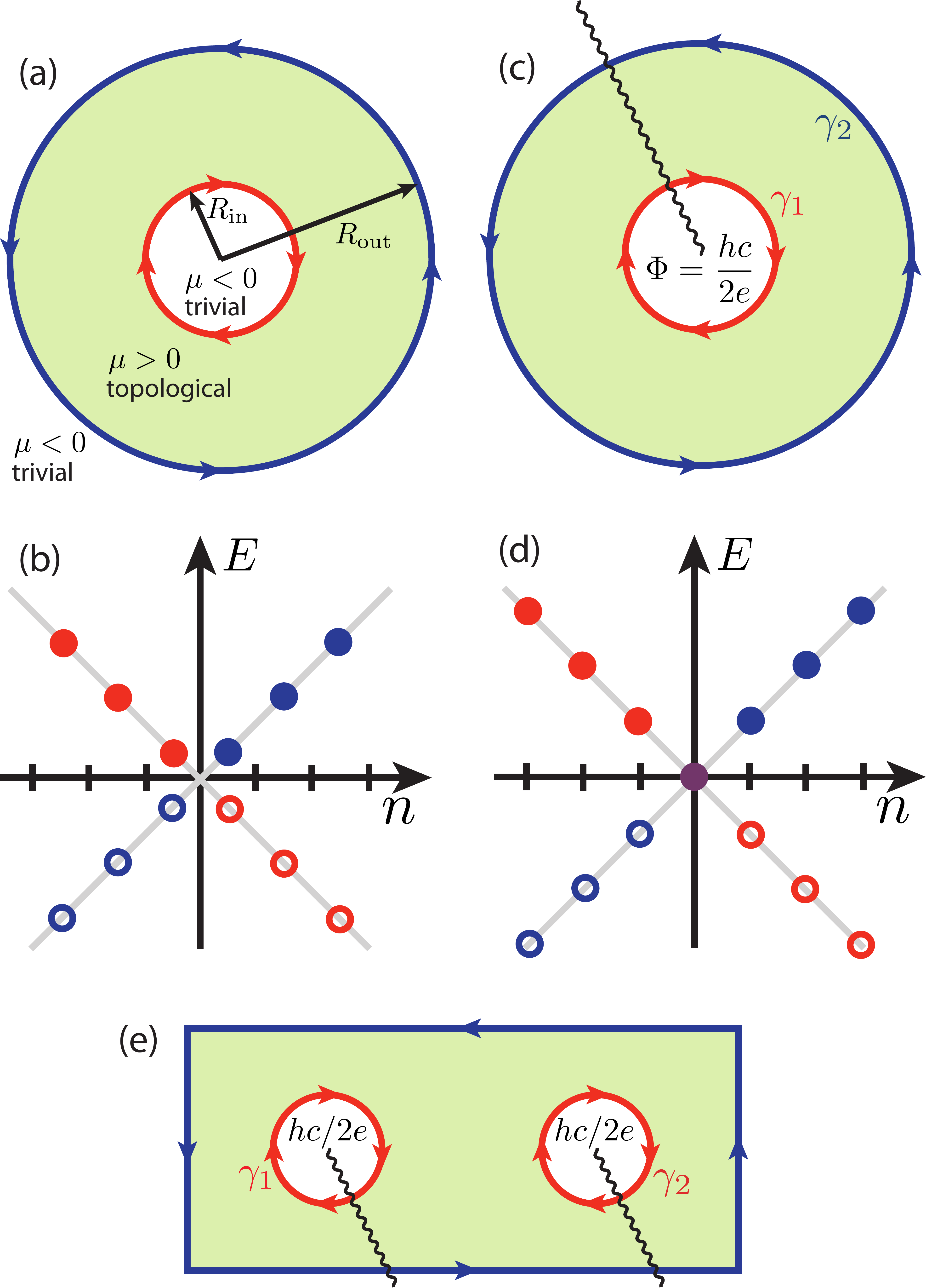}
\caption{(a) A topological $p+ip$ superconductor on an annulus supports chiral Majorana edge modes at its inner and outer boundaries.  (b) Energy spectrum versus angular momentum $n$ for the inner (red circles) and outer (blue circles) edge states in the setup from (a).  Here $n$ takes on half-integer values because the Majorana modes exhibit anti-periodic boundary conditions on the annulus.  An $\frac{hc}{2e}$ flux piercing the central trivial region as in (c) introduces a branch cut (wavy line) which, when crossed, leads to a sign change for the Majorana edge modes.  The flux therefore changes the boundary conditions to periodic and shifts $n$ to integer values.  This leads to the spectrum in (d), which includes Majorana zero-modes $\gamma_1$ and $\gamma_2$ localized at the inner and outer edges.  The two-vortex setup in (e) supports one Majorana zero-mode localized around each puncture, while the outer boundary remains gapped.}
\label{EdgeStates}
\end{figure}

We will now explore the physical consequences of the nontrivial Chern number uncovered in the topological weak pairing phase.  Consider the geometry of Fig.\ \ref{EdgeStates}(a), where a topological $p+ip$ superconductor occupies the annulus and a trivial phase forms elsewhere.  We will model this geometry by $H$ in Eq.\ (\ref{Hp+ip}) with a spatially dependent $\mu(r)$ that is positive inside the annulus and negative outside.  Since these regions realize topologically distinct phases one generically expects edge states at their interface, which we would like to now understand following various authors\cite{ReadGreen,StoneRoy,FendleyFisherNayak,GrosfeldStern}.  Focusing on low-energy edge modes and assuming that $\mu(r)$ is slowly varying, one can discard the $-\nabla^2/(2m)$ kinetic term in $H$.  A minimal Hamiltonian capturing the edge states can then be written in polar coordinates $(r,\theta)$ as
\begin{eqnarray}
  H_{\rm edge} &=& \int d^2{\bf r} \bigg{\{}-\mu(r)\psi^\dagger \psi 
  \nonumber \\
  &+&  \left[\frac{\Delta}{2}e^{i\phi}e^{i \theta}\psi\left(\partial_r +  \frac{i\partial_\theta}{r}\right)\psi + H.c.\right]\bigg{\}}.
  \label{Hp+ipPolar}
\end{eqnarray}
Because of the $e^{i\theta}$ factor above, the $p+ip$ pairing field couples states with orbital angular momentum quantum numbers of different magnitude.  In what follows it will be convenient to gauge this factor away by defining $\psi = e^{-i\theta/2}\psi'$.  (Note that $i\partial_\theta \rightarrow i\partial_\theta + 1/2$ under this change of variables, though the constant shift vanishes in the pairing term by Fermi statistics.)  Crucially, the new field $\psi'$ must exhibit \emph{anti-periodic} boundary conditions upon encircling the annulus.   

In terms of $\Psi'^\dagger({\bf r}) = [\psi'^\dagger({\bf r}),~ \psi'({\bf r})]$, the edge Hamiltonian becomes
\begin{eqnarray}
  H_{\rm edge} &=& \frac{1}{2} \int d^2{\bf r}\Psi'^\dagger({\bf r}) \mathcal{H}({\bf r}) \Psi'({\bf r}),
\nonumber \\
    \mathcal{H}({\bf r}) &=& \left( \begin{array}{cc}
  -\mu(r) & \Delta e^{-i\phi}(-\partial_r +\frac{i\partial_\theta}{r}) \\
  \Delta e^{i\phi}(\partial_r +\frac{i\partial_\theta}{r}) & \mu(r) 
  \end{array}\right).
  \label {Hedge} 
\end{eqnarray}
To find the edge state wavefunctions satisfying $\mathcal{H}({\bf r})\chi({\bf r}) = E\chi({\bf r})$, it is useful to parametrize $\chi({\bf r})$ as
\begin{eqnarray}
  \chi_n({\bf r})  =  e^{in\theta} \left( \begin{array}{c}
  e^{-i\phi/2}[f(r)+i g(r)]  \\
  e^{i\phi/2}[f(r)-i g(r)]  
  \end{array}\right),
\end{eqnarray}
where $n$ is a \emph{half-integer} angular momentum quantum number to ensure the proper anti-periodic boundary conditions.  The functions $f$ and $g$ obey
\begin{eqnarray}
  (E+n\Delta/r)f &=& -i[\mu(r) - \Delta\partial_r]g
  \nonumber \\
  (E-n\Delta/r )g &=& i[\mu(r) + \Delta\partial_r]f.
  \label{EdgeModeEqs}
\end{eqnarray}
For modes well-localized at the inner/outer annulus edges, it suffices to replace $r\rightarrow R_{\rm in/out}$ on the left-hand side of Eqs.\ (\ref{EdgeModeEqs}).  Within this approximation one finds that the energies of the outer edge states are 
\begin{equation}
  E_{\rm out} = \frac{n\Delta}{R_{\rm out}}, 
  \label{Eout}
\end{equation}
while the corresponding wavefunctions follow from $f = 0$ and $[\mu(r)-\Delta\partial_r]g = 0$.  The latter equations yield
\begin{eqnarray}
  \chi_n^{\rm out}({\bf r})  =  e^{in\theta}e^{\frac{1}{\Delta}\int_{R_{\rm out}}^r dr'\mu(r')} \left( \begin{array}{c}
  i e^{-i\phi/2} \\
  -i e^{i\phi/2}
  \end{array}\right),
  \label{chi_out}
\end{eqnarray}
which indeed describes modes exponentially localized around the outer edge.  Similarly, the inner-edge energies and wavefunctions are given by
\begin{eqnarray}
  E_{\rm in} &=& -\frac{n\Delta}{R_{\rm in}}
  \label{Ein} \\
  \chi_n^{\rm in}({\bf r})  &=&  e^{in\theta}e^{-\frac{1}{\Delta}\int_{R_{\rm in}}^r dr'\mu(r')} \left( \begin{array}{c}
  e^{-i\phi/2}  \\
  e^{i\phi/2}
  \end{array}\right).
  \label{chi_in}
\end{eqnarray}
Figure \ref{EdgeStates}(b) sketches the energies versus angular momentum $n$ for the inner (red circles) and outer (blue circles) edge states.  

These edge modes exhibit several remarkable features.  First, they are chiral---the inner modes propagate clockwise while the outer modes propagate counterclockwise, as is clear from Fig.\ \ref{EdgeStates}(b).  (For a $p-ip$ superconductor, the chiralities are reversed.)
While this is reminiscent of edge states found in the integer quantum Hall effect, there is an important distinction.  The edge states captured above correspond to 
\emph{chiral Majorana modes} which, roughly, comprise `half' of an integer quantum Hall edge state.  To be more precise let us expand $\Psi'({\bf r})$ in terms of edge-mode operators $\Gamma^{\rm in/out}_n$:
\begin{equation}
  \Psi'({\bf r}) = \sum_n [\chi^{\rm in}_n({\bf r})\Gamma^{\rm in}_n + \chi^{\rm out}_n({\bf r})\Gamma^{\rm out}_n].
  \label{EdgeModeExpansion}
\end{equation}
Since the upper and lower components of $\Psi'({\bf r})$ are related by Hermitian conjugation, Eqs.\ (\ref{chi_out}-\ref{EdgeModeExpansion}) imply that $\Gamma^{\rm in/out}_n = (\Gamma^{\rm in/out}_{-n})^\dagger$.  This property in turn implies that ($i$) only edge modes with energy $E\geq 0$ [solid circles in Fig.\ \ref{EdgeStates}(b)] are physically distinct, and $(ii)$ the real-space operators 
\begin{equation}
  \Gamma^{\rm in/out}(\theta) = \sum_n e^{i n\theta}\Gamma^{\rm in/out}_n =  [\Gamma^{\rm in/out}(\theta)]^\dagger
\end{equation}
are in fact Majorana fermions.  

While these chiral Majorana edge modes become gapless when the topological and trivial regions are thermodynamically large, in any finite system there remains a unique ground state.  This is a direct consequence of the anti-periodic boundary conditions on $\Psi'({\bf r})$ which led to half-integer values of $n$ and hence minimum edge-excitation energies of $\Delta/(2R_{\rm in/out})$.  The physics changes qualitatively when a flux quantum $\Phi = \frac{hc}{2e}$ threads the central trivial region as shown in Fig.\ \ref{EdgeStates}(c).  This flux induces a vortex in the superconducting pair field so that (say) $\Delta \rightarrow \Delta e^{-i\theta}$ in Eq.\ (\ref{Hp+ipPolar}).  The edge Hamiltonian in the presence of this vortex can be written in terms of our original fermion fields $\Psi^\dagger({\bf r}) = [\psi^\dagger({\bf r}),~ \psi({\bf r})]$ (which exhibit \emph{periodic} boundary conditions) as
\begin{equation}
  H^{\rm v}_{\rm edge} = \frac{1}{2} \int d^2{\bf r}\Psi^\dagger({\bf r}) \mathcal{H}({\bf r}) \Psi({\bf r})
\end{equation}
with $\mathcal{H}({\bf r})$ again given by Eq.\ (\ref{Hedge}).  The Hamiltonians with and without a vortex appear identical, so the edge-state energies and wavefunctions again take the form of Eqs.\ (\ref{Eout}-\ref{chi_in}), with one critical difference.  Since $\Psi({\bf r})$ exhibits periodic boundary conditions, the angular momentum quantum number $n$ now takes on \emph{integer} values.  The edge state spectrum sketched in Fig.\ \ref{EdgeStates}(d) then includes two zero-energy Majorana modes $\gamma_1$ and $\gamma_2$, one localized at each interface.  These Majorana zero-modes are the counterpart of the Majorana end-states discussed in Sec.\ \ref{1D_toy_model} and similarly result in a two-fold ground state degeneracy for the $p+ip$ superconductor.  (Technically, the edge-state wavefunctions overlap if the topological region is finite, splitting this ground-state degeneracy by an energy that is exponentially small in the width of the annulus.  Throughout we will neglect such a splitting unless specified otherwise.)

The shift in boundary conditions underlying the formation of Majorana zero-modes can be intuitively understood as follows.  First, note that sending $\psi\rightarrow e^{i\delta\phi/2}\psi$ is equivalent to changing the phase of the superconducting pair field by $\delta\phi$.  Thus a $\delta \phi = 2\pi$ shift in the superconducting phase, while irrelevant for Cooper pairs, effectively leads to a sign change for \emph{unpaired} fermions [such as the edge mode operators $\Gamma_n^{\rm in/out}$; see the wavefunctions in Eqs.\ (\ref{chi_out}) and (\ref{chi_in})].\cite{Ivanov}  To account for such sign changes it is useful to take the superconducting phase in the interval $[0,2\pi)$ and introduce branch cuts indicating where the phase jumps by $2\pi$.  The wavy line in Fig.\ \ref{EdgeStates}(c), for instance, represents the branch cut arising due to the $\frac{hc}{2e}$ flux.  A Majorana fermion crossing that branch cut acquires a minus sign, thereby changing the anti-periodic boundary conditions to periodic as we found above in our analytic solution.  This perspective is exceedingly valuable partly because it allows one to immediately deduce where Majorana zero-modes form even when an analytic treatment is unavailable.  In the two-vortex setup of Fig.\ \ref{EdgeStates}(e), for example, chiral Majorana edge states at the inner boundaries exhibit periodic boundary conditions and therefore host zero-modes, whereas the outer edge modes suffer anti-periodic boundary conditions and exhibit a finite-size gap.  Furthermore, this picture will prove essential for understanding interferometry experiments and non-Abelian statistics later in this review.

So far we have discussed chiral Majorana modes residing at fixed boundaries of a topological $p+ip$ superconductor.  An interface between topological and trivial regions can also form dynamically when a magnetic flux penetrates the bulk of a (type II) topological superconductor.  In this case the vortex core---which has a size of order the coherence length $\xi \sim v_F/(k_F\Delta)$, with $v_F$ and $k_F$ the Fermi velocity and momentum---forms the trivial region.  Adapting Eq.\ (\ref{Ein}) to this situation, the energies of the chiral Majorana modes bound to an $\frac{hc}{2e}$ vortex are given roughly by
\begin{equation}
  |E_{\rm vortex}| \sim \frac{|n|\Delta}{\xi}\sim \frac{|n|(k_F\Delta)^2}{E_F},
  \label{Evortex}
\end{equation}
where $k_F \Delta$ is the bulk gap, $E_F$ is the Fermi energy, and $n$ takes on integer values.\footnote{Actually the pairing field $\Delta$ rather than the chemical potential acquires spatial dependence near the core of a field-induced vortex, but the energies nevertheless still scale like $E_{\rm vortex}$ in Eq.\ (\ref{Evortex}).}  The spectrum of Eq.\ (\ref{Evortex}) reflects the $p+ip$ analog\cite{Kopnin} of Caroli-de Gennes-Matricon states\cite{Caroli} bound to vortices in $s$-wave superconductors.\cite{Ivanov}  Since $n$ is an integer the vortex binds a single Majorana zero-mode (unlike the $s$-wave case where all bound states have finite energy).  It is important to observe, however, that this zero-mode is separated by a `mini-gap' $E_{\rm mini-gap}\sim (k_F\Delta)^2/E_F$ from the next excited state.  In a `typical' superconductor $E_{\rm mini-gap}$ can easily be a thousand times smaller than the bulk gap, which can pose challenges for some of the proposals we will review later on.  In this regard, an appealing feature of the 1D $p$-wave superconductor discussed in Sec.\ \ref{1D_toy_model} is that there the Majorana zero-modes are generally separated from excited states by an energy comparable to the bulk gap.

Because we considered a spinless $p+ip$ superconductor above, each $\frac{hc}{2e}$ vortex threading a topological region binds a single localized Majorana zero-mode.  Remarkably, stable isolated Majorana zero-modes can also form in a \emph{spinful} 2D electron system exhibiting spin-triplet $p+ip$ superconductivity.  For such a superconductor the pairing term in Eq.\ (\ref{Hp+ip}) generalizes to\cite{VolovikBook}
\begin{equation}
  H_{\rm triplet} = \int d^2{\bf r}\frac{\Delta}{2}\left[e^{i\phi} \psi \sigma^y({\bf \hat{d}}\cdot {\bm \sigma})(\partial_x + i \partial_y)\psi + H.c.\right],
  \label{Htriplet}
\end{equation}
where $\psi_\alpha^\dagger({\bf r})$ creates an electron with spin $\alpha = \uparrow,\downarrow$ and spin indices are implicitly summed.  Note that $H_{\rm triplet}$ is invariant under arbitrary spin rotations about the ${\bf \hat{d}}$ direction, $\psi \rightarrow e^{i\frac{\theta}{2}{\bf \hat{d}}\cdot {\bm \sigma}}\psi$, but transforms nontrivially under all other spin rotations, reflecting the spin-triplet nature of Cooper pairs.  In the presence of an ordinary $\frac{hc}{2e}$ vortex, the superconducting phase $\phi$ rotates by $2\pi$ around the vortex core.  This vortex binds a pair of Majorana zero-modes (one for each electron spin) which generically hybridize and move to finite energy upon including spin-mixing perturbations such as spin-orbit coupling.  

The order parameter in Eq.\ (\ref{Htriplet}), however, supports additional stable topological defects.\cite{VolovikBook,HQVvolovik,HQVcross,SalomaaVolovik,Ivanov,SrRu}  This is tied to the fact that $H_{\rm triplet}$ is invariant under combined shifts of $\phi \rightarrow \phi + \pi$ and ${\bf \hat{d}}\rightarrow -{\bf \hat{d}}$, which allows for $\frac{hc}{4e}$ \emph{half quantum vortices} in which the superconducting phase $\phi$ and ${\bf \hat{d}}$ both rotate by $\pi$ around a vortex core.  As a concrete example, consider the order parameter configuration\cite{SrRu}
\begin{equation}
  e^{i \phi({\bf r})} = ie^{-i\theta/2},~~~~{\bf \hat{d}}({\bf r}) = \cos(\theta/2){\bf \hat{x}} + \sin(\theta/2){\bf \hat{y}}, 
\end{equation}
where $(r,\theta)$ are polar coordinates.  Inserting this form into Eq.\ (\ref{Htriplet}), one finds
\begin{eqnarray}
  H_{\rm triplet} &\rightarrow& \int d^2{\bf r}\frac{\Delta}{2}[\psi_\uparrow(\partial_x + i \partial_y) \psi_\uparrow 
  \nonumber \\
  &-&e^{-i\theta}\psi_\downarrow(\partial_x+i\partial_y)\psi_\downarrow + H.c.],
\end{eqnarray}
revealing a key feature of half quantum vortices---these defects are equivalent to configurations in which only one spin component `sees' an ordinary $\frac{hc}{2e}$ vortex.\cite{Ivanov}  Thus a half quantum vortex binds a single zero-energy Majorana mode, just as for vortices in the spinless $p+ip$ superconductor discussed earlier.  Typically, however, nucleating half quantum vortices costs more energy than ordinary $\frac{hc}{2e}$ vortices due to spin-orbit coupling, though clever routes of avoiding this outcome have been proposed\cite{HQVstabilization1,HQVstabilization2,SrRu,HQVstabilization3}.  In fact evidence of half quantum vortices in mesoscopic Sr$_2$RuO$_4$ samples was very recently reported experimentally\cite{SrRuO_HalfQuantumVortex} (see Sec.\ \ref{SrRuO}).    

Finally, we note in passing that it is also in principle possible for a time-reversal-invariant 2D superconductor to form such that one spin undergoes $p+ip$ pairing while its Kramer's partner exhibits $p-ip$ pairing\cite{TRItopologicalSC1,TRItopologicalSC2,TRItopologicalSC3,TRItopologicalSC4}. Provided time-reversal symmetry is present, such phases support stable counter-propagating chiral Majorana modes at the boundaries between topological and trivial regions.  These can be viewed as a superconducting analog of 2D topological insulators, where counter-propagating edge states formed by Kramer's pairs are similarly stable due to time-reversal symmetry\cite{KaneMele}.

\section{Practical realizations of Majorana modes in 1D $p$-wave superconductors}
\label{1D_toy_model_realizations}

\subsection{Preliminary Remarks}
\label{Preliminaries1D}

We will now survey several ingenious schemes that have been proposed to realize Majorana fermions in topological phases similar to that of Kitaev's model for a 1D spinless $p$-wave superconductor reviewed in Sec.\ \ref{1D_toy_model}.  To put the problem in perspective, it is useful to highlight the basic challenges involved in realizing Kitaev's model experimentally.  First, there is a `fermion doubling problem' of sorts that must be overcome---since electrons carry spin-1/2 one must freeze out half of the degrees of freedom so that the 1D system appears effectively `spinless'.  Stabilizing $p$-wave superconductivity for such a `spinless' system poses a still more serious challenge.  Not only are $p$-wave superconductors exceedingly rare in nature, but an attractively interacting 1D electron system that conserves particle number can at best exhibit power-law superconducting correlations in contrast to the long-range ordered superconductivity assumed in Kitaev's model.  (Remarkably, power-law superconducting order \emph{can} be sufficient to stabilize Majorana modes\cite{InteractionsMeng,InteractionsSau,InteractionsFidkowski}, though the splitting of the degenerate ground states in such cases scales as a power-law of the system size rather than exponentially.)  The proposals we review below employ the same three basic ingredients to cleverly overcome these challenges: superconducting proximity effects, time-reversal symmetry breaking, and spin-orbit coupling.  

The essence of the first ingredient is that a 1D system can \emph{inherit} Cooper pairing from a nearby long-range-ordered superconductor.  Fluctuations of the resulting superconducting order parameter for the 1D system are largely controlled by the parent bulk superconductor, and can thus remain unimportant even at finite temperature despite the low dimensionality of the parasitic material.  Since superconducting proximity effects are central to much of this review, we will digress briefly to elaborate on the physics in greater detail.  Consider for the moment some 1D electron system with a Hamiltonian of the form
\begin{equation}
  H_{1D} = \int \frac{dk}{2\pi}\psi^\dagger_k \mathcal{H}_k\psi_k
\end{equation}
and a conventional bulk $s$-wave superconductor described by 
\begin{eqnarray}
  H_{SC} = \int \frac{d^3{\bf k}}{(2\pi)^3}[\epsilon_{sc}(k)\eta^\dagger_{\bf k}\eta_{\bf k}
  + \Delta_{sc}(\eta_{\uparrow{\bf k}} \eta_{\downarrow-{\bf k}} + H.c.)].
\end{eqnarray}
Here $\psi^\dagger_{\sigma k}$ and $\eta^\dagger_{\sigma {\bf k}}$ add electrons with spin $\sigma$ to the 1D system and superconductor, respectively, while $\epsilon_{sc}(k) = k^2/(2m_{sc})-\mu_{sc}$ and $\Delta_{sc}$ are the superconductor's kinetic energy and pairing amplitude.  When the 1D system is brought into intimate contact with the superconductor [as in Fig.\ \ref{1D_wires_fig}(a)], the resulting structure can be described by
\begin{equation}
  H = H_{1D} + H_{SC} + H_\Gamma, 
  \label{H_proximity_effect}
\end{equation}
where $H_\Gamma$ encodes single-electron tunneling between the two subsystems with amplitude $\Gamma$.  Taking the 1D system to lie along the line $(x,y,z) = (x,0,0)$, one can explicitly write
\begin{equation}
  H_\Gamma = -\Gamma \int dx[\psi^\dagger_x \eta_{(x,0,0)} + H.c.].
\end{equation}

The effect of the hybridization term $H_\Gamma$ can be crudely deduced using perturbative arguments and dimensional analysis.  Suppose that the superconductor's Fermi wavevector $k_F^{sc}$ greatly exceeds that of the 1D system.  Intuitively, in this regime (which is relevant for all of the setups of interest) the hybridization between the two subsystems should be primarily controlled by $\Gamma$ and properties of the superconductor.  When $\Gamma k_{F}^{sc} \ll \Delta_{sc}$, it suffices to treat $H_\Gamma$ perturbatively since in this limit single electron tunneling is strongly suppressed due to the parent superconductor's gap.  At second order one generates an effective \emph{Cooper-pair} hopping term which, using dimensional analysis, takes the form
\begin{equation}
  \delta H \propto \frac{ \Gamma^2}{k_F^{sc}\Delta_{sc}}\int dx \left(\psi_{\uparrow x}\psi_{\downarrow x} \eta^\dagger_{\downarrow (x,0,0)}\eta^\dagger_{\uparrow (x,0,0)} + H.c.\right).
\end{equation}
At low energies one can replace $\eta^\dagger_{\downarrow}\eta^\dagger_{\uparrow} \rightarrow \langle \eta^\dagger_{\downarrow}\eta^\dagger_{\uparrow}\rangle \propto \rho_{sc}\Delta_{sc}$, where the brackets denote a ground state expectation value and $\rho_{sc}$ is the superconductor's density of states at the Fermi level.  In this way one arrives at the following effective Hamiltonian for the 1D system, 
\begin{eqnarray}
  H_{\rm eff} &=& H_{1D} + H_\Delta
  \nonumber \\
  H_\Delta &=& \Delta \int dx \left(\psi_{\uparrow x}\psi_{\downarrow x} + H.c.\right),
\end{eqnarray}
with $\Delta \propto \frac{ \Gamma^2}{k_F^{sc}\Delta_{sc}}(\rho_{sc}\Delta_{sc}) \propto \rho_{2D}\Gamma^2$ and $\rho_{2D} = m_{sc}/(2\pi)$ the superconductor's 2D density of states at $k_x = 0$.  

The treatment above captures a simple effective Hamiltonian for the 1D system that incorporates proximity-induced pairing.  Similar models appear frequently in the literature and will be employed often here as well.  Several authors have, however, emphasized the need to treat the proximity effect more rigorously to obtain a quantitative understanding of the devices we will explore below\cite{StanescuProximityEffect,SauProximityEffect,SauReview,AnnicaProximityEffect,LababidiProximityEffect,TewariProximityEffect,Disorder4,KhaymovichProximityEffect,GreinProximityEffect,SauGapEstimates}.  A more accurate way forward involves constructing the Euclidean action corresponding to $H$ in Eq.\ (\ref{H_proximity_effect}) and then integrating out the parent superconductor's degrees of freedom.  Appendix \ref{EffectiveActionDerivation} sketches the calculation and yields the following effective action for the 1D system,
\begin{eqnarray}
  S_{\rm eff} &=& \int \frac{d\omega}{2\pi}\frac{dk}{2\pi} Z^{-1}(\omega)\{\psi^\dagger_{(k,\omega)}[-i\omega + Z(\omega)\mathcal{H}_k]\psi_{(k,\omega)}
  \nonumber \\
  &+& \Delta_{sc}[1-Z(\omega)][\psi_{\uparrow (k,\omega)} \psi_{\downarrow (-k,-\omega)} + H.c.]\}
  \label{Seff1D}
\end{eqnarray}
As in our perturbative analysis, an effective Cooper pairing term (now frequency dependent) once again appears.  This more rigorous procedure, however, reveals that the tunneling $\Gamma$ also generates a reduced quasiparticle weight $Z(\omega)$ for electrons in the 1D system given approximately by\cite{Disorder4}
\begin{equation}
  Z(\omega) \approx \left[1 + \frac{\pi \rho_{2D} \Gamma^2}{\sqrt{\omega^2 + \Delta_{sc}^2}}\right]^{-1}.
  \label{Z}
\end{equation}
The physics underlying Eqs.\ (\ref{Seff1D}) and (\ref{Z}) is that by enhancing $\Gamma$ the wavefunctions for electrons in the 1D system bleed farther into the parent superconductor, thereby reducing their quasiparticle weight $Z(\omega)$ and \emph{enhancing} the pairing amplitude that they inherit [which can reach a maximum of $\Delta_{sc}$ as $Z(\omega) \rightarrow 0$].  The reduced quasiparticle weight also, however, effectively rescales the original Hamiltonian $\mathcal{H}_k$ and diminishes the energy scales intrinsic to the 1D system.\cite{Disorder4}  [Poles in the electron Green's function follow from $Z(\omega) \mathcal{H}_k$, rather than $\mathcal{H}_k$.]  In other words, in an effective 1D description of the hybrid structure, parameters such as spin-orbit coupling, Zeeman splitting, \emph{etc}.\ do not take on the values one would measure in the absence of the superconductor, but rather are renormalized \emph{downward} due to the hybridization.  This aspect of the proximity effect is often neglected, but as we will see later can lead to important and counterintuitive consequences.  We should note that even at this level the modeling of the proximity effect remains rather crude.  More sophisticated treatments where one treats the pairing self-consistently are also possible\cite{AnnicaProximityEffect,LababidiProximityEffect,TewariProximityEffect} but will not be discussed here.  

Remarkably, most proposals for engineering Kitaev's model for a 1D spinless $p$-wave superconductor in fact exploit proximity effects with ordinary $s$-wave superconductors like we treated above.  (It is hard to overemphasize the importance of this feature insofar as experimental prospects are concerned, given the many thousands of known $s$-wave superconductors.)  While this naively appears somewhat paradoxical, spin-orbit coupling---typically in conjuction with time-reversal symmetry breaking---can \emph{effectively} convert such a 1D system into a $p$-wave superconductor.  We will now explore a variety of settings in which such a mechanism appears.

\subsection{2D Topological Insulators}
\label{2DTIproposal}

In 2005, a revolution in our understanding of a seemingly well-understood phase of matter---the band insulator---began to emerge\cite{KaneMele,KaneReview,MooreReview,QiReview}.  It is now appreciated that such states need not be trivial in the sense of having no available low-energy degrees of freedom at zero temperature.  Rather, there exists a class of \emph{topological} band insulators that while inert in the bulk necessarily possess novel conducting states at their boundary.  These topological phases can appear in either two- or three-dimensional crystals and, remarkably, merely require appreciable spin-orbit coupling and time-reversal symmetry.  The numerous fascinating developments that grew out of the discovery of topological insulators include Fu and Kane's pioneering proposals\cite{FuKane,MajoranaQSHedge} for generating Majorana fermions at their edges (in 2D crystals) or surfaces (in 3D).  In this section we will describe how one can engineer a topological superconducting state similar to that of Kitaev's model using the edge of a 2D topological insulator; the 3D case will be reviewed in Sec.\ \ref{3DTIproposal}.  (See also Sec.\ \ref{3DTInanowires} for a proposal involving nanowires built from 3D topological insulators.)

\begin{figure}
\includegraphics[width = 8cm]{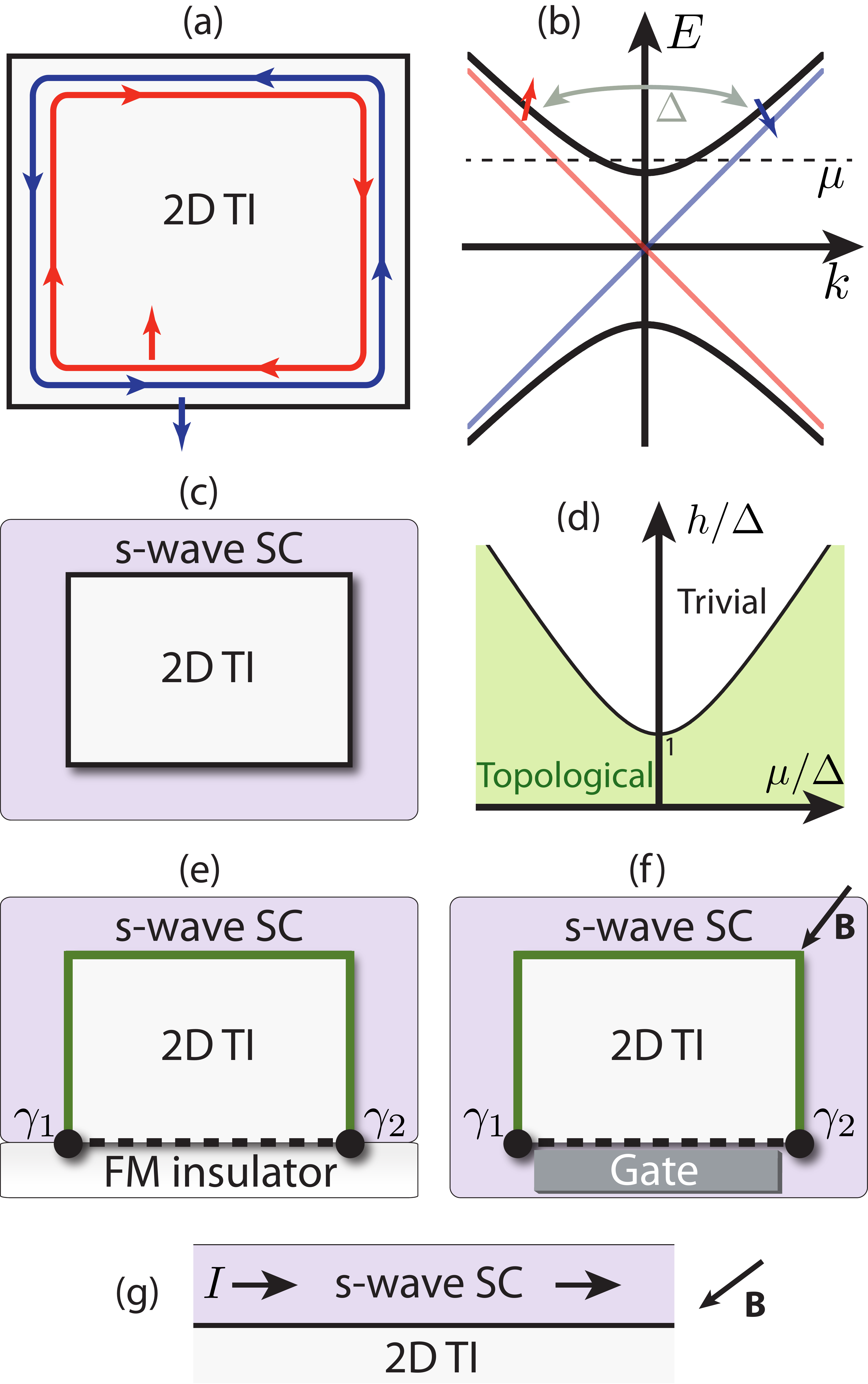}
\caption{(a) Schematic of counter-propagating, spin-filtered edge states in a 2D topological insulator.  (b) Edge-state dispersion when time-reversal symmetry is present (red and blue lines) and with a Zeeman field $h$ of the form in Eq.\ (\ref{HTIedge3}) (solid curves).  (c) A proximate $s$-wave superconductor drives the edge into a topological phase similar to the weak-pairing phase in Kitaev's toy model for a 1D spinless $p$-wave superconductor.  When the Zeeman field $h$ is present, the topological phase survives provided $h < \sqrt{\Delta^2 + \mu^2}$ leading to the phase diagram in (d).  Domain walls between topological (green lines) and trivial regions (dashed lines) on the edge trap localized Majorana zero-modes.  As described in the text these can be created with (e) a ferromagnetic insulator, (f) a Zeeman field combined with electrostatic gating, or (g) applying supercurrents near the edge.}
\label{2D_TI_fig}
\end{figure}

The hallmark of a 2D topological insulator is the presence of counter-propagating, spin-filtered edge states that are connected by time-reversal symmetry.  In an oversimplified picture that is adequate for our purposes, one can envision spin up electrons propagating clockwise around the edge while their Kramer's partners with spin down circulate counterclockwise as shown in Fig.\ \ref{2D_TI_fig}(a).  These low-energy edge modes can be described by the Hamiltonian
\begin{equation}
  H_{\rm 2DTI} = \int dx \psi^\dagger(-i v\partial_x \sigma^z-\mu)\psi,
  \label{HTIedge}
\end{equation}
where $v$ is the edge-state velocity, $\mu$ is the chemical potential, and $\psi^\dagger_{\sigma x}$ adds an electron with spin $\sigma$ at position $x$ along the edge.  The blue and red lines of Fig.\ \ref{2D_TI_fig}(b) sketch their dispersion.  Provided time-reversal symmetry is preserved (elastic) backscattering between the counter-propagating edge states is prohibited even in the presence of strong non-magnetic disorder.  Consequently these modes are robust against localization that plagues conventional 1D systems.  Furthermore, by focusing on these edge states one immediately beats the fermion doubling problem noted earlier---the spectrum supports only a single pair of Fermi points as long as the Fermi level does not intersect the bulk bands, and in this sense the system appears `spinless'.  

Realizing topological superconductivity then simply requires gapping out the edge via Cooper pairing.  Since the counter-propagating edge modes carry opposite spins, this can be achieved by interfacing the topological insulator with an ordinary $s$-wave superconductor\cite{MajoranaQSHedge}; see Fig.\ \ref{2D_TI_fig}(c).  As discussed in Sec.\ \ref{Preliminaries1D} the superconducting proximity effect on the edge can be crudely modeled with a Hamiltonian 
\begin{eqnarray}
  H &=& H_{\rm 2DTI} + H_\Delta
   \label{HTIedge2} \\
  H_\Delta &=& \int dx \Delta(\psi_\uparrow \psi_\downarrow + H.c.),
\end{eqnarray}
where $\Delta$ is the pairing amplitude inherited from the nearby superconductor.  Equation (\ref{HTIedge2}) yields quasiparticle energies
\begin{equation}
  E_\pm(k) = \sqrt{(\pm v k-\mu)^2 + \Delta^2},
  \label{edge_quasiparticle_energies}
\end{equation}
with $k$ the momentum, and describes a gapped topological superconductor that is a time-reversal-symmetric relative of the weak-pairing phase in Kitaev's toy model.\cite{MajoranaQSHedge}  Let us now make this connection more precise and elucidate how the spin-singlet pairing $\Delta$ mediates $p$-wave superconductivity.

To this end it is instructive to violate time-reversal symmetry by introducing a Zeeman field that cants the spins away from the $z$ direction:
\begin{eqnarray}
  H' &=& H_{\rm 2DTI} + H_Z + H_\Delta
  \label{HTIedge3} \\
  H_Z &=& -h \int dx \psi^\dagger \sigma^x \psi
  \label{H_B}  
\end{eqnarray}
with $h\geq 0$ the Zeeman energy.  When $\Delta = 0$ the edge-state spectrum becomes $\epsilon_\pm(k) = -\mu \pm \sqrt{(v k)^2 + h^2}$, and as shown by the solid black lines in Fig.\ \ref{2D_TI_fig}(b) exhibits a gap at $k = 0$ due to the broken time-reversal symmetry.  To understand the influence of proximity-induced pairing, we first note that the effect of $\Delta$ is obscured by the fact that Eq.\ (\ref{HTIedge3}) contains a standard spin-singlet pairing term but an unconventional kinetic energy form (due to the interplay of spin-momentum locking and the field).  The physics becomes much more transparent upon expressing $H'$ in terms of operators $\psi_\pm^\dagger(k)$ that add electrons with energy $\epsilon_\pm(k)$ to the edge.  In this basis $H'$ reads
\begin{eqnarray}
  H' &=& \int \frac{dk}{2\pi}\big{\{}\epsilon_+(k)\psi_+^\dagger(k)\psi_+(k) + \epsilon_-(k)\psi_-^\dagger(k)\psi_-(k)
  \nonumber \\
  &+& \frac{\Delta_p(k)}{2}[\psi_+(-k)\psi_+(k) + \psi_-(-k)\psi_-(k) + H.c.]
  \nonumber \\
  &+& \Delta_s(k)[\psi_-(-k)\psi_+(k) + H.c.]\big{\}},
  \label{HTIedge4}
\end{eqnarray}
where the pairing functions are 
\begin{equation}
  \Delta_p(k) = \frac{v k\Delta}{\sqrt{(v k)^2 + h^2}},~~~\Delta_s(k) = \frac{h\Delta}{\sqrt{(v k)^2 + h^2}}.
  \label{pairing_functions}
\end{equation}
The first line of Eq.\ (\ref{HTIedge4}) simply describes the band energies while the third captures interband $s$-wave pairing.  Most importantly, the second line encodes \emph{intraband $p$-wave pairing}.  This emerges because, as shown schematically in Fig.\ \ref{2D_TI_fig}(b), electrons at $k$ and $-k$ in a given band have misaligned spins and can thus form  Cooper pairs in response to $\Delta$.  By Fermi statistics, the effective potential $\Delta_p(k)$ that pairs these electrons must exhibit \emph{odd} parity since they derive from the same band.  (Physically, the odd parity reflects the fact that the electron spins rotate as one sweeps the momentum from $k$ to $-k$.)  This is the first of many instances we will encounter in which an $s$-wave order parameter effectively generates $p$-wave pairing by virtue of spin-orbit coupling.

The connection to Kitaev's model becomes explicit in the limit where $h \gg \Delta$ and $\mu$ resides near the bottom of the upper band as in Fig.\ \ref{2D_TI_fig}(b).  In this case the lower band plays essentially no role and can be projected away by simply sending $\psi_-\rightarrow 0$.  Furthermore, only momenta near $k= 0$ are important here so it suffices to expand $\epsilon_+(k)\approx -(\mu-h) + \frac{v^2}{2h}k^2 \equiv -\mu_{\rm eff} + k^2/(2m_{\rm eff})$ and $\Delta_p(k) \approx \frac{v\Delta}{h}k \equiv \Delta_{\rm eff}k$.  With these approximations, one obtains an effective Hamiltonian that in real space reads
\begin{eqnarray}
  H_{\rm eff} &=& \int dx \bigg{[}\psi_+^\dagger\left(-\frac{\partial_x^2}{2m_{\rm eff}} -\mu_{\rm eff}\right)\psi_+ 
  \nonumber \\
  &+& \frac{\Delta_{\rm eff}}{2}(-\psi_+ i\partial_x\psi_+ + H.c.)\bigg{]},
  \label{HTIedge5}
\end{eqnarray}
which describes Kitaev's model for a 1D spinless $p$-wave superconductor in the low-density limit [\emph{i.e.}, near $\mu = -t$ in Fig.\ \ref{BandStructureFigKitaevModel}(a)].  A similar mapping can be implemented for $\mu$ near the top of the lower band in Fig.\ \ref{2D_TI_fig}(b).  These considerations show that for $h \gg \Delta$, the edge forms a trivial strong pairing phase when $|\mu| \lesssim h$ and a topological weak pairing phase at $|\mu| \gtrsim h$.  

A more accurate phase diagram valid at any $h,\Delta$ can be deduced by studying the unprojected Hamiltonian in Eq.\ (\ref{HTIedge4}), which yields quasiparticle energies 
\begin{eqnarray}
  E_\pm'(k) = \sqrt{\Delta^2 + \frac{\epsilon_+^2 + \epsilon_-^2}{2} \pm (\epsilon_+ -\epsilon_-)\sqrt{\Delta_s^2 + \mu^2}}.
  \label{Epm}
\end{eqnarray}
The quasiparticle gap extracted from Eq.\ (\ref{Epm}) vanishes only when $h^2 = \Delta^2 + \mu^2$.  Matching onto the $h \gg \Delta$ results derived above, we then conclude that the edge forms a topological superconductor provided 
\begin{equation}
  h < \sqrt{\Delta^2 + \mu^2}~~({\rm topological~criterion}).  
  \label{topological_criterion}
\end{equation}
Physically, the edge forms a topological phase if superconductivity dominates the gap but a trivial phase if the gap is driven by time-reversal symmetry breaking.  Figure \ref{2D_TI_fig}(d) illustrates the resulting phase diagram.  Note that topological superconductivity persists even in the time-reversal-symmetric limit with $h = 0$; this has important physical consequences as we discuss shortly.  We also note that electrons on the edge are additionally subject to Coulomb repulsion, which have dramatic consequences in 1D and have so far been neglected.  References \onlinecite{MajoranaInteractions} and \onlinecite{MajoranaInteractionsQSHedge} find that while strong interactions (with a Luttinger parameter $g < 1/2$) can destroy the topological phase, milder repulsion ($1/2 < g < 1$) leaves the phase diagram of Fig.\ \ref{2D_TI_fig}(d) qualitatively intact.

Thus far we have only shown how to utilize the edge to construct a 1D topological superconductor on a ring, with no ends.  Stabilizing localized Majorana zero-modes requires introducing a domain wall between gapped topological and trivial phases on the edge\cite{MajoranaQSHedge}.  The setups of Figs.\ \ref{2D_TI_fig}(e) and (f) trap Majorana modes $\gamma_{1,2}$ by gapping three sides with superconductivity and the fourth with a Zeeman field $h$ of the form in Eq.\ (\ref{H_B}).  Topological and trivial regions are respectively indicated by green and dashed lines in these figures.  In (e) electrons on the lower edge `inherit' the Zeeman field via a proximity effect with a ferromagnetic insulator, just as a pairing field $\Delta$ is inherited from a superconductor.\cite{SauReview}  Note that the chemical potential for the bottom edge must reside within the field-induced spectral gap [recall Fig.\ \ref{2D_TI_fig}(b)]; otherwise that region remains gapless despite the broken time-reversal.  In (f) both superconductivity and the Zeeman field $h$ are uniformly generated on all four edges, the latter by applying a magnetic field.  Provided $h > \Delta$, the topological and trivial regions form simply by adjusting the chemical potential $\mu$ via gating so that $h < \sqrt{\Delta^2 + \mu^2}$ on three sides while $h > \sqrt{\Delta^2 + \mu^2}$ on the fourth.\cite{Nate}  We emphasize here that one can simultaneously have $h > \Delta$ \emph{and} still be well below the superconductor's critical field since the Zeeman energy for the edge can greatly exceed that in the superconductor due to spin-orbit enhancement of the $g$-factor\cite{WinklerBook}.  

Majorana zero-modes can also be trapped by selectively driving supercurrents near the edge of the sample\cite{Romito}.  To understand the underlying principle, let us revisit the Hamiltonian in Eq.\ (\ref{HTIedge3}) when a supercurrent $I$ flows as in Fig.\ \ref{2D_TI_fig}(g).  The current generates a phase twist in the superconducting order parameter so that $H_\Delta$ becomes
\begin{equation}
  H_\Delta \rightarrow \int dx \Delta[e^{i\phi(x)}\psi_\uparrow\psi_\downarrow + H.c.],
\end{equation}
with $I \propto \partial_x \phi(x)$.  It is convenient to gauge away the phase factor above by sending $\psi_\sigma \rightarrow e^{-i\phi(x)/2}\psi_\sigma$; defining $h_z \equiv v \partial_x \phi/2$, the full Hamiltonian then reads
\begin{eqnarray}
  H' &\rightarrow& \int dx\big{\{} \psi^\dagger\left[-(iv \partial_x+h_z)\sigma^z  - \mu -h\sigma^x\right]\psi
  \nonumber \\
   &+& \Delta(\psi_\uparrow\psi_\downarrow + H.c.)\big{\}}.
   \label{HwithCurrent}
\end{eqnarray}
Equation (\ref{HwithCurrent}) shows that the supercurrent mimics the effect of a Zeeman field $h_z$ directed along the $z$ direction.  Contrary to $h$, this does not open a gap but rather breaks the resonance between electrons with momentum $k$ and $-k$, thereby suppressing their ability to Cooper pair.  Suppose now that $|\mu| < h < \sqrt{\Delta^2 + \mu^2}$.  When $I = 0$ the edge then forms a topological phase where $\Delta$ dominates the gap.  At large $I$ (such that $h_z \gg \Delta$), however, the pair-breaking effect of $h_z$ essentially kills $\Delta$ and the edge forms a trivial state with a gap arising from $h$.  Supercurrents therefore allow one to turn a topological portion of the edge into a trivial state, similar to the gate in Fig.\ \ref{2D_TI_fig}(f), providing yet another means for stabilizing Majorana-carrying domain walls.

In our view 2D topological insulators hold great promise as a potential venue for Majorana fermions, particularly in the long term.  For one, their reduced dimensionality should allow for bulk carriers---which usually bedevil 3D topological insulators---to be removed relatively easily by electrostatic gating.  
The topological superconducting phase hosted by the edge also exhibits several remarkable features.  First, this phase is `easy' to access in the sense that its appearance requires the chemical potential $\mu$ to satisfy the inequality in Eq.\ (\ref{topological_criterion}) while not intersecting the bulk bands; this chemical potential window is therefore limited by the bulk gap for the topological insulator which can reach the $\sim 0.1$eV scale (see, \emph{e.g.}, Ref.\ \onlinecite{OxideInterfaceTI}).  By contrast the trivial gapped state requires positioning $\mu$ inside of the Zeeman-induced gap of Fig.\ \ref{2D_TI_fig}(b), which likely requires greater care.  While ultimately the ability to access both kinds of states is essential, the comparative ease for forming the topological phase greatly facilitates the Josephson-based Majorana detection schemes discussed in Sec.\ \ref{Fractional_Josephson_effect}.  More strikingly, as a consequence of Anderson's theorem the gap protecting the time-reversal-invariant topological superconductor that forms when $h = 0$ is \emph{unaffected} by non-magnetic disorder\cite{AndersonTheorem,Disorder4}.  We emphasize that one needn't work at $h = 0$ to enjoy this protection: with $h \neq 0$ but $\mu$ far from the Zeeman-induced gap, electrons near the Fermi energy are weakly perturbed by the field and hence `almost' obey Anderson's theorem\cite{Disorder4}.

A final noteworthy feature pertains to how large the topological superconductor's gap can be in principle.  Addressing this question requires the more rigorous treatment of the proximity effect discussed in Sec.\ \ref{Preliminaries1D}.  Recall that increasing the tunneling $\Gamma$ between the superconductor and topological insulator enhances $\Delta$ but reduces the energy scales intrinsic to the edge.  When $h = 0$ [such as in the setup of Fig.\ \ref{2D_TI_fig}(e)] it follows from Eq.\ (\ref{edge_quasiparticle_energies}) that the gap is simply $E_{\rm gap} = \Delta$, which is independent of the quantities $v,\mu$ that $\Gamma$ suppresses.  Thus in this case the gap increases monotonically with $\Gamma$,  reaching a maximum of $\Delta_{sc}$ for the parent superconductor.\cite{SauProximityEffect,Disorder4}  In other words, it is in principle possible for the edge to inherit the \emph{full pairing gap} exhibited by the parent superconductor.  This issue becomes subtler in setups such as Fig.\ \ref{2D_TI_fig}(f) where $h \neq 0$, for in this case increasing $\Gamma$ supresses the Zeeman-induced gap in the spectrum, making it more difficult to stabilize the trivial phase to trap Majoranas.  How large a hybridization is desirable then depends on details such as the tolerable fields one can apply, sample purity, \emph{etc}.  

Despite these virtues this platform for Majorana fermions faces the hurdle that experimental progress on 2D topological insulators has to date remained rather limited.  Though numerous candidate materials have been proposed\cite{KaneMele,BismuthQSH,BHZ,QuantumWellQSH,SiliceneTI,GrapheneAdatoms,OxideInterfaceTI} only predictions for HgTe have so far been confirmed experimentally\cite{Konig,EdgeTransportHgTe}.  Some evidence for a topological insulator phase in InAs/GaSb quantum wells also appeared recently\cite{InAsGaSbExpt1,InAsGaSbExpt2}, though the signatures are less clear cut due to persistence of bulk carriers in the samples.  The situation, however, already shows signs of improvement---very recently topological insulator behavior in HgTe has been independently confirmed by the Yacoby group, and experimental efforts to introduce a proximity effect at the edge are underway.  It will be very interesting to see how this avenue progresses in the near future.

\subsection{Conventional 1D wires}
\label{1Dwiresproposal}

\begin{figure*}
\includegraphics[width = 16cm]{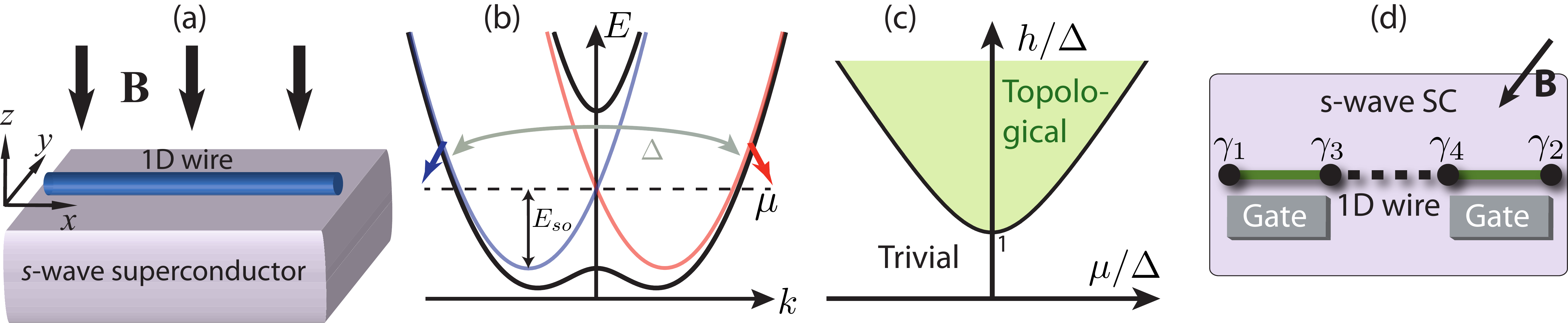}
\caption{(a) Basic architecture required to stabilize a topological superconducting state in a 1D spin-orbit-coupled wire.  (b) Band structure for the wire when time-reversal symmetry is present (red and blue curves) and broken by a magnetic field (black curves).  When the chemical potential lies within the field-induced gap at $k = 0$, the wire appears `spinless'.  Incorporating the pairing induced by the proximate superconductor leads to the phase diagram in (c).  The endpoints of topological (green) segments of the wire host localized, zero-energy Majorana modes as shown in (d).}
\label{1D_wires_fig}
\end{figure*}

Two seminal works (Lutchyn \emph{et al}.\cite{1DwiresLutchyn} and Oreg \emph{et al}.\cite{1DwiresOreg}) recently established that one can engineer the topological phase in Kitaev's toy model by judiciously combining three exceedingly simple and widely available ingredients: a 1D wire with appreciable spin-orbit coupling, a conventional $s$-wave superconductor, and a modest magnetic field.  Figure \ref{1D_wires_fig}(a) illustrates the basic architecture required, which can be modeled by the following Hamiltonian,
\begin{eqnarray}
  H &=& H_{\rm wire} + H_\Delta
  \label{Hwire}
  \\
  H_{\rm wire} &=& \int dx \psi^\dagger\left(-\frac{\partial_x^2}{2m}-\mu - i \alpha \sigma^y \partial_x + h \sigma^z\right)\psi
  \\
  H_\Delta &=& \int dx \Delta(\psi_\uparrow\psi_\downarrow + H.c.).
\end{eqnarray}
Here $\psi_\sigma^\dagger$ adds an electron with effective mass $m$, chemical potential $\mu$, and spin $\sigma$ to the wire; $\alpha > 0$ denotes the strength for spin-orbit coupling that favors aligning spins along or against the $y$ direction depending on the momentum; and $h\geq 0$ is the Zeeman energy arising from a magnetic field applied along $z$.  (The precise spin-orbit and magnetic field axes are unimportant so long as they are perpendicular.)  For concreteness one can envision $H_{\rm wire}$ describing an electron-doped semiconducting wire such as InAs with Rashba coupling\cite{Rashba}, in the limit where only the lowest transverse subband is relevant.  The pairing term $H_\Delta$ crudely models the proximity effect on the wire arising from the adjacent $s$-wave superconductor.  Note that $H$ above takes on the same form as the topological insulator edge Hamiltonian $H'$ in Eq.\ (\ref{HTIedge3}), with the sole addition of an ordinary $k^2/(2m)$ kinetic energy contribution.  This important modification underlies many of the qualitative distinctions between the two setups.  

Let us first consider $\Delta = 0$ and elucidate how the Hamiltonian $H_{\rm wire}$ overcomes our fermion doubling problem.  The red and blue curves in Fig.\ \ref{1D_wires_fig}(b) illustrate the wire's band structure in the limit where $h = 0$.  Due to spin-orbit coupling, the blue and red parabolas respectively correspond to electronic states whose spin aligns along $+y$ and $-y$.  Clearly no `spinless' regime is possible here---the spectrum always supports an \emph{even} number of pairs of Fermi points for any $\mu$.  The magnetic field remedies this problem by lifting the crossing between these parabolas at $k = 0$, producing band energies
\begin{equation}
  \epsilon_\pm(k) = \frac{k^2}{2m}-\mu \pm \sqrt{(\alpha k)^2 + h^2}
  \label{wire_band_energies}
\end{equation}
sketched by the solid black curves of Fig.\ \ref{1D_wires_fig}(b).  When the Fermi level resides within this field-induced gap (\emph{e.g.}, for $\mu$ shown in the figure) the wire appears `spinless' as desired.  

The influence of the superconducting proximity effect on this band structure can be intuitively understood by focusing on this `spinless' regime and projecting away the upper unoccupied band, which is legitimate provided $\Delta \ll h$.  Crucially, because of competition from spin-orbit coupling the magnetic field only partially polarizes electrons in the remaining lower band as Fig.\ \ref{1D_wires_fig}(b) indicates schematically.  Turning on $\Delta$ weakly compared to $h$ then effectively $p$-wave pairs these carriers, driving the wire into a topological superconducting state that connects smoothly to the weak-pairing phase of Kitaev's toy model (see Ref.\ \onlinecite{AliceaBraiding} for an explicit mapping).  

More formally, one can proceed as we did for the topological insulator edge and express the full, unprojected Hamiltonian in terms of operators $\psi^\dagger_\pm(k)$ that add electrons with energy $\epsilon_\pm(k)$ to the wire.  The resulting Hamiltonian is again given by Eqs.\ (\ref{HTIedge4}) and (\ref{pairing_functions}) [but with $v \rightarrow \alpha$ and band energies $\epsilon_\pm(k)$ from Eq.\ (\ref{wire_band_energies})], explicitly demonstrating the intraband $p$-wave pairing mediated by $\Delta$.  Furthermore, Eq.\ (\ref{Epm}) provides the quasiparticle energies for the wire with proximity-induced pairing and again yields a gap that vanishes only when $h = \sqrt{\Delta^2 + \mu^2}$.  For fields below this critical value the wire no longer appears `spinless', resulting in a trivial state, while the topological phase emerges at higher fields,
\begin{equation}
  h > \sqrt{\Delta^2 + \mu^2}~~({\rm topological~criterion}).  
  \label{topological_criterion_wires}
\end{equation}
Figure \ref{1D_wires_fig}(c) summarizes the phase diagram for the wire.  Notice that this is inverted compared to the topological insulator edge phase diagram in Fig.\ \ref{2D_TI_fig}(d).  This important distinction arises because the $k^2/(2m)$ kinetic energy for the wire causes an upturn in the lower band of Fig.\ \ref{1D_wires_fig}(b) at large $|k|$, thereby either adding or removing one pair of Fermi points relative to the edge band structure.

Since a wire in its topological phase naturally forms a boundary with a trivial state (the vacuum), Majorana modes $\gamma_1$ and $\gamma_2$ localize at the wire's ends when the inequality in Eq.\ (\ref{topological_criterion_wires}) holds.  Majorana-trapping domain walls between topological and trivial regions can also form at the wire's interior by applying gate voltages to spatially modulate the chemical potential\cite{Hassler,AliceaBraiding} or by driving supercurrents through the adjacent superconductor\cite{Romito} (using the same mechanism discussed in Sec.\ \ref{2DTIproposal}).  Figure \ref{1D_wires_fig}(d) illustrates an example where four Majoranas form due to a trivial region in the center of a wire.  

It is useful address how one optimizes the 1D wire setup to streamline the route to experimental realization of this proposal.  This issue is subtle, counterintuitive, and difficult even to define precisely given several competing factors.  First, how well should the wire hybridize with the parent superconductor?  The naive guess that the hybridization should ideally be as large as theoretically possible to maximize the pairing amplitude $\Delta$ imparted to the wire is incorrect.  One practical issue is that exceedingly good contact between the two subsystems may lead to an enormous influx of electrons from the superconductor into the wire, pushing the Fermi level far above the Zeeman-induced gap of Fig.\ \ref{1D_wires_fig}(b) where the topological phase arises.  Restoring the Fermi level to the desired position by gating will then be complicated by strong screening from the superconductor.  

Reference \onlinecite{Disorder4} emphasized a more fundamental issue related to the optimal hybridization.  The topological phase's stability is determined not only by the pairing gap induced at the Fermi momentum, $E_{k_F} \propto \Delta$, but also the field-induced gap at zero momentum, $E_{0} = |h-\sqrt{\Delta^2 + \mu^2}|$, required to open a `spinless' regime.  The minimum excitation gap for the topological phase is set by the smaller of these two energies.  As reviewed in Sec.\ \ref{Preliminaries1D}, increasing the tunneling $\Gamma$ between the wire and superconductor indeed enhances $\Delta$ but simultaneously reduces the Zeeman energy $h$.  From the effective action in Eq.\ (\ref{Seff1D}) we explicitly have $h = Z h_{\rm bare}$ and $\Delta = (1-Z)\Delta_{sc}$, where $h_{\rm bare}$ is the Zeeman energy for the wire when the superconductor is absent, $\Delta_{sc}$ is the parent superconductor's gap, and $Z$ is the quasiparticle weight defined in Eq.\ (\ref{Z}) (for simplicity we neglect the frequency dependence).  Suppose now that $\Gamma$ increases from zero, thereby reducing $Z$ from unity.  As $Z$ decreases the topological phase's gap initially increases due to an enhancement of $E_{k_F}$.  Eventually, however, the gap \emph{decreases} due to a suppression of $E_{0}$, and the topological phase disappears entirely beyond a critical value of $\Gamma$ at which $E_{0}$ vanishes.  The maximum achievable gap depends sensitively on details such as the spin-orbit coupling strength, applied field amplitude, mobility for the wire, \emph{etc.}\cite{Disorder4,SauGapEstimates}

But should one ideally design the setup to achieve this maximum gap?  This, too, is not necessarily the case.  As an illustrative example, consider a $200,000$cm$^2$/Vs mobility wire with parameters relevant for InAs, adjacent to a superconductor with $\Delta_{sc} = 2$K.  Reference \onlinecite{SauGapEstimates} predicted (including short-range disorder) that such a wire realizes an optimized gap of $E_{\rm max} \approx 0.3$K when the Zeeman energy is $h_{\rm bare} \approx 1$K.  Realizing the topological phase in a meaningful way then requires positioning the chemical potential within a rather narrow $\sim 1$K window over distances long compared to the wire's coherence length, which could prove challenging experimentally due to disorder-induced chemical potential fluctuations.  Thus it may be desirable to apply larger magnetic fields to soften these constraints at the expense of reducing the gap somewhat.

The question of how large this Zeeman field should be is also rather delicate.  On one hand, enhancing $h$ indeed alleviates the need to fine tune $\mu$.  But on the other, increasing $h$ suppresses superconductivity in the parent superconductor, reduces the effective $p$-wave pairing amplitude for electrons in the wire due to a further alignment of their spins [see Eq.\ (\ref{pairing_functions})], and makes the wire more susceptible to disorder\cite{Disorder4,SauGapEstimates}.  As discussed in the topological insulator context, the first effect can be rather minor if the $g$ factor in the wire greatly exceeds that in the superconductor.  The effect of disorder warrants more serious consideration.  Since the topological phase appears only at finite magnetic fields Anderson's theorem does not protect the gap against disorder in the wire---which is always pair-breaking in this context as many studies have shown\cite{ZeroBiasAnomaly4,Disorder2,Disorder3,Disorder4,Disorder5,Disorder6,Disorder7,Disorder8,SauGapEstimates}.  (Fortunately though, Refs.\ \onlinecite{Disorder4erratum} and \onlinecite{LutchynDisorderSC} conclude that disorder native to the proximate $s$-wave superconductor is benign.)  

The sensitivity of the topological phase to disorder is determined by how severely time-reversal symmetry is broken.  One can quantify this by the ratio of the Zeeman energy $h$ to the spin-orbit energy\cite{Disorder4,SauGapEstimates}, which we define here by $E_{so} = \frac{1}{2}m\alpha^2$.  Physically, $E_{so}$ is the Fermi energy measured relative to the bottom of the bands when $\mu = h = 0$; see Fig.\ \ref{1D_wires_fig}(b).  At large $h/E_{so}$ spins near the Fermi level are fairly well polarized, so time reversal is strongly violated and hence disorder can efficiently suppress the gap.  In contrast, at small $h/E_{so}$ spins at $k_F$ and $-k_F$ are nearly antiparallel due to the dominance of spin-orbit coupling.  Time-reversal is then `almost' present insofar as carriers near the Fermi level are concerned, thereby sharply suppressing the impact of disorder on the topological phase.\cite{Disorder4,SauGapEstimates}  

The strength of spin-orbit coupling is thus a crucial materials parameter.  `Large' spin-orbit values allow one to operate at relatively high Zeeman fields---where the topological phase occurs over a broad chemical potential range---while maintaining some degree of robustness against disorder.  (Though one should bear in mind that increasing spin-orbit coupling can reduce the mobility, thus at least partially offsetting this advantage.\cite{SauGapEstimates})  Furthermore, the maximum gap that the topological phase can in principle exhibit increases with the spin-orbit strength,\cite{Disorder4,SauGapEstimates}  approaching a value of $\Delta_{sc}$ in the limit where\cite{Disorder4} $E_{so} \gg h \gg \Delta_{sc}$.  

In light of this discussion it is interesting to ask how electron-doped InAs and InSb wires fare as platforms for Majorana fermions.  At present these are the most commonly discussed wires for this proposal, and for good reason.  Both exhibit exceptionally large $g$ factors ($g_{\rm InAs} \approx 15$ and $g_{\rm InSb} \approx 50$ for bulk crystals), and can be synthesized with high mobility and long mean free paths.  Good superconducting proximity effects have also been measured in both systems.\cite{InAsProximityEffect,InAsProximityEffect2,InSbProximityEffect}  One challenge with these materials, however, is that while they are often lauded as having strong spin-orbit coupling, the energy scale $E_{so}$ is typically of order 1K for both InAs and InSb\cite{InSbCharacterization}.  Rashba coupling is gate-tunable to some extent,\cite{RashbaTuning} but it may nonetheless prove difficult to access the topological phase in a spin-orbit dominated regime where $h/E_{so} \ll 1$.  Disorder is thus likely to play a nontrivial role in these settings.  Still, it is hard not to be optimistic about the prospects of success given the high level and rapid pace of ongoing experimental activity.  

Several subsequent works have pursued variations on this proposal in an effort to mitigate the challenges involved with realizing the topological phase experimentally.  One issue that has been explored is whether one can reduce the applied magnetic field required to access the topological phase.  Reference \onlinecite{MajoranaInteractions} proposed the use of nuclear spins to generate the Zeeman energy\cite{NuclearSpins}, thus removing the external magnetic field altogether.  Repulsive interactions---which are inevitably present in the wire---also allow the topological phase to be accessed at weaker magnetic fields and over a broader chemical potential window due partly to exchange enhancement of the Zeeman field\cite{MilesInteractions} (but if the repulsion is too strong this state can disappear\cite{MajoranaInteractions,MilesInteractions}).  In principle, interactions can even stabilize topological superconductivity at zero magnetic field due to \emph{spontaneous} time-reversal symmetry breaking.\cite{MilesInteractions}  Generating the topological phase by applying supercurrents as described in Ref.\ \onlinecite{Romito} also lowers the critical Zeeman field needed.

Multichannel wires have been shown by numerous studies to support a 1D topological superconducting state away from the lowest-subband limit.\cite{Multichannel0,Multichannel1,Multichannel2,Multichannel3,Multichannel4,Multichannel5,Multichannel6,Multichannel7,Multichannel8,Disorder3,Disorder7}  In this more general setting one simply needs an \emph{odd} number of partially occupied bands and a wire whose width does not exceed the coherence length.  The former criterion is rather natural: starting from the topological phase in a single-channel wire, pairs of partially occupied bands can always be adiabatically introduced without closing the gap.  The latter criterion ensures that the induced superconducting phase remains quasi-1D, which is required for the wire to exhibit a substantial gap.\cite{Multichannel1,Multichannel4,Multichannel7}  These works are significant in part because multiple subbands are usually occupied in semiconducting wires such as InAs and InSb.  Gating these wires into the lowest subband regime may be nontrivial particularly when a superconductor is nearby, but fortunately is unnecessary.  Furthermore, multichannel wires open the door to realizing a topological phase in a variety of other settings, such as gate-defined channels in quantum wells or surface states featuring large spin-orbit coupling\cite{Multichannel1,BiAgSurfaceStates,NaCoO}.

Even in a multichannel wire accessing the topological phase will require some degree of gating.  An interesting possible route to enhancing gate-tunability is to employ periodically modulated structures in which a regular array of superconducting islands contacts the wire.\cite{SauPeriodic1,SmithaPeriodic,SauPeriodic2}  The basic idea is that gating the wire in the regions between adjacent superconductors may be relatively easy.  Although the Hamiltonian is rather different from that of the uniform structure considered previously, a topological phase can still arise, which can be understood simply in two limits.  First, suppose that the chemical potential varies only modestly along a single-channel wire.  In the limit where the Fermi wavelength exceeds the spacing between superconducting islands the electrons effectively `see' only the average induced pairing potential.  Periodic modulations are then essentially smeared out, and a topological phase arises under similar conditions to the uniform case.  

Second, suppose that a large potential barrier formed at the boundary between the gated and superconducting regions of the wire, effectively creating a chain of quantum dots bridged by superconducting islands.\cite{SauPeriodic2}  By introducing a magnetic field and fine-tuning the gates, one can in principle bring a single level into resonance on each dot.  Effectively each dot then behaves as a single site in a `spinless' chain.  Electrons from neighboring sites communicate indirectly via the superconductor, which can mediate both the nearest-neighbor hopping and $p$-wave pairing in Kitaev's toy model.  This setup can be particularly promising if the hopping and pairing amplitudes can be tuned equal to one another; this limit corresponds to the $t = \Delta$ case discussed in Sec.\ \ref{1D_toy_model} where Majorana end-states localize at a \emph{single} site.  Majorana zero-modes might then be observable in an array consisting of relatively few quantum dots.  One obvious challenge here is the high level of fine-tuning required to reach this regime.  Furthermore, strong randomness may pose an issue given that the hopping and pairing parameters presumably depend exponentially on factors such as the width of the superconducting islands, barrier heights, \emph{etc}.  

Numerous other interesting variants have been introduced.  Hole-doped semiconducting wires---which benefit from a greatly enhanced spin-orbit energy $E_{so}$ relative to their electron-doped counterparts---comprise one very promising alternative for realizing Majorana modes.\cite{HoleDopedWires,HoleDopedWires2,SauGapEstimates}  Carbon nanotubes can also in principle host Majoranas despite the fact that obtaining a `spinless' regime with proximity-induced pairing is rather nontrivial.\cite{CNTmajoranas1,CNTmajoranas2,CNTmajoranas3}  Another interesting proposal involves a half-metallic ferromagnetic wire in which only one spin species conducts.  While a `spinless' regime emerges trivially here, our standard trick for inducing $p$-wave superconductivity via a conventional $s$-wave superconductor no longer works.  (A spin-singlet order parameter can not pair spins that are fully aligned).  This problem can be solved by coupling the half-metal to a non-centrosymmetric superconductor with spin-orbit coupling.  Such a superconductor generically contains both spin-singlet \emph{and} spin-triplet Cooper pairing\cite{SpinOrbitSC} and can therefore induce a proximity effect in the wire to generate topological superconductivity.\cite{HalfMetallicWires}  Remarkably, clever routes to engineering a topological phase in systems \emph{without} spin-orbit coupling were even devised recently.\cite{NoSOC1,NoSOC2}  The key idea can be understood by rewriting our original wire Hamiltonian in Eq.\ (\ref{Hwire}) in terms of rotated operators $\tilde \psi = e^{i m\alpha x\sigma^y}\psi$:
\begin{eqnarray}
  H &=& \int dx \bigg{\{}\tilde\psi^\dagger\left[-\frac{\partial_x^2}{2m}-(\mu+E_{so}) + {\bf h}_{\rm eff}(x)\cdot\bm{\sigma}\right]\tilde\psi
  \nonumber \\
  &+& \Delta(\tilde \psi_\uparrow \tilde\psi_\downarrow + H.c.)\bigg{\}},
  \\
  {\bf h}_{\rm eff}(x) &=& -\sin(2m\alpha x){\bf \hat{x}} + \cos(2m\alpha x){\bf \hat{z}}.
\end{eqnarray}
It follows that the Hamiltonian for a spin-orbit-coupled wire subjected to a uniform magnetic field is unitarily equivalent to that of a spin-orbit-free wire in a spatially rotating Zeeman field---so either system can support a topological phase.  The required non-uniform Zeeman field in the latter setup can be generated using an array of magnetic nanoparticles deposited on a superconductor\cite{NoSOC1} or by coupling a set of magnetic gates to a wire\cite{NoSOC2}.  Finally, we note that one can engineer a 1D topological superconducting state with cold fermionic atoms using optical Raman transitions (to generate effective spin-orbit coupling and Zeeman fields) and a proximity effect with a bulk molecular BEC.\cite{ColdAtomMajoranas}  This route is especially tantalizing given the recent pioneering experiments by Lin \emph{et al}.\cite{SOCbosons}, where Raman lasers created a band structure for bosons similar to Fig.\ \ref{1D_wires_fig}(b).

\subsection{3D topological insulator nanowires}
\label{3DTInanowires}

Three-dimensional topological insulators, like their 2D counterparts, are strongly spin-orbit-coupled materials in which electrons are insulating in the bulk but due to topology form novel metallic states at their boundary.\cite{KaneReview,MooreReview,QiReview}  (Remarkably, these boundary states were captured very early on in Refs.\ \onlinecite{EarlySurfaceStates1} and \onlinecite{EarlySurfaceStates2}.)  In the simplest cases each surface hosts a \emph{single} Dirac cone described by
\begin{equation}
  H_{\rm 3DTI} = \int d^2{\bf r}\psi^\dagger[-i v {\bf \hat{n}}\cdot(\nabla \times \bm{\sigma})-\mu]\psi,
  \label{Hsurf}
\end{equation}
where ${\bf \hat{n}}$ is the surface normal, $\psi_{\sigma{\bf r}}^\dagger$ adds an electron with spin $\sigma$ at position ${\bf r}$ on the surface, $\mu$ is the chemical potential, and $v$ is the surface state velocity.  This Hamiltonian favors orienting the electron spins along the surface, but perpendicular to the momentum, similar to Rashba coupling in a 2D electron gas.  

\begin{figure}
\includegraphics[width = 8cm]{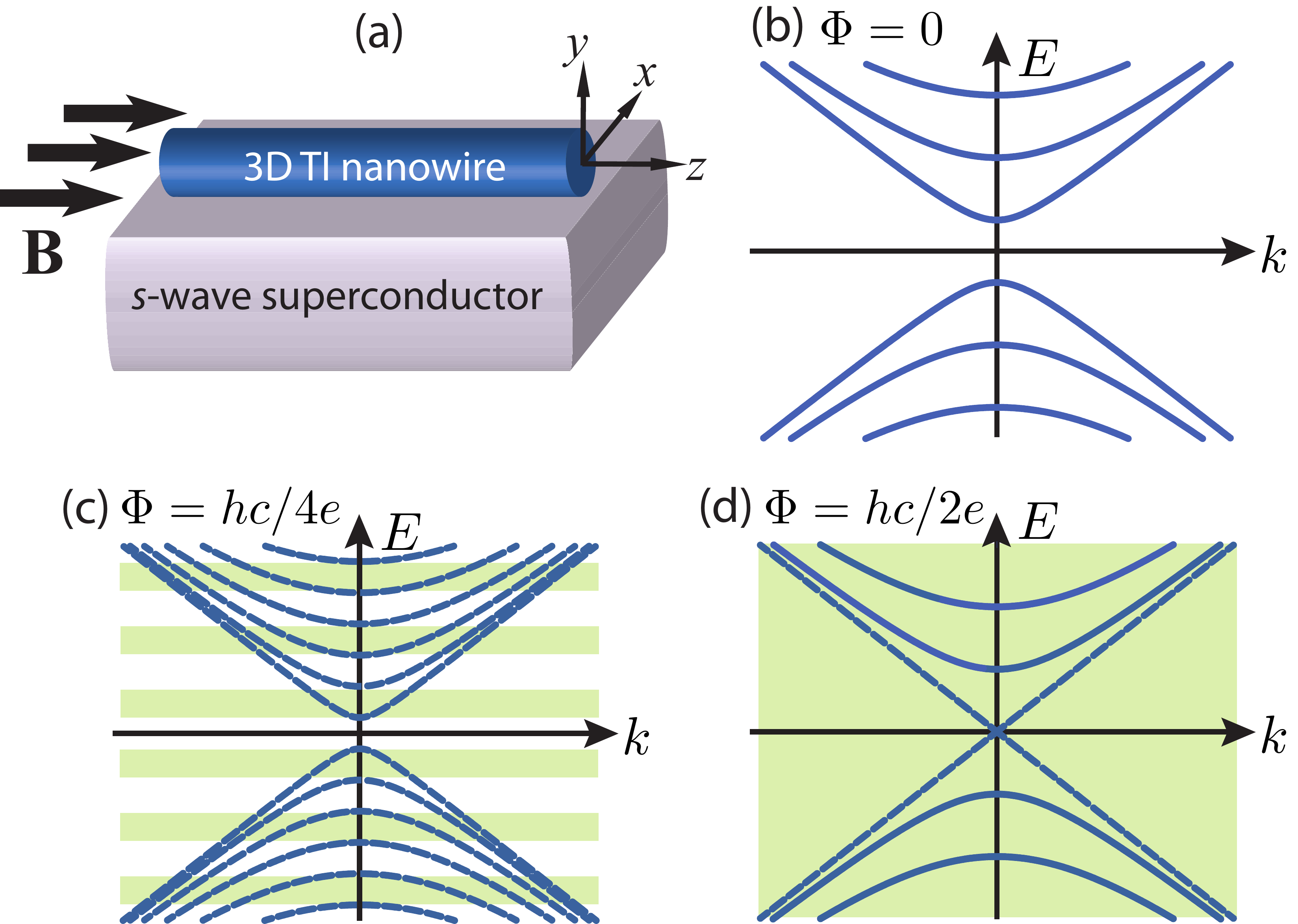}
\caption{(a) A 3D topological insulator nanowire with a magnetic field applied along its axis can realize a topological superconducting state when in contact with an $s$-wave superconductor.  (b)-(d) Nanowire band structure when the flux $\Phi$ passing through its center is (b) $0$, (c) $\frac{hc}{4e}$, and (d) $\frac{hc}{2e}$.  Solid and dashed curves respectively denote doubly degenerate and non-degenerate bands.  Green shaded regions indicate chemical potential windows in which an odd number of channels is occupied, as required for generating topological superconductivity.  }
\label{3D_TI_nanowire_fig}
\end{figure}

Cook and Franz recently showed that nanowires fabricated from 3D topological insulators can host a topological superconducting state with several advantageous features.\cite{CookFranz}  Consider a cylindrical nanowire of radius $R$, whose axis orients along the $z$ direction as sketched in Fig.\ \ref{3D_TI_nanowire_fig}(a).  Rewriting Eq.\ (\ref{Hsurf}) in cylindrical coordinates and using ${\bf \hat{n}} = {\bf \hat{r}}$ for the surface normal, one obtains the following nanowire Hamiltonian,\cite{NanowireHamiltonian,CookFranz}
\begin{equation}
  H_{\rm TI~wire} = \int dz d\theta \psi^\dagger\left[-i v \left( \frac{\sigma^z}{R}\partial_\theta-{\hat{\bm\theta}}\cdot{\bm \sigma}\partial_z\right)-\mu\right]\psi.
  \label{HsurfNanowire}
\end{equation}
It is convenient to remove the angular dependence in the $\partial_z$ term by defining a new field $\tilde \psi \equiv e^{i\theta\sigma^z/2}\psi$ that exhibits anti-periodic boundary conditions, \emph{i.e.}, $\tilde\psi(\theta+2\pi,z) = -\tilde\psi(\theta,z)$.  After absorbing a constant into the definition of $\mu$ the Hamiltonian then reads
\begin{equation}
  H_{\rm TI~wire} = \int dz d\theta \tilde\psi^\dagger\left[-i v \left( \frac{\sigma^z}{R}\partial_\theta-\sigma^y\partial_z\right)-\mu\right]\tilde\psi
  \label{HsurfNanowire2}
\end{equation}
and can be trivially diagonalized.  Equation (\ref{HsurfNanowire2}) admits band energies $\epsilon_{n\pm}(k) = \pm v\sqrt{(n/R)^2+k^2}-\mu$, where $k$ is the momentum along the cylinder axis and $n$ is a half-integer angular momentum quantum number due to the boundary conditions on $\tilde\psi$.  Figure \ref{3D_TI_nanowire_fig}(b) illustrates the spectrum; each band is doubly degenerate so our fermion doubling problem remains here.  

Applying a magnetic field along the cylinder axis as in Fig.\ \ref{3D_TI_nanowire_fig}(a) heals this problem in an interesting way.\cite{CookFranz}  Upon incorporating the field by sending $-i\partial_\theta \rightarrow -i\partial_\theta -\Phi/\Phi_0$, where $\Phi_0 = hc/e$ and $\Phi$ is the flux piercing the cylinder, the band energies become
\begin{equation}
  \epsilon_{n\pm}(k) \rightarrow \pm v\sqrt{[(n-\Phi/\Phi_0)/R]^2+k^2}-\mu.
\end{equation}
Note that we have neglected the Zeeman term $\frac{1}{2}g\mu_B B\sigma^z$ since this contribution merely renormalizes $\Phi$.  Generally, the flux lifts the band degeneracies present at zero field as Figs.\ \ref{3D_TI_nanowire_fig}(c) and (d) respectively illustrate for $\Phi = \frac{hc}{4e}$ and $\frac{hc}{2e}$; here dashed curves represent non-degenerate bands while solid curves are doubly degenerate.  This produces chemical potential windows (shaded green regions in Fig.\ \ref{3D_TI_nanowire_fig}) in which the electrons partially occupy an odd number of bands as desired.  In these odd-channel regimes Cooper pairing states at the Fermi level can drive the nanowire into a 1D topological superconducting phase.\cite{CookFranz}  Because of spin-orbit coupling this can be achieved in the standard way using the proximity effect with an $s$-wave superconductor as in Fig.\ \ref{3D_TI_nanowire_fig}(a).

The case of $\frac{hc}{2e}$ flux is particularly appealing.  Here an odd number of surface-state bands is occupied for \emph{any} $\mu$ that resides in the bulk band gap.  Accessing the topological superconducting phase is then `easy' and automatically results in localization of Majorana zero-modes at the ends of the nanowire.  Recall that topological superconductivity is similarly `easy' to obtain in a 2D topological insulator edge, but there the formation of Majoranas is less trivial; see Figs.\ \ref{2D_TI_fig}(e) and (f).  Moreover, in the (fictitious) limit where the $\frac{hc}{2e}$ flux is confined to a thin solenoid passing through the nanowire's center, the bulk retains time-reversal symmetry and thus maintains immunity against non-magnetic disorder.\cite{AndersonTheorem,CookFranz}  (Time-reversal is always lifted at the nanowire ends where the solenoid enters and exits the wire, which is essential for the formation of Majorana end-states.)  In an actual experiment additional ingredients---such as the Zeeman effect, fluctuations in the nanowire radius that cause the flux to deviate locally from $\frac{hc}{2e}$, \emph{etc.}---will inevitably remove this exact protection, though some robustness against disorder likely survives.  Other flux values are also interesting despite the absence of these features.  For example, near $\Phi = \frac{hc}{4e}$ the topological and trivial superconducting phases appear over chemical potential windows of roughly equal size, so tuning between these phases by gating may be ideal here.  

To realize this proposal the nanowire's radius $R$ should surpass the surface state penetration depth (to have well-defined surface states) but not exceed the induced superconducting coherence length in the wire (so that the superconducting state remains quasi-1D).  Since penetration depths of a few unit cells can arise\cite{PenetrationDepth} while the coherence length is typically on the micron scale, these criteria rather loosely constrain the wire size.  
Another relevant scale is the spacing $v/R$ between adjacent bands in Fig.\ \ref{3D_TI_nanowire_fig}(b)-(d).  Enhancing $v/R$ allows one to suppress unwanted interband coupling generated, \emph{e.g.}, by disorder, so reducing $R$ significantly below the coherence length will likely prove worthwhile.  Wires with $R$ of a few tens of nanometers should allow one to achieve a sizable interband spacing ($\sim100$K) and reach $\frac{hc}{2e}$ flux with sub-Tesla fields.  As experiments on topological insulator nanowires and nanoribbons are now steadily progressing, this approach certainly warrants further attention.  The prospect of using a weak magnetic field to stabilize a topological phase and Majorana modes \emph{without} requiring careful control over the chemical potential is well worth pursuing.

\section{Practical realizations of Majorana modes in 2D $p+ip$ superconductors}
\label{2D_toy_model_realizations}

\subsection{Preliminary remarks}

The proposals we review below for experimentally realizing the physics of topological 2D $p+ip$ superconductivity discussed in Sec.\ \ref{2D_toy_model} loosely fall into two categories.  The first corresponds to `intrinsic' realizations, where $p+ip$ pairing emerges by virtue of a material's internal dynamics.  We briefly discuss two classic systems of this type: the fractional quantum Hall state at filling factor $\nu = 5/2$ and the layered spinful triplet superconductor Sr$_2$RuO$_4$.  

Most of our discussion will center around the second category---`engineered' topological phases---wherein `spinless' $p+ip$ superconductivity is driven by forming appropriate heterostructures with various kinds of 2D electron systems.  In this approach one faces the same basic hurdles that arose in realizing Kitaev's 1D model: removing the spin degeneracy so that the system appears effectively `spinless' and inducing long-range-ordered $p+ip$ pairing in the remaining Fermi surface.  As we will see, time-reversal-symmetry breaking, spin-orbit coupling, and superconducting proximity effects (usually with conventional $s$-wave superconductors) again provide the key to overcoming these challenges.  Note that our discussion from Sec.\ \ref{Preliminaries1D} on the proximity effect induced in a 1D system applies to the 2D case with only trivial modifications.  We again stress that while the influence of a nearby superconductor can be crudely modeled by simply adding a pairing term to the Hamiltonian for the 2D electron system, at this level one misses important physics.  Hybridization with the parent superconductor also effectively reduces the energy scales intrinsic to the 2D system\cite{StanescuProximityEffect,SauProximityEffect,SauReview,Disorder4}; see Sec.\ \ref{Preliminaries1D} and Appendix \ref{EffectiveActionDerivation}.   Both effects are important to keep in mind for all of the engineered heterostructures discussed below.

\subsection{$\nu = 5/2$ fractional quantum Hall effect}
\label{FiveHalves}

Although our focus is on reviewing new routes to Majorana fermions, we would be remiss to not at least briefly discuss the fractional quantum Hall state observed\cite{FiveHalvesDiscovery} in GaAs quantum wells at filling factor $\nu = 5/2$---which remains a leading candidate in this search.  The basic question we would like to explore is how a 2D electron gas (2DEG) subjected to a strong perpendicular magnetic field can realize the physics of a topological 2D spinless $p+ip$ superconductor.  The connection between these very different systems was first elucidated in highly influential work by Read and Green.\cite{ReadGreen}  

Because the field quenches the kinetic energy and induces a Zeeman splitting, obtaining a spinless regime in such a 2DEG is rather natural (though not guaranteed).  The onset of $p+ip$ `superconductivity', by contrast, is far subtler.  To see how this arises we will first examine a 2DEG with a half-filled lowest Landau level ($\nu = 1/2$). Assuming perfect spin polarization, the system may be described by the Euclidean action
\begin{equation}
  S = \int d^2{\bf r}d\tau \psi^\dagger\left[\partial_\tau + \frac{(-i\nabla-\frac{e}{c}{\bf A})^2}{2m}\right]\psi + S_{\rm int},
\end{equation}
where $\psi$ is a `spinless' fermion operator for electrons with mass $m$, $\nabla \times {\bf A} = B{\bf \hat{z}}$ is the applied field, and $S_{\rm int}$ encodes Coulomb interactions. The physics of $\nu = 1/2$ can be elegantly captured by decomposing the electron $\psi$ in terms of a `composite fermion' $f$ bound to two $\Phi_0 = \frac{hc}{e}$ flux quanta using Chern-Simons theory.\cite{HLR}  In this framework the action becomes
\begin{eqnarray}
  S &=& \int d^2{\bf r}d\tau \bigg{\{}f^\dagger\left[(\partial_\tau-i a^0) + \frac{(-i\nabla-\frac{e}{c}{\bf A} - \frac{e}{c}{\bf a})^2}{2m^*}\right]f 
  \nonumber \\
  &-& \frac{i}{4\Phi_0} a^\mu \epsilon_{\mu\nu\lambda}\partial_\nu a^\lambda\bigg{\}} + S_{\rm int}.
\end{eqnarray}
Here $m^*$ is an effective mass and $a^\mu$ is the Chern-Simons field.  The temporal component $a^0$ serves as a Lagrange multiplier that pins the statistical Chern-Simons flux to the composite fermion density via $\nabla \times {\bf a} = -2\Phi_0 f^\dagger f{\bf \hat{z}}$.  Since the mean density of a half-filled Landau level is $\langle f^\dagger f\rangle = B/(2\Phi_0)$, on average the attached flux exactly cancels the applied magnetic field.  Composite fermions then behave similarly to electrons at $B = 0$: they form a `composite Fermi sea' and can propagate in straight lines over long distances despite the strong magnetic field.\cite{HLR}  

This composite Fermi sea can in principle undergo a BCS instability, just as for a conventional metal.  Moore and Read\cite{MooreRead} explored the possibility of $p+ip$ composite fermion pairing and proposed the following lowest Landau level wavefunction for such a state,
\begin{equation}
  \Psi_{MR} = {\rm Pf}\left(\frac{1}{z_i-z_j}\right)\prod_{i<j}(z_i-z_j)^2 e^{-\sum_k\frac{|z_k|^2}{4\ell_B^2}},
\end{equation}
with $\ell_B$ the magnetic length and $z_j$ the complex coordinate for particle $j$.  The $(z_i-z_j)^2$ factors (roughly) correspond to the attached Chern-Simons flux, whereas the Pfaffian is the real-space wavefunction for $p+ip$-paired composite fermions in the weak-pairing phase\cite{ReadGreen}.  Universal topological properties, such as the existence of chiral Majorana edge states and Majorana zero-modes bound to vortices, are shared by the Moore-Read state and the 2D spinless $p+ip$ superconductor explored earlier.\cite{ReadGreen}

At $\nu = 1/2$, such a pairing instability does not arise experimentally---the composite Fermi sea is stable and underlies the formation of an interesting compressible `composite Fermi liquid' phase.\cite{HLR}  A compelling body of theoretical evidence\cite{TQCreview}, however, indicates that the Moore-Read state (or its particle-hole conjugate\cite{antiPfaffian1,antiPfaffian2}) provides an energetically very competitive candidate for the measured plateau in the half-filled second Landau level.  Very likely, either the Moore-Read state or its particle-hole conjugate emerge as the ground state over some range of density, quantum well width, mobility, \emph{etc}., and a growing set of experiments\cite{FiveHalvesCharge,FiveHalvesTunneling,Willett1,Willett2,FiveHalvesNeutralModes,FiveHalvesKang,FiveHalvesSpinPolarization} indeed support this possibility.  For more details on this interesting subject we refer readers to Read and Green\cite{ReadGreen} and the comprehensive review by Nayak \emph{et al.}\cite{TQCreview}

\subsection{`Intrinsic' $p+ip$ superconductivity: Sr$_2$RuO$_4$}
\label{SrRuO}

In rare cases, $p+ip$ superconductivity can emerge `intrinsically' through interactions in a material.  At present Sr$_2$RuO$_4$---a layered compound with a somewhat complex, spin-degenerate Fermi surface deriving from Ru $d$-orbitals\cite{SrRuO_FermiSurface,SrRuO_Review}---constitutes the best experimental candidate for such a superconductor.  While the precise nature of the superconducting state that appears below $T_c = 1.5$K remains unsettled (see, \emph{e.g.}, Ref.\ \onlinecite{SrRuO_Sri}), a variety of experiments support the onset of spin-triplet Cooper pairing and spontaneous time-reversal symmetry breaking in this system.\cite{SrRuO_Review,SrRuO_OddParity,SrRuO_OrderParameter,SrRuO_BrokenT,SrRuO_HalfQuantumVortex,SrRuO_EdgeStates}  

Recall from Sec.\ \ref{2D_toy_model} that spinful 2D $p+ip$ superconductors allow for $\frac{hc}{4e}$ half quantum vortices that bind stable Majorana zero-modes.  In this context, the recent experiments of Jang \emph{et al}.\cite{SrRuO_HalfQuantumVortex} are particularly fascinating.  These authors employed torque magnetometry to measure the magnetization of annular, mesoscopic Sr$_2$RuO$_4$ samples as a function of an applied magnetic field $B$.  With the field oriented perpendicular to the layers, increasing $B$ produced discrete jumps in the magnetization at certain field values associated with nucleation of an ordinary $\frac{hc}{2e}$ vortex in the sample.  Remarkably, repeating the same experiment in the presence of a fixed in-plane field component `fractionalized' these magnetization jumps into steps half as large---consistent with the entry of half quantum vortices.  Precisely why the in-plane field should stabilize these defects is presently unclear, though Ref.\ \onlinecite{SrRuO_HalfQuantumVortex} discusses one possible scenario.  (Note that Ref.\ \onlinecite{SrRu} proposed applying \emph{perpendicular} fields to stabilize half quantum vortices.)

A few cautionary remarks are in order regarding Sr$_2$RuO$_4$---and likely any `intrinsic' $p+ip$ superconductor---as a setting for Majorana physics.  First, since time-reversal symmetry is broken spontaneously $p+ip$ and $p-ip$ pairings are degenerate, and domains featuring both chiralities will generally exist in a given crystal (see, \emph{e.g.}, Refs.\ \onlinecite{SrRuO_Domains1,SrRuO_Domains2,SrRuO_Domains3}).  These domains will complicate the edge-state structure relative to the toy model discussed in Sec.\ \ref{2D_toy_model}.  Second, half quantum vortices need not trap Majorana zero-modes in Sr$_2$RuO$_4$ crystals consisting of $N>1$ layers.  Consider, for instance, a half quantum vortex threading a Sr$_2$RuO$_4$ bilayer at $T = 0$ where phase fluctuations can be neglected.  In the artificial limit where the layers decouple, the vortex binds one Majorana zero-mode in each layer; restoring the interlayer coupling hybridizes these modes and produces an ordinary, finite-energy state.  For larger $N$ a chain of Majorana modes will hybridize and broaden into a gapless `band' in the $N\rightarrow \infty$ limit.  Strictly speaking, for any odd $N$ a single Majorana zero-mode must survive the interlayer coupling but in practice may prove difficult to disentangle from other low-energy modes.  Even in a single-layer sample Majorana zero-modes are protected only by a `mini-gap' in the spectrum of vortex bound states [Eq.\ (\ref{Evortex})], which for Sr$_2$RuO$_4$ falls in the milliKelvin range since the Fermi energy exceeds the pairing gap by orders of magnitude.

As an aside, we briefly mention a clever idea proposed in Ref.\ \onlinecite{SrRuO_VortexLines} for realizing Kitaev's 1D toy model along an ordinary $\frac{hc}{2e}$ vortex line threading a layered spinful $p+ip$ superconductor such as Sr$_2$RuO$_4$.  Neglecting spin-orbit interactions and interlayer coupling, the vortex binds a pair of Majorana zero-modes in each layer.  One can view each pair as comprising a single site in Kitaev's 1D toy model (recall Fig.\ \ref{KitaevModelFig}).  When coupling between nearby Majorana zero-modes is restored, Ref.\ \onlinecite{SrRuO_VortexLines} predicts that the topological phase of Kitaev's model emerges upon driving a supercurrent perpendicular to the layers.  The small mini-gap associated with the vortex, however, still poses a challenge for such a setup.

\subsection{3D topological insulators}
\label{3DTIproposal}

\begin{figure}
\includegraphics[width = \linewidth]{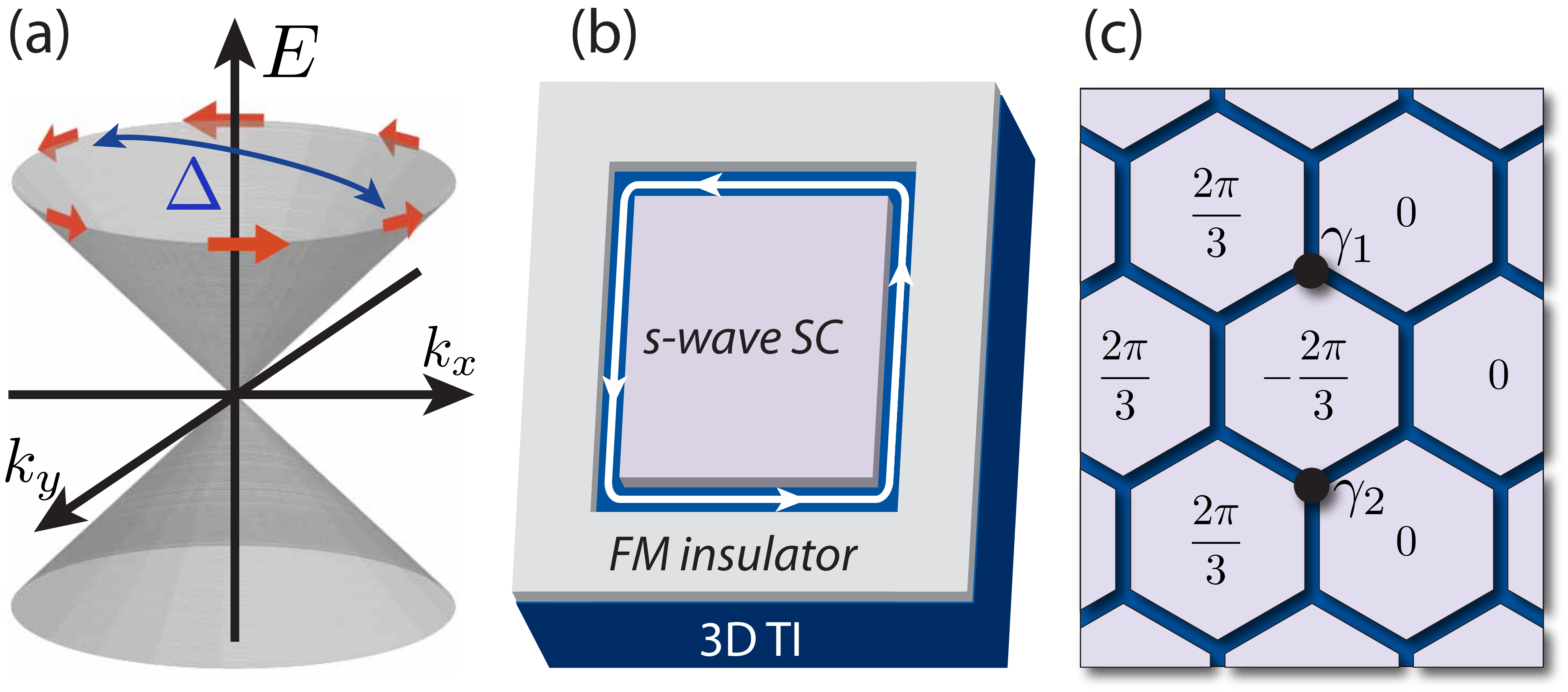}
\caption{(a) Single Dirac cone describing the surface states of a 3D topological insulator.  Because the electron spins wind by $2\pi$ upon encircling the Dirac cone, the proximity effect with a conventional $s$-wave superconductor drives the surface into a time-reversal invariant relative of a topological 2D spinless $p+ip$ superconductor.  Chiral Majorana edge states form at the boundary between superconductivity- and magnetically-gapped regions of the surface as shown, for example, in (b).  Introducing an $\frac{hc}{2e}$ vortex through a topologically superconducting portion of the surface binds a stable Majorana zero-mode.  This can be achieved either by applying a magnetic field, or by adjusting the phases on superconducting islands (hexagons) as shown in (c).}
\label{3D_TI_fig}
\end{figure}

In Sec.\ \ref{3DTInanowires} we described how one can engineer a 1D topological superconductor using 3D topological insulator nanowires.  Here we turn to Fu and Kane's groundbreaking proposal for stabilizing 2D `spinless' $p+ip$ superconductivity using the surface of a macroscopic 3D topological insulator.\cite{FuKane}  We will continue to focus on materials such as Bi$_2$Se$_3$\cite{KaneReview,MooreReview,QiReview} whose boundary hosts a single Dirac cone described by Eq.\ (\ref{Hsurf}).  For a surface located in the $(x,y)$ plane, the Hamiltonian reads
\begin{equation}
    H_{\rm 3DTI} = \int d^2{\bf r}\psi^\dagger[-i v (\partial_x \sigma^y-\partial_y \sigma^x)-\mu]\psi.
  \label{H3DTI}
\end{equation}
Equation (\ref{H3DTI}) yields band energies $\epsilon_\pm({\bf k}) = \pm v |{\bf k}|-\mu$ which correspond to the upper and lower branches of the massless Dirac cone sketched in Fig.\ \ref{3D_TI_fig}(a).  This band structure is ideal for forming a 2D topological superconducting phase.  First, accessing a `spinless' regime is trivial here: for any $\mu$ that resides within the material's bulk band gap there exists only a single Fermi surface as desired (rather than two as ordinarily arises due to spin degeneracy).  Furthermore, since the electrons along this Fermi surface are not spin-polarized, $p+ip$ pairing can be effectively induced using the proximity effect with a conventional $s$-wave superconductor.

One can see this explicitly by examining an effective Hamiltonian for the surface with proximity-induced spin-singlet pairing:
\begin{eqnarray}
  H &=& H_{\rm 3DTI} + H_\Delta
  \label{H3DTI2}
  \\
  H_\Delta &=& \int d^2{\bf r} \Delta(\psi_\uparrow \psi_\downarrow + H.c.),
\end{eqnarray}
which describes a superconductor with a fully gapped quasiparticle spectrum given by
\begin{equation}
  E_\pm({\bf k}) = \sqrt{\epsilon_\pm({\bf k})^2 + \Delta^2}.  
  \label{3DTI_BdG_spectrum}
\end{equation}  
To understand the nature of this state it is instructive to perform a unitary transformation that diagonalizes the kinetic energy in $H$.  In terms of operators $\psi_\pm^\dagger({\bf k})$ that add electrons to the upper and lower half of the Dirac cone, the Hamiltonian can be written as
\begin{eqnarray}
  H &=& \sum_{s = \pm}\int \frac{d^2{\bf k}}{(2\pi)^2}\bigg{\{}\epsilon_s({\bf k})\psi_s^\dagger({\bf k})\psi_s({\bf k}) 
  \nonumber \\
  &+& \left[\frac{\Delta}{2}\left(\frac{k_x+i k_y}{|{\bf k}|}\right)\psi_s({\bf k}) \psi_s(-{\bf k}) + H.c.\right]\bigg{\}}.
  \label{H3DTI3}
\end{eqnarray}
It is clear in this basis that the proximate $s$-wave superconductor mediates $p+ip$ pairing for electrons at the Fermi level.\cite{FuKane}  Figure \ref{3D_TI_fig}(a) illustrates the physical origin of this effect: $\Delta$ can pair resonant electrons with momenta ${\bf k}$ and $-{\bf k}$ since they carry opposite spins, while the nontrivial Cooper pair angular momentum arises because the spins rotate by $2\pi$ upon encircling the Dirac cone.  

The relation between Eq.\ (\ref{H3DTI3}) and the toy model discussed in Sec.\ \ref{2D_toy_model} becomes manifest when $\mu$ resides far from the Dirac point.  For a heavily electron-doped surface, for instance, one can safely project out the Dirac cone's lower half by sending $\psi_-\rightarrow 0$.  Equation (\ref{H3DTI3}) then maps precisely onto the Hamiltonian for a 2D `spinless' $p+ip$ superconductor---albeit with a non-standard kinetic energy and pairing potential---in the topological weak pairing phase.  By continuity, the surface forms a topological superconductor for \emph{any} chemical potential that does not intersect the bulk bands since the quasiparticle spectrum of Eq.\ (\ref{3DTI_BdG_spectrum}) is always fully gapped.  In addition to being `easy' to access in this sense, we emphasize this phase can also in principle persist up to relatively high temperatures.  For one, the surface can potentially inherit the parent superconductor's full pairing gap\cite{SauProximityEffect,Disorder4} for the same reasons discussed in the 2D topological insulator context; see Sec.\ \ref{2DTIproposal}.  Moreover, the topological phase captured here preserves time-reversal symmetry\cite{FuKane} [which is obvious in the original $\psi_{\uparrow,\downarrow}$ basis but somewhat hidden in Eq.\ (\ref{H3DTI3})].  This feature guarantees that the topological superconductor's gap enjoys immunity against non-magnetic disorder.\cite{AndersonTheorem,Disorder4}  Time-reversal symmetry breaking of some kind, however, is required to uncover the seeds of Majorana physics encoded in this state.  

Similar to the toy model of Sec.\ \ref{2D_toy_model}, chiral Majorana edge states form at the boundary between topologically superconducting and magnetically gapped regions of the surface.\cite{FuKane,MajoranaModesWithChargeTransport,3DTI_Majorana_edge_states1,3DTI_Majorana_edge_states2,3DTI_Majorana_edge_states3}   
Figure \ref{3D_TI_fig}(b) illustrates one possible architecture supporting such an interface.  There, an $s$-wave superconductor generates topological superconductivity, while a surrounding ferromagnetic insulator imparts the surface beneath it with a Zeeman field that we assume cants the spins out of the $(x,y)$ plane.  The surface state Hamiltonian governing the latter region then becomes $H = H_{\rm 3DTI} + H_Z$, with $H_Z = -h\int d^2{\bf r} \psi^\dagger \sigma^z \psi$.   The Zeeman energy modifies the spectrum to $\epsilon_\pm({\bf k}) = \pm \sqrt{(v|{\bf k}|)^2 + h^2}-\mu$, so that the surface forms a magnetically gapped state when the Fermi level lies within the resulting field-induced gap in the Dirac cone.  Interestingly, since the topological superconductor induced at the center retains time-reversal symmetry, the edge-state chirality depends on whether the ferromagnet cants the spins along $+{\bf \hat{z}}$ or $-{\bf \hat{z}}$.\cite{MajoranaModesWithChargeTransport,3DTI_Majorana_edge_states1,3DTI_Majorana_edge_states3}  This will be important to keep in mind when we discuss interferometry in Sec.\ \ref{Interferometry}.

Another useful way to lift time-reversal symmetry is to introduce an $\frac{hc}{2e}$ vortex in a topologically superconducting region of the surface; exactly as for an ordinary spinless $p+ip$ superconductor, a single zero-energy Majorana mode localizes at the vortex core.\cite{FuKane}  While vortices can always be induced by applying a magnetic field, Fu and Kane invented an alternative, more versatile method for creating and manipulating Majorana zero-modes.\cite{FuKane}  Figure \ref{3D_TI_fig}(c) illustrates their proposed setup, consisting of an array of superconducting islands (hexagons) deposited on a 3D topological insulator surface.  Here vortices appear when the superconducting phases on the islands wind by $\pm 2\pi$ around a trijunction where three islands meet.  For example, the pattern of phases in Fig.\ \ref{3D_TI_fig}(c) traps two Majorana modes $\gamma_{1,2}$.  Manipulating the superconducting phases (by, say, driving currents across the islands) allows one to controllably create, transport, and remove vortices---all crucial ingredients for the topological quantum information processing schemes that we highlight in Sec.\ \ref{NonAbelianStatistics}.  

One commonly noted obstacle to realizing Fu and Kane's proposal experimentally is that most 3D topological insulators  studied to date do not actually insulate in the bulk.  Rather, they contain a substantial concentration of gapless bulk carriers---which must be removed entirely for the physics described above to survive.  (Section \ref{Majorana_tunneling} discusses the fate of Majorana modes coupled to gapless degrees of freedom.)  While important progress towards rectifying this issue has recently been reported---see, \emph{e.g.}, Refs.\ \onlinecite{HgTe_3DTI} and \onlinecite{ReducedBulkCarriers}---these carriers may ultimately prove to be a feature rather than a bug.  A number of recent experiments observe superconductivity in metallic 3D topological insulator materials upon doping or under pressure.\cite{Superconducting3DTI1,Superconducting3DTI2,Superconducting3DTI3,Superconducting3DTI4,Superconducting3DTI5,Superconducting3DTI6,Superconducting3DTI7,Superconducting3DTI8}  The nature of the resulting superconducting state is itself a fascinating problem, and it has been suggested that these systems may constitute the first realization of a class of exotic 3D topological superconductors\cite{KitaevClassification,RyuClassification,TRItopologicalSC2,TRItopologicalSC3,FuBerg,Sato3DTopologicalSuperconductor,TopologicalSuperconductivityExpt,TopologicalSuperconductivityExpt2,HsiehFu,Yamakage}.  Here we will concentrate on the simplest possibility wherein conventional $s$-wave superconductivity emerges in the bulk.  Naively, it is tempting to conclude that employing such materials kills two birds with one stone: the problematic bulk carriers are gapped by Cooper pairing and simultaneously play the role of the proximate $s$-wave superconductor in Fu and Kane's proposal.  Upon closer inspection, however, the validity of this physical picture is suspect.  In the metallic phase well-defined surface states need not exist at the Fermi energy once the chemical potential intersects the bulk bands.\cite{DoronTopologicalMetals}(Generally, the surface state penetration depth diverges due to hybridization with resonant bulk extended states.  Even when the bulk and surface-state Fermi surfaces are well-separated in momentum space disorder can still induce hybridization.)  So when the bulk becomes superconducting, do Majorana modes still localize at the surface when a vortex is present?

An important work by Hosur \emph{et al}.\cite{Hosur,HosurViewpoint} showed that they can and provided the following appealing picture for the physics.  Consider first the limit where the bulk superconducts but the Fermi level resides within the bulk band gap.  Well-defined surface states then appear and realize a topological superconducting phase via proximity with the bulk.  A vortex line penetrating the material binds a pair of localized Majorana zero-modes, one at each end, effectively realizing Kitaev's toy model for a 1D topological superconductor.  Upon raising the Fermi level the vortex line eventually transitions into the trivial strong pairing phase of Kitaev's model, at a critical chemical potential that depends on the bulk band structure and the vortex orientation.  The crucial point is that this transition can occur well \emph{after} the Fermi level first intersects the bulk bands.  In doped Bi$_2$Se$_3$, for instance, Hosur \emph{et al}.\ predict that a vortex line binds Majorana zero-modes up until the chemical potential lies roughly $0.2-0.3$eV above the conduction band minimum.  As with any `intrinsic' superconductor where the Fermi energy greatly exceeds the pairing gap, the small mini-gap associated with the vortex will complicate the identification of the zero-modes.  Nevertheless, as the material science of 3D topological insulators is perfected the superconducting variety of these compounds provides one very promising venue for the exploration of Majorana physics.

\subsection{Conventional 2D electron systems}

\begin{figure}
\includegraphics[width = \linewidth]{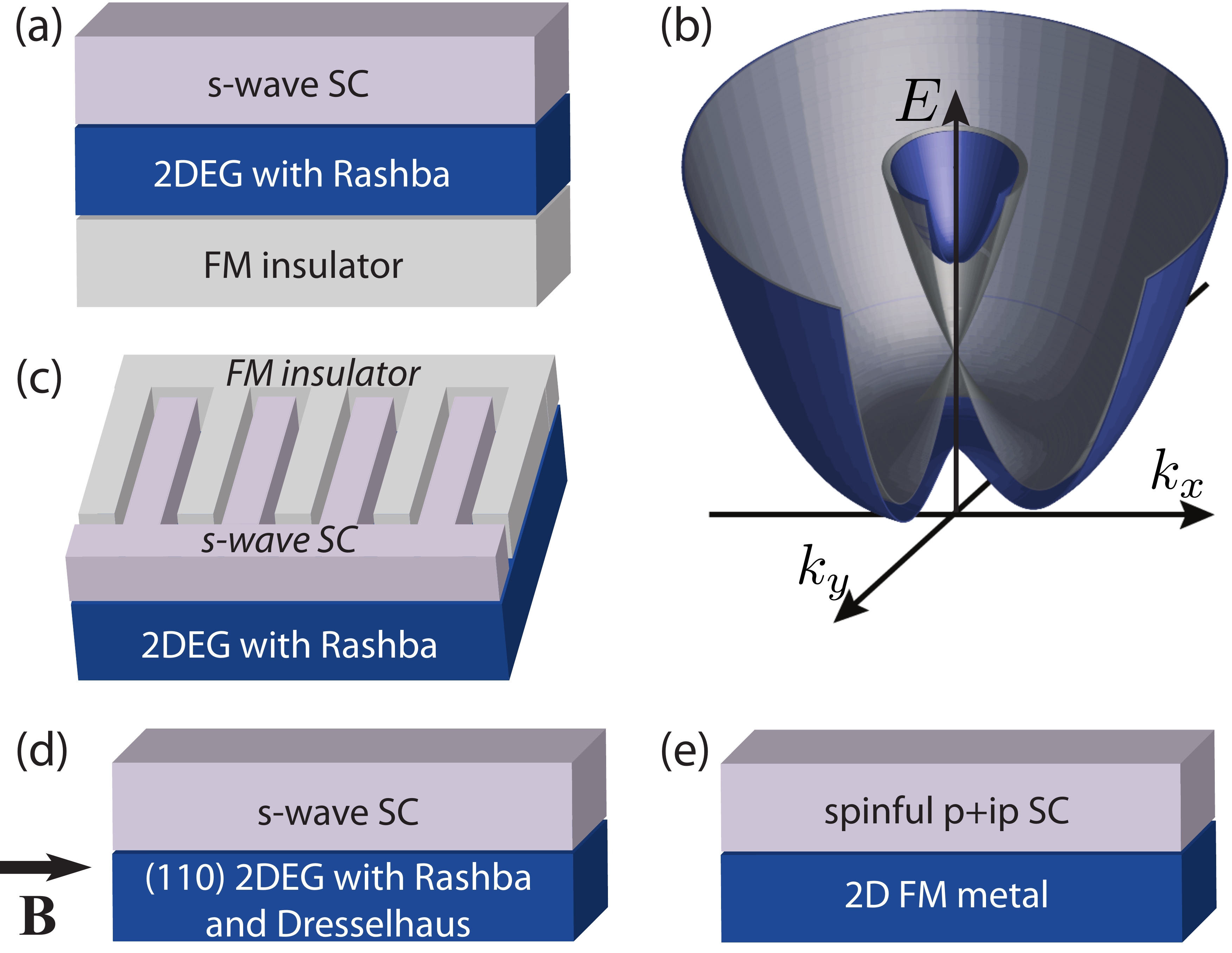}
\caption{(a) A 2DEG with Rashba spin-orbit coupling can effectively realize a topological 2D `spinless' $p+ip$ superconductor when in contact with a ferromagnetic insulator and conventional $s$-wave superconductor.  (b) Band structure for the 2DEG when time-reversal symmetry is present (gray) and with a non-zero Zeeman field (blue) that opens up a `spinless' regime.  (The break in the spectrum appears for clarity.)  Alternative devices that support topological phases appear in (c)-(e).}
\label{2D_semiconductor_fig}
\end{figure}

Following Fu and Kane's work a number of authors pursued alternative routes to engineering 2D spinless $p+ip$ superconductivity using more conventional materials.  We will initially explore the semiconductor-based proposal of Sau \emph{et al}.\cite{Sau}, though we note that similar results in related contexts appear in some earlier works\cite{FujimotoTSC,TopologicalSF,FujimotoColdAtoms}.  Consider the architecture of Fig.\ \ref{2D_semiconductor_fig}(a) consisting of a Rashba-coupled, electron-doped semiconductor 2DEG (such as InAs or InSb) sandwiched by an $s$-wave superconductor and ferromagnetic insulator.  We take the 2DEG to lie in the $(x,y)$ plane and assume that the ferromagnetic insulator's magnetization aligns along the $z$ direction.  The following effective Hamiltonian crudely captures the dynamics of electrons in the semiconductor,
\begin{eqnarray}
  H &=& H_{\rm 2DEG} + H_Z + H_\Delta
  \label{Hsandwich}
  \\
  H_{\rm 2DEG} &=& \int d^2{\bf r}\psi^\dagger\left[-\frac{\nabla^2}{2m}-\mu -i \alpha (\partial_x \sigma^y-\partial_y \sigma^x)\right]\psi
  \label{H2DEG}
  \\
  H_Z &=& -h\int d^2{\bf r}\psi^\dagger \sigma^z \psi
  \label{H_Z_2DEG}
  \\
  H_\Delta &=& \int d^2{\bf r}\Delta(\psi_\uparrow \psi_\downarrow + H.c.),
  \label{H_Delta_2DEG}
\end{eqnarray}
where $\psi_{\sigma{\bf r}}^\dagger$ adds an electron with effective mass $m$ and spin $\sigma$ to the 2DEG.  Let us first understand the Hamiltonian $H_{\rm 2DEG}$ describing the 2DEG's intrinsic couplings.  Here $\alpha$ denotes the strength of Rashba spin-orbit interactions\cite{Rashba} which favor orienting the electron spins within the 2DEG plane, perpendicular to their momentum.  The gray curves in Fig.\ \ref{2D_semiconductor_fig}(b) illustrate the band structure obtained from $H_{\rm 2DEG}$.  Note that the energies resemble a Dirac cone at small ${\bf k}$, similar to a 3D topological insulator surface, but whose lower half eventually bends upward at large momenta.  Because of this property two Fermi surfaces appear for any $\mu$ above the bottom of the conduction band---one of which we would like to eliminate.

As we have seen several times before time-reversal symmetry breaking, now through the Zeeman term $H_Z$ inherited from the ferromagnetic insulator, again overcomes this problem.  (The Zeeman field $h$ arises primarily from electron tunneling between the two subsystems and \emph{not} the magnetic field emanating from the ferromagnet;\cite{SauReview} this justifies the neglect of orbital effects in the Hamiltonian above.  Incidentally, for the 1D wire proposals discussed in Sec.\ \ref{1Dwiresproposal} orbital effects from an applied magnetic field are generically subdominant to the Zeeman energy.  Ultimately this key advantage allows for simpler setups employing wires compared to 2DEGs.)  With $h > 0$ the `Dirac-like' portion of the 2DEG's band structure near ${\bf k} = 0$ acquires a gap as shown in the blue curves of Fig.\ \ref{2D_semiconductor_fig}(b).  When the Fermi level resides within this gap and one turns on a weak proximity-induced $\Delta$ from Eq.\ (\ref{H_Delta_2DEG}), the resulting `spinless' metal enters a topological superconducting phase supporting chiral Majorana edge states and Majorana zero-modes localized at vortex cores.  The physics here is nearly identical to the 3D topological insulator proposal reviewed previously: the $s$-wave pair field mediates $p+ip$ Cooper pairing because Rashba coupling causes the in-plane spin components to wind by $2\pi$ upon encircling the Fermi surface.\cite{FujimotoTSC,TopologicalSF}  (Reference \onlinecite{Alicea} provides an explicit mapping to a spinless 2D $p+ip$ superconductor.)  A quantitative analysis analogous to that carried out for 1D wires in Sec.\ \ref{1Dwiresproposal} reveals that the topological phase appears provided\cite{FujimotoColdAtoms,Sau}
\begin{equation}
  h > \sqrt{\Delta^2 + \mu^2}~~({\rm topological~criterion}),
  \label{topological_criterion_2DEG}
\end{equation}
which is the same criterion given in Eq.\ (\ref{topological_criterion_wires}).  The phase diagram is thus again given by Fig.\ \ref{1D_wires_fig}(c).  

Our discussion regarding optimization of the 1D wire proposal from Sec.\ \ref{1Dwiresproposal} applies to the present case with almost no modification, so here we will simply highlight a few important points.  For concreteness suppose that the parent superconductor in the device of Fig.\ \ref{2D_semiconductor_fig}(a) exhibits a pairing gap $\Delta_{sc}$ of a few Kelvin while the ferromagnetic insulator's Curie temperature is of order 100K.  With these rough energy scales in mind it is interesting to ask how large the gap for the topological superconductor formed in the 2DEG can be in principle.  The answer depends on the spin-orbit energy $E_{so} = \frac{1}{2}m\alpha^2$ and varies from a small fraction of $\Delta_{sc}$ when $E_{so}/\Delta_{sc} \ll 1$ up to the full value of $\Delta_{sc}$ in the opposite limit $E_{so}/\Delta_{sc} \gg 1$, highlighting the importance of sizable spin-orbit coupling.\cite{Disorder4}  (Capturing this physics requires a more rigorous treatment of the proximity effect as discussed in Sec.\ \ref{Preliminaries1D} and Appendix \ref{EffectiveActionDerivation}.)  Large values of $E_{so}$ also endow the topological phase with some robustness against disorder despite Anderson's theorem not applying here.  In the limit $h/E_{so} \ll 1$ electrons at the Fermi surface in the `spinless' regime are weakly perturbed by the induced Zeeman field, thus suppressing the pair-breaking effect of disorder.\cite{Disorder4}  Operating in this limit, however, may be neither possible nor desirable due to competing physics.  Satisfying the topological criterion in Eq.\ (\ref{topological_criterion_2DEG}) requires that $h$ exceed the inherited pairing field $\Delta$, and in practice it will likely prove advantageous to engineer the interface with the ferromagnetic insulator such that $h \gg \Delta$.  Though this might reduce the topological phase's gap below its theoretical maximum, large Zeeman fields facilitate tuning of the chemical potential into the `spinless' regime while simultaneously increasing the tolerance to long-range potential fluctuations.

Despite the exceptional purity with which they can be fabricated GaAs quantum wells are (unfortunately) unsuitable 2DEG candidates.  Their Rashba spin-orbit energy $E_{so}$ falls in the milliKelvin range\cite{GaAsSpinOrbit1,GaAsSpinOrbit2} due to the lightness of the constituent elements; consequently disorder is almost certain to dominate at the extraordinarily low densities required to access a topological phase in this material.\cite{Alicea}  More promising are InAs quantum wells, which feature $E_{so}$ values of a few tenths of a Kelvin\cite{InAsSpinOrbit} and also contact well with superconductors\cite{InAsProximityEffect2D}.  The spin-orbit energy can be increased further still by employing heavier materials such as InSb, or hole-doped semiconductors which generally exhibit larger effective masses and spin-orbit coupling strengths compared to their electron-doped analogs.  Interestingly, in the latter systems the `heavy holes' can be driven into a topological $f+if$-paired state that also supports Majorana modes.\cite{HoleDopedMajoranas}  (The extra Cooper pair angular momentum arises because as one traverses the Fermi surface in a heavy hole band the spins wind by $6\pi$ rather than $2\pi$.)  

Apart from materials considerations, experimentally demonstrating a topological phase in the proposed structure in Fig.\ \ref{2D_semiconductor_fig}(a) poses several other challenges.  Forming two high quality interfaces---one on each side of the 2DEG---presents a nontrivial fabrication problem.  Furthermore, the device exhibits limited tunability: whereas the Zeeman field in the 1D wire proposal from Sec.\ \ref{1Dwiresproposal} can be easily varied, $h$ is now largely fixed once the structure is fabricated.  And finally, the electrons in the semiconductor are effectively buried in the sandwich structure, making it difficult to manipulate or probe these carriers.  It may, fortunately, be possible to alleviate some of these challenges by employing modified setups.  

The interdigitated ferromagnet-superconductor device shown in Fig.\ \ref{2D_semiconductor_fig}(c), for example, can realize a topological phase while requiring lithography on only one side of the 2DEG.\cite{InterdigitatedDevice}  In such a structure the electrons inherit periodically modulated Zeeman and pairing fields due to the interdigitation.  Provided their Fermi wavelength exceeds the finger spacing, however, the electrons effectively feel the spatial average of these quantities, resulting in a robust topological phase under similar conditions to the sandwich structure in Fig.\ \ref{2D_semiconductor_fig}(a).  (A topological phase can appear even in the opposite limit, though with a diminished gap.)  Somewhat more elaborate interdigitated setups also allow one to \emph{electrically} generate vortices binding Majorana zero-modes by applying currents to modulate the superconducting phases along the fingers.\cite{InterdigitatedDevice}  

It is even possible to eliminate the ferromagnetic insulator altogether and drive a transition into the topological phase using an external magnetic field, similar to 1D wire setups.  As alluded to earlier the principal reason for the ferromagnetic insulator in Figs.\ \ref{2D_semiconductor_fig}(a) and (c) is that one wishes to generate a Zeeman field that cants the spins out of the 2DEG plane while avoiding orbital effects that would accompany an applied perpendicular magnetic field.  In-plane fields largely circumvent unwanted orbital effects but unfortunately do not open a `spinless' regime---at least in a semiconductor with only Rashba coupling.  Reference \onlinecite{Alicea} showed that in-plane fields can generate topological superconductivity provided one employs a 2DEG grown along the (110) direction with strong Rashba \emph{and} Dresselhaus\cite{Dresselhaus} spin-orbit coupling (such as InSb); see Fig.\ \ref{2D_semiconductor_fig}(d).  The special feature of this growth direction is that these two kinds of spin-orbit interactions conspire to rotate the plane in which the spins orient away from the 2DEG plane, so that an in-plane field plays a similar role to the ferromagnetic insulator in a Rashba-only setup.\cite{Alicea}  Another system that may eliminate the need for a ferromagnetic insulator is NaCoO$_2$.  Recent first-principles calculations predict that this material is a (conventional) bulk insulator with surface states exhibiting very strong Rashba coupling and \emph{spontaneous} time-reversal symmetry breaking that opens a broad `spinless' regime.\cite{NaCoO}  These predictions would be interesting to explore experimentally using ARPES.  

Early on Lee suggested another route to engineering topological superconductivity, using a 2D fully spin-polarized ferromagnetic metal adjacent to a spinful bulk $p+ip$ superconductor such as Sr$_2$RuO$_4$ [Fig.\ \ref{2D_semiconductor_fig}(e)].\cite{PatrickProposal}  Essentially, the ferromagnetic metal `filters out' one spin component from the parent superconductor and realizes a 2D spin\emph{less} $p+ip$ superconductor due to the proximity effect.  Later it was realized that even a conventional $s$-wave superconductor can drive a ferromagnetic metal into a topological phase provided appreciable spin-orbit coupling appears at the interface.\cite{HalfMetalMajoranas}  One virtue of these setups is that the topological phase can in principle exist without requiring fine-tuning of the chemical potential, though finding suitable 2D ferromagnets poses an experimental challenge.  A more exotic alternative proposal predicts that topological superconductivity can appear in proximity-coupled systems realizing a novel `quantum anomalous Hall' state.\cite{QAHproposal1,QAHproposal2}  This possibility provides strong motivation for experimentally pursuing this as yet undiscovered phase of matter.

\section{Experimental Detection Schemes}
\label{Detection}

One very simple (but also highly indirect) way of inferring the existence of a Majorana mode is through the detection of a topological quantum phase transition in the bulk of a 1D or 2D superconductor.  Consider for example the 1D spin-orbit-coupled wire proposal reviewed in Sec.\ \ref{1Dwiresproposal}, where one can tune between trivial and topological superconducting states simply by turning on a magnetic field.  Observing the bulk gap collapse and then reopen as the field strength increases would provide strong evidence for the onset of topological superconductivity and, by extension, the appearance of Majorana end states in the wire.  A topological phase transition in this setting also manifests itself in thermal and electrical transport\cite{TopologicalPhaseTransitionConductance,TopologicalCriticalPoints} as well as Coulomb blockade experiments\cite{CoulombBlockadeMajoranas}.  Clearly, however, more direct probes of Majorana modes are desirable.  Below we review three classes of such measurement schemes, based on tunneling, Josephson effects, and interferometry.

\subsection{Tunneling signatures of Majorana modes}
\label{Majorana_tunneling}

\begin{figure}
\includegraphics[width = 7cm]{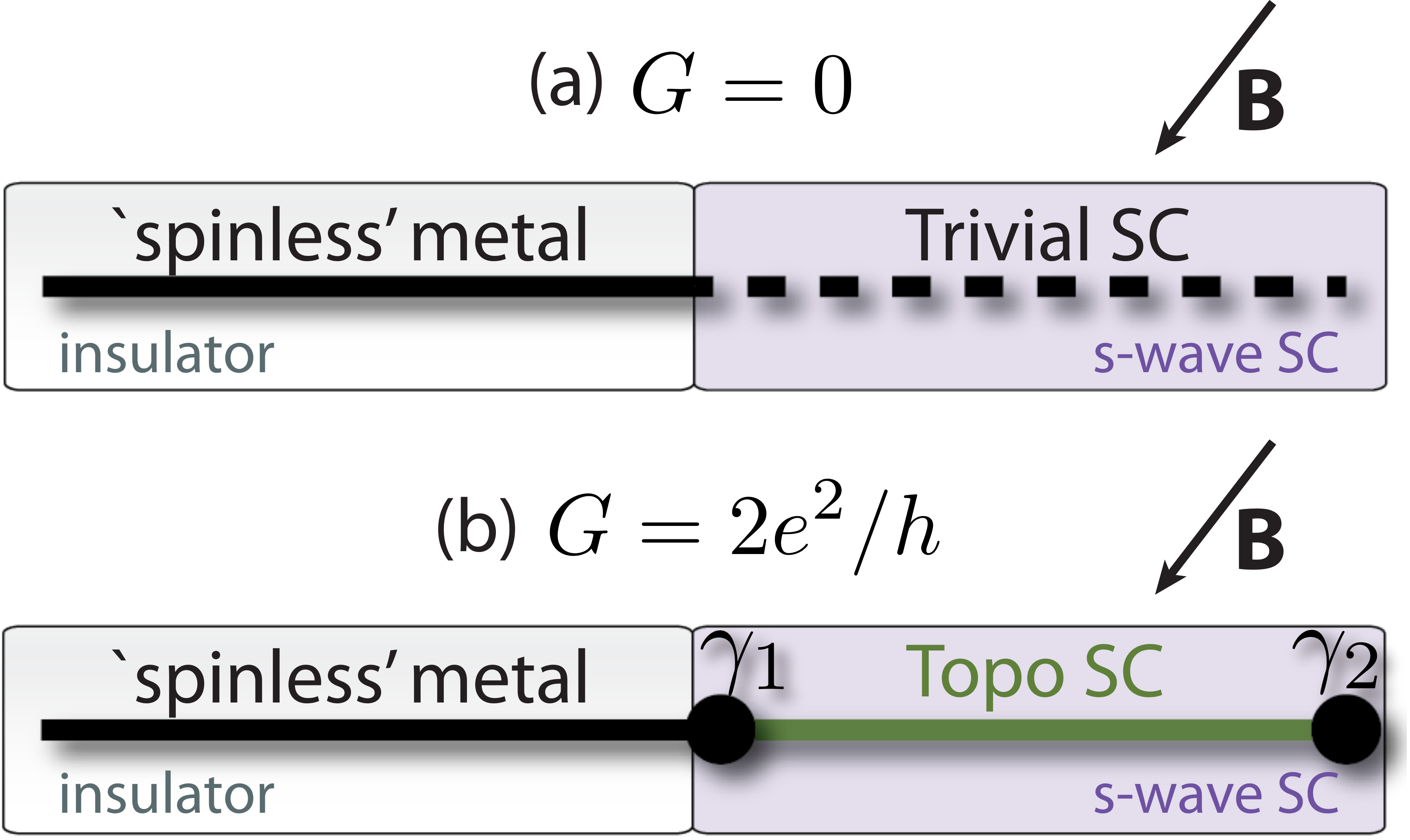}
\caption{A 1D spin-orbit-coupled wire draped across an insulator--$s$-wave superconductor junction allows one to detect Majorana modes via transport.  In (a) the wire's left half is gated into a `spinless' metallic regime while the right half forms a trivial gapped superconductor; in this setup the zero-bias conductance of the junction vanishes.  When the right half of the wire instead forms a topological phase as in (b), hybridization between the `spinless' metal and the Majorana $\gamma_1$ produces a quantized $2e^2/h$ zero-bias conductance.  The same results apply when the wire's left half forms a spinful metal, provided repulsive interactions are present.  This geometry can be readily adapted to detect isolated Majorana modes in many other settings.  }
\label{Tunneling_fig}
\end{figure}

Tunneling spectroscopy provides a powerful and conceptually appealing method for detecting Majorana zero-modes.  To illustrate the physics we will focus on the experimentally accessible geometry shown in Figs.\ \ref{Tunneling_fig}(a) and (b), where a long spin-orbit-coupled wire subjected to a magnetic field extends across an insulator--$s$-wave superconductor junction.\cite{ZeroBiasAnomaly7}  In both cases we assume that the wire's left half is gated into a `spinless' regime and remains metallic.  The proximity effect with the $s$-wave superconductor, however, drives the right half into a trivial gapped state in (a) but a topological phase supporting Majorana modes $\gamma_{1,2}$ in (b).  (One can tune between these configurations by adjusting the chemical potential in the right half.)  We would like to contrast the conductance of these two setups---particularly at zero bias---when current flows from the `spinless' metal into the superconductor.  Our approach will follow closely Refs.\ \onlinecite{ZeroBiasAnomaly3} and especially \onlinecite{MajoranaTunnelingFixedPoints} which emphasizes universal features of the problem.  

We initially attack the trivial case of Fig.\ \ref{Tunneling_fig}(a).  The only available low-energy degrees of freedom reside in the metallic region, which is taken to lie along $x<0$.  We therefore model the system by an effective Hamiltonian $H = H_{\rm metal} + H_{\rm junction}$ in which the first term describes the linearized kinetic energy near the Fermi level at $x<0$ while the second captures terms at $x = 0$ imposed by the superconductor.  Defining spinless fermion operators $\psi_{R/L}$ describing right/left-moving excitations in the wire, we have
\begin{equation}
  H_{\rm metal} = \int_{-\infty}^0 dx \left(-i v_F\psi_R^\dagger \partial_x\psi_R + i v_F \psi_L^\dagger \partial_x \psi_L\right)
\end{equation}
with $v_F$ the Fermi velocity.  One can rewrite $H_{\rm metal}$ in terms of a single chiral field $\psi(x)$ defined over all $x$ such that $\psi(x>0) = \psi_L(-x)$ and $\psi(x<0) = \psi_R(x)$, yielding 
\begin{equation}
  H_{\rm metal} = \int_{-\infty}^\infty  dx\left(-iv_F \psi^\dagger \partial_x \psi\right).
  \label{Hmetal}
\end{equation}
The leading local potential and pairing terms induced at the junction read
\begin{equation}
  H_{\rm junction} = \int_{-\infty}^\infty dx\left[u \psi^\dagger \psi + \Delta(\psi \partial_x\psi + H.c.)\right]\delta(x).
  \label{HjunctionTrivial}
\end{equation}
In particular, the $\Delta$ term (which requires a derivative by Fermi statistics) allows Cooper pairs to hop from the metallic region of the wire into the superconductor.  

The full Hamiltonian $H$ takes the form of a Bogoliubov-de Gennes equation and is therefore diagonalized with quasiparticle operators carrying energy $E$ of the form
\begin{equation}
  \Gamma_E = \int_{-\infty}^\infty dx e^{-i\frac{E x}{v_F}}[P_E(x)\psi(x) + H_E(x) \psi^\dagger(x)],
\end{equation}
where $P_E$ and $H_E$ determine the particle- and hole-like amplitudes for the quasiparticle wavefunctions.  The conductance we are after follows from the scattering matrix $S(E)$ that relates states incident on the superconductor to the reflected states.  Recalling the definition of $\psi(x)$ in terms of left and right movers, the $S$-matrix components are defined by
\begin{eqnarray}
\left[ \begin{array}{c}
  P_E(\infty) \\
  H_E(\infty)
  \end{array}\right]  
  = \left[ \begin{array}{cc}
  S_{PP}(E) & S_{PH}(E) \\
  S_{HP}(E) & S_{HH}(E)
  \end{array}\right] \left[ \begin{array}{c}
  P_E(-\infty) \\
  H_E(-\infty)
  \end{array}\right].
\end{eqnarray}
Since $|S_{PH}(E)|^2$ is the probability that an incident electron at energy $E$ Andreev reflects as a hole at the junction, passing charge $2e$ into the superconductor, a bias voltage $V$ applied across the junction generates a current
\begin{equation}
  I = \frac{2e}{h} \int_0^{eV}dE |S_{PH}(E)|^2.
\end{equation}
The differential conductance $G(V) = \frac{dI}{dV}$ follows as
\begin{equation}
  G(V) = \frac{2e^2}{h}|S_{PH}(eV)|^2.
\end{equation}

Very general arguments adapted from Refs.\ \onlinecite{MajoranaModesWithChargeTransport} and \onlinecite{3DTI_Majorana_edge_states1} strongly constrain the $S$-matrix at zero energy, and hence the zero-bias conductance.  First, particle-hole symmetry of the Bogoliubov-de Gennes equation dictates that $\Gamma_E = -\Gamma_E^\dagger$, which in turn implies $S(E) = \sigma^x S^*(-E)\sigma^x$.  This relation, together with unitarity of the $S$-matrix as required by current conservation, restricts $S(0)$ to one of two possible forms:
\begin{equation}
S(0) = \left( \begin{array}{cc}
  e^{i\alpha} & 0 \\
  0 &  e^{-i\alpha} 
  \end{array}\right) ~~~{\rm or} ~~~ \left( \begin{array}{cc}
  0 & e^{i\beta} \\
  e^{-i\beta} & 0
  \end{array}\right)
  \label{Smatrix_possibilities}
\end{equation}
for some phases $\alpha,\beta$.  The purely diagonal case corresponds to the onset of perfect normal reflection---with unit probability an electron reflects as an electron and similarly for holes---and hence a vanishing zero-bias conductance.  In contrast, the off-diagonal possibility yields perfect \emph{Andreev} reflection; here electrons scatter perfectly into holes and vice versa, yielding a quantized zero-bias conductance $G = 2e^2/h$.  These very different limits represent renormalization group fixed points at which the superconductor imposes either perfect normal reflecting or perfect Andreev reflecting boundary conditions on the metal at low energies.\cite{MajoranaTunnelingFixedPoints}  We stress that Eq.\ (\ref{Smatrix_possibilities}) relies only on the metal being `spinless', and holds even when Majorana zero-modes are present as in Fig.\ \ref{Tunneling_fig}(b).

By explicitly calculating the $S$-matrix for Eqs.\ (\ref{Hmetal}) and (\ref{HjunctionTrivial}) it is straightforward to show that $S(0)$ is purely diagonal so that the zero-bias conductance vanishes in Fig.\ \ref{Tunneling_fig}(a).  The $\Delta$ term in $H_{\rm junction}$ does permit Cooper pairs to tunnel into the superconductor, but because the metal is `spinless' the probability vanishes at zero energy due to Pauli blocking.  In other words the non-interacting `spinless' metal we have treated so far generically flows at low energies to the perfect normal reflection fixed point.  It is, however, essential to understand the impact of interactions---which transform the metal into a Luttinger liquid---on this result.  Reference \onlinecite{MajoranaTunnelingFixedPoints} demonstrates via bosonization that this fixed point remains stable even in the interacting case except when the wire exhibits very strong attractive interactions.  Remarkably, for such an attractive wire a pair of asymptotically decoupled Majorana modes emerges \emph{dynamically} at the junction and drives the system to the perfect Andreev reflection fixed point with quantized conductance.\cite{MajoranaTunnelingFixedPoints}

Dramatically different physics arises for the topological setup in Fig.\ \ref{Tunneling_fig}(b) due to the Majorana modes.  We will assume zero overlap between $\gamma_2$ and $\gamma_1$ and now focus on a junction Hamiltonian 
\begin{equation}
  H_{\rm junction} = t\int_{-\infty}^\infty dx \gamma_1(\psi^\dagger-\psi)\delta(x)
  \label{HjunctionTopological}
\end{equation}
that hybridizes $\gamma_1$ and the `spinless' metal.  [The $u$ and $\Delta$ terms from Eq.\ (\ref{HjunctionTrivial}) are qualitatively unimportant so we neglect them for simplicity.]  For any $t\neq 0$ the Majorana $\gamma_1$ no longer represents a zero-energy mode since $[H_{\rm junction},\gamma_1] \neq 0$.  A single Majorana zero-mode can never exist on its own, however, so where is $\gamma_2$'s partner?  The answer is that the zero-mode described by $\gamma_1$ when $t = 0$ gets absorbed into the metal where it becomes a delocalized plane wave; explicitly, one can readily verify that for $t \neq 0$ $\gamma_2$'s partner is $\tilde \gamma_1 \propto \int_{-\infty}^\infty dx(\psi + \psi^\dagger)$.\footnote{The tendency for Majorana zero-modes to delocalize when coupled to extended gapless fermionic modes is rather general.  Incidentally, for this reason our view is that proposals that invoke proximity effects with gapless superconductors such as the cuprates to drive topological superconducting states are most likely ill-suited for exploring Majorana physics.}  The hybridization with $\gamma_1$ leading to this delocalized Majorana plane-wave mode mediates perfect Andreev reflection at the junction at low energies.  Indeed, extracting the $S$-matrix from the Hamiltonian given by Eqs.\ (\ref{Hmetal}) and (\ref{HjunctionTopological}) yields a Lorentzian conductance 
\begin{equation}
  G(V) = \frac{2e^2}{h}\left[\frac{1}{1 + \left(\frac{e V v_F}{2 t^2}\right)^2}\right]
\end{equation}
that collapses to $2e^2/h$ in the zero-bias limit.  From a renormalization group perspective, coupling to $\gamma_1$ drives the non-interacting `spinless' metal to the perfect Andreev reflection fixed point characterized by a quantized zero-bias conductance, in stark contrast to the trivial setup of Fig.\ \ref{Tunneling_fig}(a).  This result survives even for a Luttinger liquid unless strongly repulsive interactions with a Luttinger parameter $g < 1/2$ are present; there the repulsion obliterates the zero-bias conductance peak entirely.\cite{MajoranaTunnelingFixedPoints} 

What happens when the metallic region of the wires in Fig.\ \ref{Tunneling_fig} carry spin?  In this case the arguments leading to Eq.\ (\ref{Smatrix_possibilities}), which underlies the conductance dichotomy for the setups in (a) and (b), provide much weaker constraints on the $S$-matrix.  Furthermore, a local pairing term at the junction $\Delta_0 [\psi_\uparrow(x = 0)\psi_\downarrow(x = 0)+H.c.]$ now allows singlet Cooper pairs to tunnel into the superconductor \emph{without} Pauli blocking at zero energy.  When the spinful metal impinges on a topological superconductor as in Fig.\ \ref{Tunneling_fig}(b) a quantized $2e^2/h$ zero-bias conductance nevertheless emerges from coupling to the Majorana $\gamma_1$, both at the non-interacting level and over a range of interactions.\cite{MajoranaTunnelingFixedPoints}  For a non-interacting spinful metal adjacent to a trivial superconductor as in Fig.\ \ref{Tunneling_fig}(a), however, $\Delta_0$ produces an $S$-matrix that yields a zero-bias conductance ranging from 0 to $4e^2/h$ depending on parameters---potentially making it difficult to differentiate with the topological case.  Fortunately this result is non-generic.  Arbitrarily weak repulsive interactions drive the system to the perfect normal reflection fixed point at low energies, leading to a vanishing zero-bias conductance just as for a `spinless' metal.\cite{MajoranaTunnelingFixedPoints}  Thus sharp tunneling signatures of Majorana zero-modes appear also in the spinful case.  Adding more channels, however, obscures these signatures since the conductance in both the trivial and topological setups is then non-universal, at least in the free-fermion limit.\cite{ZeroBiasAnomaly6}

While our discussion so far centered on a specific geometry involving 1D wires, the conclusions apply far more generally.  Nowhere in this analysis did we use the fact that the Majorana $\gamma_1$ derived from a topological region of a 1D wire.  In fact the quantized `zero-bias anomaly' captured above has been discussed in numerous contexts from several different perspectives\cite{ZeroBiasAnomaly1,ZeroBiasAnomaly2,ZeroBiasAnomaly3,ZeroBiasAnomaly4,ZeroBiasAnomaly5,ZeroBiasAnomaly6,ZeroBiasAnomaly7,MajoranaTunnelingFixedPoints} and is a general property of spinless or spinful 1D metals tunneling onto an isolated Majorana mode.  The setups of Fig.\ \ref{Tunneling_fig} can be readily adapted to probe Majorana modes in 2D topological insulator edges, half-quantum vortices in a 2D spinful $p+ip$ superconductor, chiral Majorana edge states\cite{ZeroBiasAnomaly3}, \emph{etc}.  (Though one should keep in mind that the small mini-gap typically associated with vortices and edge states can place stringent limits on temperature and resolution.)

One might view the tunneling signatures of Majorana modes discussed above as somewhat less than a `smoking gun' detection method since zero-bias anomalies can arise from unrelated sources in mesoscopic systems.  We are somewhat sympathetic to this perspective but stress the following points.  First, whereas Majorana modes produce a zero-bias conductance of $2e^2/h$ this quantized value need not appear when tunneling into a conventional `accidental' low-energy mode.  A second, more important point is that in many of the proposals we reviewed this conductance peak can be controllably brought in and out of resonance by inducing a transition between topological and trivial phases (by, say, adjusting the magnetic field or gating).  While measuring a zero-bias peak may not by itself constitute a smoking-gun signature of a Majorana mode, observing this peak collapse and revive in accordance with theoretical expectations arguably would.  

Majorana zero-modes provide several other noteworthy transport signatures, including through current noise.\cite{ZeroBiasAnomaly1,ZeroBiasAnomaly2,ZeroBiasAnomaly3,ZeroBiasAnomaly5}  In particular, coupling a \emph{pair} of 1D metallic wires to a topological superconductor introduces Majorana-mediated `crossed Andreev reflection'; such processes arise when a Cooper pair enters the superconductor by combining a single electron from each metal and generate maximally correlated current noise in the two wires.\cite{ZeroBiasAnomaly2,ZeroBiasAnomaly3}  The tunneling conductance through a quantum dot has also been predicted to change qualitatively when the dot couples to a Majorana mode,\cite{QuantumDotMajoranaDetection1,QuantumDotMajoranaDetection2} and a related setup allows one to probe the lifetime of the ordinary fermionic state formed by a pair of Majorana modes in the presence of `quasiparticle poisoning'\cite{MajoranaLifetime}.  Additional transport signatures arise when a Majorana mode couples to a nanomechanical resonator.\cite{MajoranaNanomechanicsDetection}  Spin-polarized scanning tunneling microscopy may also be used to identify fingerprints of Majoranas.\cite{Disorder8}  Finally, in a mesoscopic topological superconductor where charging energy is important, spatially separated Majorana modes can mediate non-local electron `teleportation' that can be detected with transport.\cite{MajoranaTransportWithInteractions1,MajoranaTransportWithInteractions2}

\subsection{Fractional Josephson effects}
\label{Fractional_Josephson_effect}

Recall from the introduction that a system with $2N$ well-separated Majorana zero-modes $\gamma_{1,\ldots,2N}$ exhibits $2^N$ degenerate ground states.  By defining conventional fermion and number operators 
\begin{equation}
  f_j = (\gamma_{2j-1} + i \gamma_{2j})/2, ~~~~ n_j = f_j^\dagger f_j
  \label{fn}
\end{equation}
the ground-state manifold can be conveniently labeled by $|n_1,\ldots,n_N\rangle$, where $n_j = 0,1$ specify topologically protected qubit states.  [The pairing of Majoranas in Eq.\ (\ref{fn}) is completely arbitrary but always sufficient; ground states labeled with different pairings are simply connected by unitary transformations.]  The experimental detection methods discussed so far allow one to deduce the \emph{existence} of Majorana modes but provide no information about the qubits they encode.  One way to extract this information is to prepare the system into a ground state and then adiabatically bring two Majorana modes---say $\gamma_1$ and $\gamma_2$---in close proximity so that their wavefunctions overlap appreciably.  The resulting hybridization of these modes can be modeled by a Hamiltonian $H_\epsilon = i \frac{\epsilon}{2} \gamma_1 \gamma_2 = \epsilon(n_1-1/2)$.  Taking $\epsilon >0$ for concreteness, the system remains in a ground state if $n_1 = 0$ whereas the fusion of $\gamma_1$ and $\gamma_2$ yields an extra finite-energy quasiparticle if $n_1 = 1$.  One can thus read out the state of $n_1$ by detecting the presence or absence of such a quasiparticle.  (Note that multiple measurements may be required since the system can form a superposition of $n_1 = 0$ and 1 states.)  Fusing Majoranas across a Josephson junction both enables qubit readout along these lines\cite{FuKane} \emph{and} provides an unambiguous fingerprint of these modes.  This approach is particularly well-suited for 1D topological superconductors, so we will confine our discussion to this class of systems. 

\begin{figure}
\includegraphics[width = \linewidth]{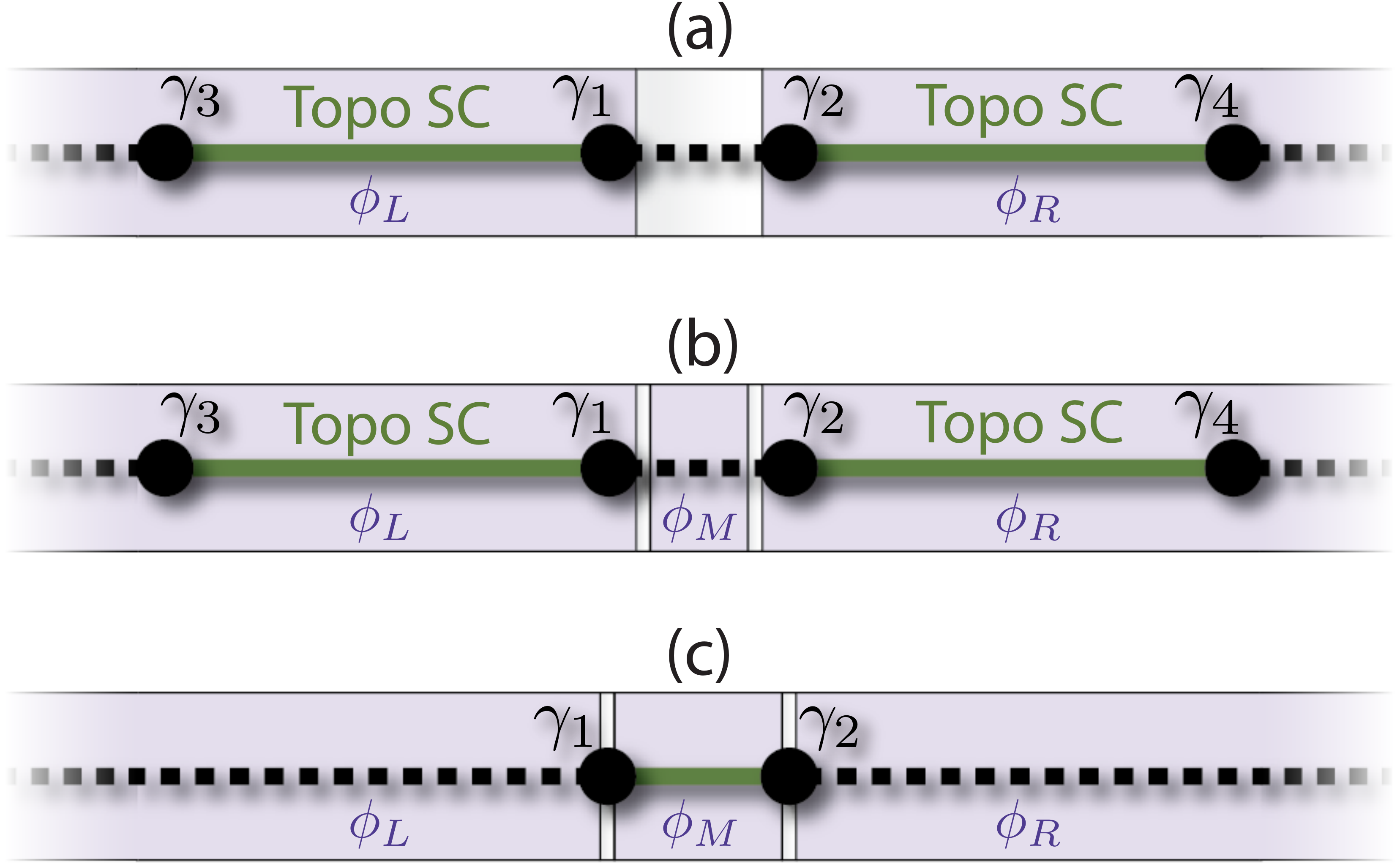}
\caption{(a) Basic experimental setup required to observe the fractional Josephson effect stemming from Majorana modes fused across a superconductor-insulator-superconductor junction.  Purple regions indicate $s$-wave superconductors (with phases $\phi_{L,R}$) that drive the green regions into a 1D topological state; dashed regions are trivially gapped.  The Majoranas $\gamma_{1,2}$ mediate a component of the Josephson current that is $4\pi$ periodic in $\phi_R-\phi_L$.  When the barrier in the junction is replaced by a superconductor with phase $\phi_M$ as in (b), $\gamma_{1,2}$ mediate a second type of unconventional current that is $4\pi$ periodic in $\phi_{L,R}$ and can be isolated with Shapiro step measurements.  The setup in (c), while superficially similar to (a) and (b), yields only conventional Josephson physics with $2\pi$ periodicity in $\phi_{L,R}$.}
\label{Josephson_fig}
\end{figure}

Figure \ref{Josephson_fig} sketches the basic type of setup required.  Here two 1D topologically superconducting regions emerge due to the proximity effect with $s$-wave superconductors that are separated by an insulating barrier, while dashed regions are assumed to be trivially gapped.  The topological segments can arise from any number of systems---a 2D topological insulator edge, 1D spin-orbit-coupled wires, 3D topological insulator nanowires, \emph{etc.}---but should be long enough that the outer Majoranas $\gamma_{3,4}$ overlap negligibly with the central Majoranas $\gamma_{1,2}$.\footnote{To avoid Kramer's degeneracy in the topological insulator realizations, we assume that time-reversal symmetry is always broken.  We also note that the presence of $\gamma_3$ and $\gamma_4$ simplifies the interpretation of the Josephson physics we will describe but is not necessary.  If desired one can eliminate these without affecting the physics at the Josephson junction by connecting the outer ends of the two topological segments shown in Fig.\ \ref{Josephson_fig}(a).  See Ref.\ \onlinecite{MajoranaQSHedge}.}  The insulating barrier, however, should be sufficiently narrow that $\gamma_1$ and $\gamma_2$ hybridize strongly.  Given this setup, our objective is to understand the zero-bias current $I$ flowing across the Josephson junction as one varies the phase difference $\Delta\phi \equiv \phi_R-\phi_L$ between the right and left $s$-wave superconductors.  

This current consists of two contributions, $I = I_{2e} + I_e$.  The first, $I_{2e}$, denotes the conventional Josephson current that arises from Cooper-pair tunneling across the insulating barrier and is $2\pi$-periodic in $\Delta\phi$.  As originally shown by Kitaev\cite{1DwiresKitaev} the hybridized Majorana modes $\gamma_{1,2}$ mediate a new contribution $I_e$ which is our primary focus.  To extract the salient universal features of $I_e$ in a very direct way we will model the two topological regions of Fig.\ \ref{Josephson_fig}(a) as $N$-site chains described by Kitaev's toy lattice model given in Eq.\ (\ref{Hkitaev}).  Furthermore, we fine-tune $\mu = 0$ and $t = \Delta$ for each region so that the Majorana zero-modes can be trivially identified.  Defining operators $c_{L/Rx}^\dagger$ that add fermions to the left/right topological segments, the full Hamiltonian for the junction is then taken to be 
\begin{eqnarray}
  H &=& \sum_{a = L/R}H_a + H_\Gamma,
  \label{Hjunction}
  \\
  H_{a} &=& - \frac{t}{2}\sum_{x = 1}^{N-1}(c_{ax}^\dagger c_{ax+1} + e^{i\phi_a} c_{ax} c_{ax+1}+ h.c.),
  \\
  H_\Gamma &=& -\Gamma(c_{LN}^\dagger c_{R1} + H.c.),
\end{eqnarray}
where $H_\Gamma$ describes single-electron tunneling across the barrier with strength $\Gamma >0$.  

Let us recall the following two facts derived in Sec.\ \ref{1D_toy_model}:  $(i)$ $H_a$ supports Majorana zero-modes localized at sites 1 and $N$ of each chain [see Fig.\ \ref{KitaevModelFig}(b)] and $(ii)$ the zero-modes $\gamma_{1,2}$ at the junction are related to the lattice fermion operators by $c_{LN} = e^{-i \phi_L/2}(\gamma_{1} + i \gamma_1')/2$ and $c_{R1} = e^{-i \phi_R/2}(\gamma_{2}' + i \gamma_2)/2$.  Here $\gamma_1'$ and $\gamma_2'$ hybridize with Majorana fermions at neighboring sites and form conventional finite-energy fermions.  One can therefore project $H$ onto the zero-energy subspace of $H_a$ by sending
\begin{equation}
  c_{LN} \rightarrow \frac{1}{2}e^{-i \phi_L/2}\gamma_{1}, ~~~~~ c_{R1} \rightarrow \frac{i}{2}e^{-i \phi_R/2}\gamma_{2},
  \label{ZeroModeProjection}
\end{equation}
which yields an effective low-energy Hamiltonian
\begin{eqnarray}
  H_{\rm eff} &=& -\frac{\Gamma}{2}\cos\left(\frac{\Delta\phi}{2}\right)i \gamma_1\gamma_2 
  \nonumber \\
  &=& 
  -\Gamma\cos\left(\frac{\Delta\phi}{2}\right)(n_1-1/2).
  \label{H_Gamma_projected}
\end{eqnarray}
Crucially, the occupation number $n_1$ is a  conserved quantity since $[H_{\rm eff},n_1] = 0$.  Thus if the system begins in a state with $n_1 = n_1^{i}$, then 
varying the phase difference $\Delta\phi$ across the junction yields a Majorana-mediated current
\begin{equation}
  I_e = \frac{2e}{\hbar}\frac{\langle H_{\rm eff}\rangle}{d\Delta\phi} = \frac{e\Gamma}{2\hbar}\sin\left(\frac{\Delta\phi}{2}\right)(2n_1^i-1).
  \label{I_e}
\end{equation}
(If the system does not form an $n_1$ eigenstate, then the current $I_e$ above emerges with a probability determined by the relative amplitude for $n_1^i = 0,1$ states.)  

Equation (\ref{I_e}) reflects a \emph{fractional Josephson effect}---$I_e$ originates from tunneling `half' of a Cooper pair across the junction and exhibits $4\pi$ periodicity in $\Delta\phi$.  The first property is easy to understand: Cooper-pair hopping dominates the Josephson current in conventional $s$-wave superconductors because the bulk gap suppresses single-electron tunneling, but since $H_\Gamma$ couples zero-modes this suppression disappears.  The $4\pi$ periodicity is much subtler.  Indeed, the original Hamiltonian in Eq.\ (\ref{Hjunction}) is clearly $2\pi$ periodic in both $\phi_L$ and $\phi_R$, so how can the current $I_e$ exhibit $4\pi$ periodicity?  This is possible because while the Hamiltonian is $2\pi$ periodic, the \emph{physical states are not}.  Suppose, for example, that $\Delta\phi = 0$ and $n_1^i = 1$ so that the system begins in a ground state of $H_{\rm eff}$ in Eq.\ (\ref{H_Gamma_projected}) with energy $E_i = -\Gamma/2$.  Because $n_1$ is conserved, after advancing $\Delta\phi$ by $2\pi$ the system ends in a physically distinct excited state with energy $E_f = +\Gamma/2$ and hence an extra finite-energy quasiparticle at the junction.\footnote{Note that this does not mean that the ground state changes parity under $\Delta\phi\rightarrow \Delta\phi + 2\pi$.  Due to the outer Majoranas $\gamma_{3,4}$ there is always a degeneracy between even and odd parity eigenstates for any $\Delta\phi$.  The key point is that the ground state with a given fermion parity may be inaccessible.}  Global fermion parity conservation dictates that the system can decay back to a ground state only if fermions can transfer between the inner and outer Majoranas in Fig.\ \ref{Josephson_fig}(a), which we precluded due to their large spatial separation.  Advancing $\Delta\phi$ by $2\pi$ a second time restores the ground state with which we started.  This doubled $\Delta\phi$-periodicity in the physical states underlies the fractional Josephson effect uncovered above.  

Measuring a $4\pi$-periodic contribution to the Josephson current would undoubtedly qualify as a smoking-gun signature of Majorana fermions.  Moreover, because the sign of the current $I_e$ in Eq.\ (\ref{I_e}) is tied to the occupation number $n_1^i$ of the Majoranas at the junction, this technique enables qubit readout as claimed above.\cite{FuKane}  The following experimental realities should, however, be kept in mind.  First, $I_e$ must be disentangled from the (potentially much larger) $2\pi$-periodic component $I_{2e}$ that flows in parallel.  To obtain a crude order-of-magnitude estimate for $I_e$, with $\Gamma \sim 1$K in Eq.\ (\ref{I_e}) the associated critical current is $I_e^c = e \Gamma/(2\hbar) \sim 10$nA, which roughly sets the required current resolution.  Second, due to the finite extent of their wavefunctions the outer Majorana modes $\gamma_{3,4}$ of Fig.\ \ref{Josephson_fig}(a) will inevitably couple to $\gamma_{1,2}$ with a characteristic energy $\delta E \propto e^{-L/\xi}$ ($L$ denotes the size of the topological regions and $\xi$ is the topological phase's coherence length).  Though exponentially suppressed in $L/\xi$, this hybridization spoils conservation of $n_1$ and restores $2\pi$ periodicity of the current $I_e$.\cite{1DwiresKitaev,PikulinFractionalJosephson}  To circumvent this problem $\Delta\phi$ should cycle on a time scale that is short compared to $\hbar\Gamma/\delta E^2$,\cite{HeckFractionalJosephson} but long on the scale set by the inverse bulk gap.  Third, inelastic processes involving stray quasiparticles---which can appear, \emph{e.g.}, because the system was imperfectly initialized or due to thermal excitation---provide another means of switching the value of $n_1$ to relax back to the ground state of $H_{\rm eff}$.  This, too, can restore $2\pi$ periodicity of $I_e$ if $\Delta\phi$ cycles on scales much longer than the typical switching time; on shorter scales $I_e$ will exhibit `telegraph noise' where the sign of the Majorana-mediated current abruptly changes with time, which would also be remarkable to observe.\cite{MajoranaQSHedge}  Even in the presence of relaxation processes signatures of the fused Majorana modes at the junction appear through the current noise spectrum\cite{MajoranaQSHedge,MeyerFractionalJosephson} and in transients\cite{AguadoFractionalJosephson}.

A second type of fractional Josephson effect arises when the barrier material in the junction forms a superconductor with phase $\phi_M$ as shown in Fig.\ \ref{Josephson_fig}(b).\cite{JiangFractionalJosephson}  The influence of the superconducting barrier can be crudely modeled by supplementing the Hamiltonian in Eq.\ (\ref{Hjunction}) with a new term
\begin{equation}
  H_M = \Delta_M(e^{i\phi_M}c_{LN} c_{R1} + H.c.)
\end{equation}
that Cooper pairs fermions at the inner ends of the topological regions.  Using Eq.\ (\ref{ZeroModeProjection}) to project the full Hamiltonian onto the low-energy subspace formed by the Majoranas then produces a modified effective Hamiltonian
\begin{eqnarray}
  H_{\rm eff}' &=& H_{\rm eff} + \Delta_M \cos\left(\phi_M-\frac{\phi_L+\phi_R}{2}\right)(n_1-1/2).
\end{eqnarray}
The meaning of the $\Delta_M$ term above can be simply understood by promoting $e^{i\phi_{L,M,R}}$ to Cooper-pair creation operators; $\Delta_M$ then clearly reflects processes whereby a Cooper pair in the central region fractionalizes, with half entering the left topological segment and the other half entering the right.  This Cooper-pair splintering allows for a current 
\begin{equation}
  I_{e}' = \frac{e\Delta_M}{\hbar}\sin\left(\frac{\phi_L+\phi_R}{2}-\phi_M\right)(2n_1^i-1)
\end{equation}
injected into the middle region to be carried away in equal parts into the left and right topological superconductors.  Like the fractional Josephson current in Eq.\ (\ref{I_e}), $I_e'$ exhibits $4\pi$ periodicity in both $\phi_L$ and $\phi_R$ because of conservation of $n_1$, and also allows one to read out the initial occupation number $n_1^i$.  A promising feature of this setup is that Shapiro step measurements can be used to isolate the $I_e'$ contribution from the parallel components $I_{2e}$ and $I_e$.\cite{JiangFractionalJosephson}

Although the Majorana-mediated currents $I_e$ and $I_e'$ were derived in a fine-tuned toy model, their anomalous $4\pi$ periodicity has a topological origin and thus arises far more generally (subject to the caveats noted above).  Indeed, fractional Josephson effects have been captured in more realistic models for 1D $p$-wave superconductors\cite{Kwon,Kwon2}, 2D topological insulator edges\cite{MajoranaQSHedge,JiangFractionalJosephson}, 3D topological insulator surfaces\cite{FuKane,IoselevichFractionalJosephson}, single-\cite{1DwiresLutchyn,1DwiresOreg,JiangFractionalJosephson,ChengFractionalJosephson} and multi-channel\cite{Multichannel5,Multichannel8} spin-orbit-coupled wires, among other systems\cite{Kwon,Kwon2,BlackFractionalJosephson}, and have even been shown to survive when capacitive charging energy is incorporated\cite{HeckFractionalJosephson}.  We also note that a long Josepshon junction formed by 2D topological superconductors supports an unconventional Fraunhofer pattern arising from chiral Majorana edge states.\cite{InterdigitatedDevice}  

Some experimental setups more easily lend themselves to observing fractional Josephson effects than others since this phenomenon requires stabilizing extended topological regions on both sides of the junction as shown in Figs.\ \ref{Josephson_fig}(a) and (b).  For instance, topological insulator edges may be ideally suited for this type of experiment since there the topological phase is `easy' to access in the sense discussed in Sec.\ \ref{2DTIproposal}.  We should caution that Josephson physics in the setup of Fig.\ \ref{Josephson_fig}(c)---which in realizations such as 1D spin-orbit-coupled wires should be much simpler to realize than those of Figs.\ \ref{Josephson_fig}(a) and (b)---is \emph{always} $2\pi$ periodic in the superconducting phases.  Perhaps the simplest way to see this is by observing that in the limit where $\gamma_1$ and $\gamma_2$ hybridize significantly across the junction in Fig.\ \ref{Josephson_fig}(c), the central region is in no meaningful way topological.

\subsection{Interferometry}
\label{Interferometry}

Several interferometric schemes have been proposed for detecting Majorana modes in topological superconductors.\cite{SrRu,MajoranaModesWithChargeTransport,3DTI_Majorana_edge_states1,BenjaminInterferometry,VortexAharonovCasher,GrosfeldStern,ZaanenInterferometry,SemiconductorInterferometry}  While the setups required generally pose greater fabrication challenges compared to the measurement techniques reviewed earlier, the signatures that appear are unambiguous and conceptually illuminating.  For concreteness we will concentrate on interferometers employing 3D topological insulators; we stress, however, that similar ideas can be applied to many other realizations of 2D topological superconductivity reviewed in Sec.\ \ref{2D_toy_model_realizations}.  

\begin{figure}
\includegraphics[width = \linewidth]{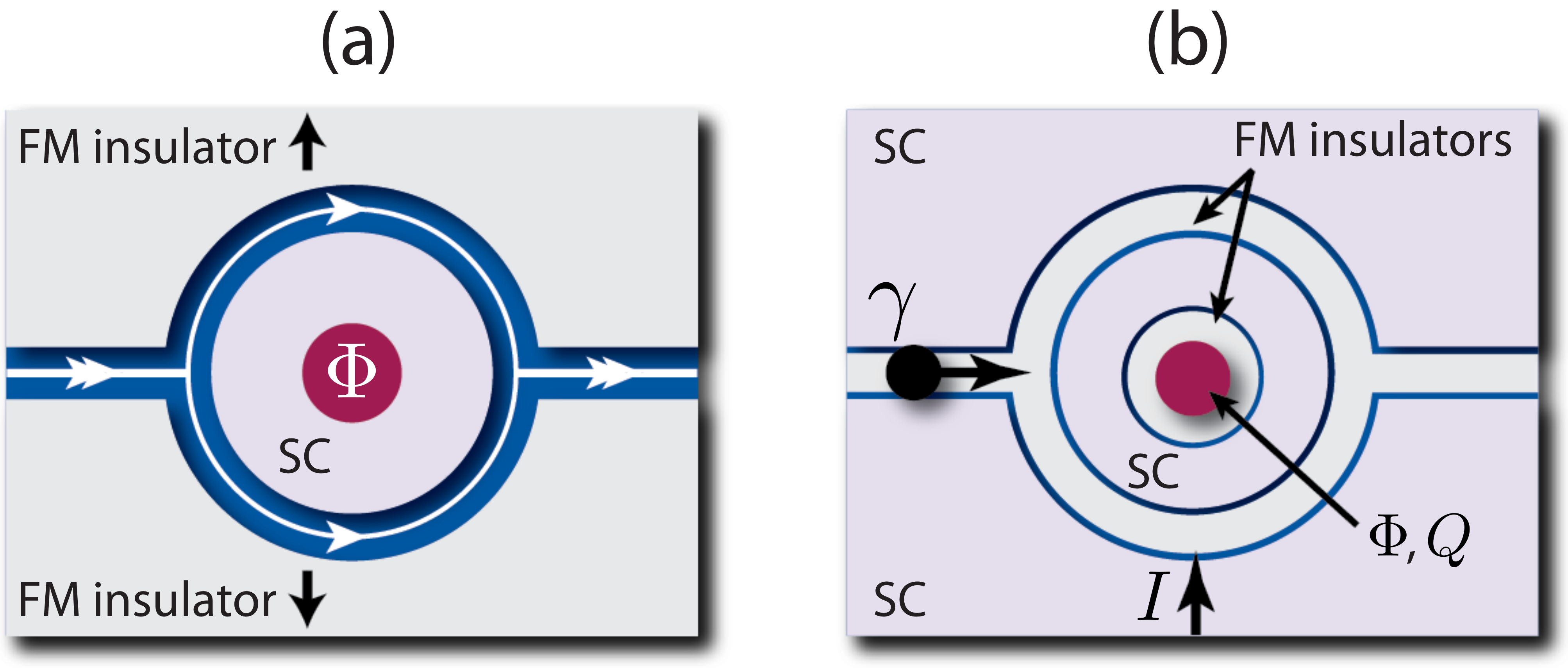}
\caption{Interferometers fabricated from ferromagnetic insulators and $s$-wave superconductors deposited on a 3D topological insulator surface.  In (a) the ferromagnets have opposite magnetizations, yielding a conventional chiral edge state (double arrows) that fractionalizes into chiral Majorana modes (single arrows) around the superconductor.  With a flux $\Phi = \frac{hc}{2e}N_v$ threading the center, the zero-bias conductance describing current flow from the left domain wall into the superconductor vanishes for even $N_v$ and is quantized at $\frac{2e^2}{h}$ for odd $N_v$.  In (b), when $\Phi = 0$ Josephson vortices flowing through the interferometer produce a vortex current that oscillates with the charge $Q$ on the central island (due to the Aharonov-Casher effect).  When $\Phi = \frac{hc}{2e}$, however, the oscillations disappear as a consequence of non-Abelian statistics.}
\label{Interferometry_fig}
\end{figure}

Consider the geometry shown in Fig.\ \ref{Interferometry_fig}(a) where an $s$-wave superconductor and two ferromagnetic insulators---importantly, with opposite magnetizations---reside on a 3D topological insulator surface.\cite{MajoranaModesWithChargeTransport,3DTI_Majorana_edge_states1}  The ferromagnets drive the upper and lower portions of the surface into gapped quantum Hall states with Hall conductivities $\sigma_{xy} = \pm \frac{e^2}{2h}$.\cite{FuKaneSurfaceQHE,TopologicalFieldTheoryTI,MajoranaModesWithChargeTransport}  Since the Hall conductivities differ by $\frac{e^2}{h}$, an `ordinary' chiral edge state (denoted by double arrows in the figure) appears at each magnetic domain wall.  As described in Sec.\ \ref{3DTIproposal} the $s$-wave superconductor drives the surface beneath it into a time-reversal-invariant 2D topological superconductor supporting chiral \emph{Majorana} edge states whose chirality follows from the neighboring ferromagnet's magnetization.  When the ordinary chiral mode at the left magnetic domain wall meets the superconductor, it therefore fractionalizes into a pair of co-propagating Majorana modes and then recombines at the right magnetic domain wall.\cite{MajoranaModesWithChargeTransport,3DTI_Majorana_edge_states1}  

Of interest here is the conductance characterizing current flow from the left magnetic domain wall into the superconductor, when a magnetic flux $\Phi$ pierces the central region of Fig.\ \ref{Interferometry_fig}(a).  Following the scattering analysis of Sec.\ \ref{Majorana_tunneling} the conductance at a bias voltage $V$ is given by $G(V) = \frac{2e^2}{h}|S_{PH}(eV)|^2$, where $S_{PH}$ is now the amplitude for an electron incident from the left to transmit a Cooper pair to the superconductor and exit as a hole at the right domain wall.  As in our treatment of a spinless metal impinging on a superconductor, particle-hole symmetry and current conservation again constrain the zero-bias conductance to $G(0) = 0$ or $\frac{2e^2}{h}$.  In other words, at zero energy an incident electron $c_{\rm in}^\dagger$ first splinters into Majoranas $\gamma_{t/b}$ localized at the top/bottom edges of the superconductor and then emerges with unit probability as either an outgoing electron $c_{\rm out}^\dagger$ or hole $c_{\rm out}$ (superpositions are forbidden).

When the flux vanishes one can deduce the conductance by adiabatically deforming the area of the superconducting region in Fig.\ \ref{Interferometry_fig}(a) to zero.\cite{MajoranaModesWithChargeTransport}  In this limit an electron incident from the left is guaranteed to exit as an electron on the right, and by continuity the same must be true when the superconducting region is finite.  This can be summarized schematically by the process
\begin{equation}
  c_{\rm in}^\dagger \rightarrow \gamma_b + i\gamma_t \rightarrow c_{\rm out}^\dagger,
\end{equation}
where $\gamma_b + i\gamma_t$ represents the intermediate fermionic state of the incident electron.  The zero-bias conductance therefore vanishes when $\Phi = 0$.\cite{MajoranaModesWithChargeTransport,3DTI_Majorana_edge_states1}  Next, suppose that one threads flux $\Phi = \frac{hc}{2e}$ to induce a single vortex in the superconducting region.  This introduces a branch cut indicating where the phase jumps by $2\pi$, which we take to emanate from the core to the top edge in Fig.\ \ref{Interferometry_fig}(a).  Since $\gamma_t$ acquires a minus sign upon crossing the branch cut (recall Sec.\ \ref{2D_toy_model}), incident electrons now exit perfectly as \emph{holes},
\begin{equation}
  c_{\rm in}^\dagger \rightarrow \gamma_b + i\gamma_t \rightarrow \gamma_b - i\gamma_t \rightarrow c_{\rm out}
\end{equation}
yielding a $\frac{2e^2}{h}$ conductance.\cite{MajoranaModesWithChargeTransport,3DTI_Majorana_edge_states1}  By generalizing this picture to flux $\Phi = \frac{hc}{2e}N_v$ one can see that the conductance oscillates between quantized values $G(0) = 0$ for even vortex number $N_v$ and $G(0) = \frac{2e^2}{h}$ for odd $N_v$.\cite{MajoranaModesWithChargeTransport,3DTI_Majorana_edge_states1}  Observing these discrete conductance oscillations (which have also been captured in semiconductor-based systems\cite{SemiconductorInterferometry}) would provide clear evidence for the chiral Majorana edge states underlying this remarkable result.  Furthermore, concepts pioneered in the quantum Hall context\cite{FradkinInterferometry,TopologicalQubits,MajoranaInterferometry1,MajoranaInterferometry2} can be employed in related experiments to implement interferometric readout of the qubit states formed by vortex Majorana zero-modes\cite{3DTI_Majorana_edge_states1,SemiconductorInterferometry}.  

References \onlinecite{SrRu} and \onlinecite{VortexAharonovCasher} proposed interesting alternative probes of Majorana modes that rely on interferometry of vortices in the bulk of a 2D topological superconductor.  Such `Abrikosov vortices' tend to behave classically, so we will discuss an elegant follow-up proposal employing `Josephson vortices' that more readily exhibit quantum phenomena.\cite{GrosfeldStern}  Figure \ref{Interferometry_fig}(b) illustrates the desired setup, consisting of $s$-wave superconductors and ferromagnets patterned on a 3D topological insulator (though any spinless $p+ip$ superconductor realization will do here).  In the center of the structure sits an island that hosts charge $Q$ and a magnetic flux $\Phi$.  As usual the innermost superconducting edge supports a Majorana zero-mode when the flux $\Phi$ induces an $\frac{hc}{2e}$ vortex.  The inner and outer superconductors in Fig.\ \ref{Interferometry_fig}(b), however, realize a Josephson junction bridged by a thin ferromagnetic barrier.  Consequently the chiral Majorana edge states at the interface hybridize across the junction and generally acquire a gap, but are not entirely inert.  Remarkably, Grosfeld and Stern showed that Josephson vortices---at which the superconducting phase difference across the junction locally winds by $2\pi$---trap a single Majorana zero-mode as in the case of an Abrikosov vortex.\cite{GrosfeldStern}  

When a Josephson vortex binding a zero-mode $\gamma$ flows rightward [which can be arranged by driving a perpendicular supercurrent $I$ as shown in Fig.\ \ref{Interferometry_fig}(b)] interference of the upper and lower trajectories depends on both the charge $Q$ and flux $\Phi$ on the center island.\cite{VortexAharonovCasher,GrosfeldStern}  The $Q$ dependence reflects the Aharonov-Casher effect\cite{AharonovCasher}---an $\frac{hc}{2e}$ flux encircling charge $Q$ acquires a phase $\phi_{AC} = \frac{hc}{2e}\frac{Q}{\hbar c} = \frac{\pi Q}{e}$, similar to the Aharonov-Bohm phase accumulated by a charge encircling a flux.  With $\Phi = 0$ a Josephson-vortex current $I_v$ flowing through the interferometer therefore oscillates with $Q$ according to
\begin{equation}
  I_v = I_v^0\left[1 + A \cos\left(\pi Q/e\right)\right],
  \label{Iv}
\end{equation}
where $I_v^0$ and $A$ denote the mean current and oscillation amplitude.\cite{VortexAharonovCasher,GrosfeldStern}  The oscillations in Eq.\ (\ref{Iv}) can be detected by measuring the transverse voltage difference induced by the vortex flow as $Q$ varies.  A flux $\Phi = \frac{hc}{2e}$ produces a Majorana zero-mode $\gamma_{\rm in}$ inside of the interferometer and changes these results qualitatively.  In this case taking the Josephson vortex in Fig.\ \ref{Interferometry_fig}(b) around the island then leads not only to an Aharonov-Casher phase, but also changes the sign of both $\gamma$ and $\gamma_{\rm in}$ due to branch cut crossings.  The latter effect causes the amplitude $A$ to vanish\cite{MajoranaInterferometry1,MajoranaInterferometry2} (which is rooted in non-Abelian statistics explored in the next section), destroying the vortex-current oscillations.\cite{VortexAharonovCasher,GrosfeldStern}  The striking dependence on $Q$ and $\Phi$ is a dramatic manifestation of Majorana modes, in particular the exotic statistics they underpin, and would be fascinating to observe.

\section{Non-Abelian statistics and quantum computation}
\label{NonAbelianStatistics}

The experimental realization of Majorana modes would pave the way to far-reaching technological innovations.  On the most basic level, a set of Majorana-carrying vortices or domain walls non-locally encodes quantum information in the degenerate ground-state space, enabling immediate applications for long-lived `topological quantum memory'.  In the longer term the prospect of manipulating that information in a manner that avoids decoherence would constitute an important breakthrough for quantum computation.  This is made possible by the most coveted manifestation of Majorana fermions: non-Abelian statistics.

Before turning to specific implementations it will be useful to discuss this phenomenon in some generality.  Consider a topological system supporting $2N$ Majorana zero-modes $\gamma_{1,\ldots,2N}$.  As in Eq.\ (\ref{fn}) we will (arbitrarily) combine the Majoranas into operators $f_j = (\gamma_{2j-1}+i \gamma_{2j})/2$ whose occupation numbers $n_j = f_j^\dagger f_j$ can be used to label the ground-state manifold.  Suppose that one prepares this system into a ground state $|\Phi_i\rangle = |n_1,\ldots,n_N\rangle$ and then adiabatically swaps the positions of any two Majoranas.  (We assume that the ground-state degeneracy is preserved throughout and that the initial and final Hamiltonians coincide.)  The exchange statistics of the defects binding these zero-modes follows from the time evolution of $|\Phi_i\rangle$; for a nice discussion, see Refs.\ \onlinecite{TQCreview} and \onlinecite{BondersonNonAbelianStatistics}.  One source of this evolution is the dynamical phase $e^{-\frac{i}{\hbar}\int_{0}^{T}E(t)dt}$ acquired by the wavefunction, where $E(t)$ is the instantaneous ground-state energy during the interval $T$ over which the interchange occurs.  This factor is irrelevant for our purposes and will henceforth be ignored.  More importantly, the degeneracy together with the fractionalized nature of the zero-modes allows the system to end in a fundamentally different ground state $|\Phi_f\rangle$ from which it began.  Proving this and the non-Abelian statistics that follows is nontrivial, and requires tracking the adiabatic evolution of the full many-body wavefunction---along with Berry matrices that can connect different ground states.  Fortunately, one can deduce the final state $|\Phi_f\rangle$ (up to an overall phase) by addressing the drastically simpler problem of how the Majorana operators transform under the exchange.  In all cases that we are aware of this procedure agrees with more rigorous approaches developed, \emph{e.g.}, in Refs.\ \onlinecite{NayakWilczekBraiding,ReadGreen,SternBraiding,StoneBraiding,ReadBraiding,BondersonNonAbelianStatistics,AliceaBraiding}.  

\begin{figure}
\includegraphics[width = 8cm]{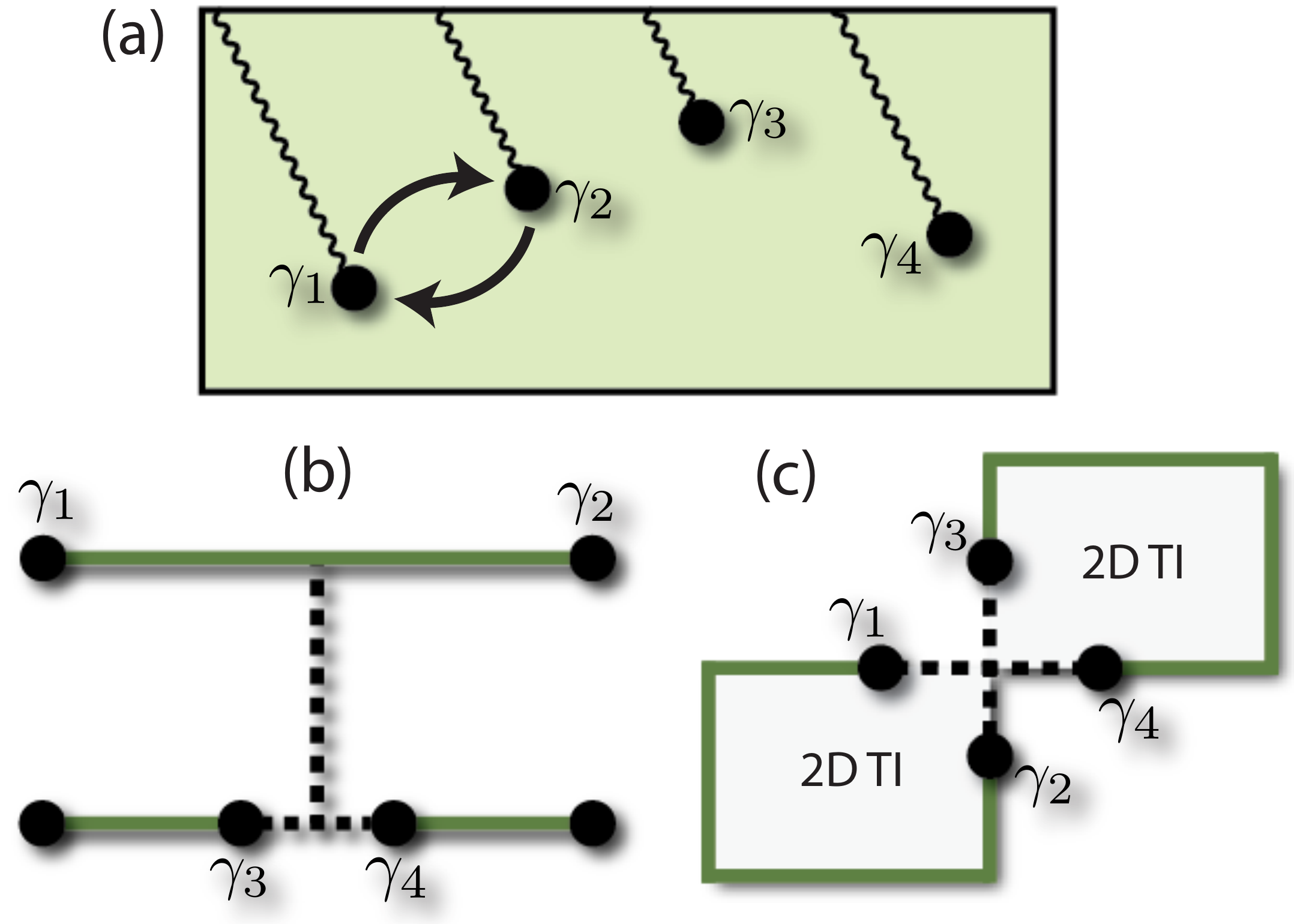}
\caption{(a) A 2D topological superconductor with four well-separated vortices binding Majorana zero-modes.  Under a clockwise braid of $\gamma_1$ and $\gamma_2$ these Majoranas not only interchange positions, but $\gamma_1$ also crosses a branch cut.  Consequently the exchange sends $\gamma_1 \rightarrow -\gamma_2$ and $\gamma_2 \rightarrow \gamma_1$.  This simple picture due to Ivanov\cite{Ivanov} leads to non-Abelian statistics as described in the text.  (b) Performing meaningful exchanges of Majorana modes arising in 1D topological superconductors (green) requires arranging wires into networks.  (c) Corner junctions formed by 2D topological insulators similarly allow Majorana modes to be interchanged along the edges.  Despite the absence of vortices the exchange statistics remains non-Abelian in these networks.}
\label{Braiding_fig}
\end{figure}

With this simplification in mind, we now review Ivanov's\cite{Ivanov} remarkably accessible picture for non-Abelian statistics in a 2D spinless $p+ip$ superconductor (this discussion applies equally well to any of the experimental realizations in Sec.\ \ref{2D_toy_model_realizations}).  It suffices to analyze the configuration of Fig.\ \ref{Braiding_fig}(a) where four well-separated vortices trap Majorana zero-modes $\gamma_{1,2,3,4}$.  Suppose that we apply local pinning potentials to adiabatically braid the left two vortices clockwise as shown in the figure.  This has two important consequences for the zero-modes---($i$) $\gamma_1$ and $\gamma_2$ swap positions and ($ii$) $\gamma_1$ crosses a branch cut and acquires an additional minus sign.  Thus the Majorana operators transform according to $\gamma_1 \rightarrow -\gamma_{2}$ and $\gamma_{2}\rightarrow \gamma_1$.  The unitary operator implementing this transformation is $U_{12} = (1+\gamma_1\gamma_{2})/\sqrt{2}$ (that is, $U_{12}\gamma_{1/2}U_{12}^\dagger = \mp \gamma_{2/1}$).  Similarly, clockwise exchange of neighboring vortices binding $\gamma_j$ and $\gamma_{j+1}$ sends $\gamma_j \rightarrow -\gamma_{j+1}$ and $\gamma_{j+1}\rightarrow \gamma_j$, which is generated by 
\begin{equation}
  U_{j,j+1} = (1+\gamma_j\gamma_{j+1})/\sqrt{2}.
  \label{BraidOps}
\end{equation}  
(For counterclockwise braids $U_{j,j+1}^{-1}$ implements the transformation.)  Up to a common overall phase factor the ground-state wavefunctions $|n_1,n_2\rangle$ therefore evolve as\cite{Ivanov} $|n_1,n_2\rangle \rightarrow U_{j,j+1}|n_1,n_2\rangle$.

For the three types of clockwise exchanges between neighboring vortices in Fig.\ \ref{Braiding_fig}(a), one explicitly finds
\begin{eqnarray}
  |n_1,n_2\rangle &\rightarrow& U_{12}|n_1,n_2\rangle = e^{i\frac{\pi}{4}(1-2n_1)}|n_1,n_2\rangle
  \label{U12}
  \\
  |n_1,n_2\rangle &\rightarrow& U_{23}|n_1,n_2\rangle 
  \label{U23}
  \\
  &=& \frac{1}{\sqrt{2}}\left[|n_1,n_2\rangle + i(-1)^{n_1}|1-n_1,1-n_2\rangle\right]
  \nonumber \\
  |n_1,n_2\rangle &\rightarrow& U_{34}|n_1,n_2\rangle = e^{i\frac{\pi}{4}(1-2n_2)}|n_1,n_2\rangle.  
  \label{U34}
\end{eqnarray}
Braiding $\gamma_{1,2}$ or $\gamma_{3,4}$ in a sense `internally rotates' the ordinary fermion operators $f_{j}$ and produces nontrivial phase factors in the states $|n_1,n_2\rangle$.  More interestingly, braiding $\gamma_{2,3}$ swaps `half' of $f_1$ with `half' of $f_2$, resulting in a nontrivial rotation of $|n_1,n_2\rangle$ within the ground-state manifold.  Together these properties give rise to non-Abelian statistics of vortices: \emph{if one performs sequential exchanges, the final state depends on the order in which they are carried out}.  Mathematically, this fascinating result follows from the nontrivial commutation relations satisfied by the operators in Eq.\ (\ref{BraidOps}).  These conclusions generalize trivially to systems supporting arbitrarily many vortices.  

If one tries to extend this analysis to Majorana zero-modes arising in 1D topological superconductors, two immediate problems arise.  The first is that exchange statistics, non-Abelian or otherwise, is never well-defined in 1D systems.  As an example, suppose we attempt to adiabatically exchange $\gamma_3$ and $\gamma_4$ in the setup of Fig.\ \ref{1D_wires_fig}(d) by moving $\gamma_3$ rightward and $\gamma_4$ leftward.  Clearly these zero-modes eventually overlap and split the ground-state degeneracy; furthermore, whether or not one actually performed an exchange when the system returns to its original configuration is completely ambiguous.  Moving away from strict one-dimensionality by fabricating networks of 1D wires\cite{AliceaBraiding} or 2D topological insulator edges\cite{Nate} circumvents this problem in a conceptually straightforward manner.  Figures \ref{Braiding_fig}(b) and (c) illustrate examples that allow Majorana modes to be meaningfully exchanged.  (As usual green denotes topological regions while dashed lines are trivial.  Note also that the structure of such networks is rather arbitrary---they can even form three-dimensional lattices.\cite{HalperinBraiding})  The zero-modes can be adiabatically transported in these setups by applying gate voltages\cite{AliceaBraiding,SauBraiding} or supercurrents\cite{Romito} to shift the domain wall locations as desired.  In this way one can exchange $\gamma_1$ and $\gamma_2$ in Fig.\ \ref{Braiding_fig}(b) by executing a `three-point-turn'\cite{AliceaBraiding,HalperinBraiding}: first moving $\gamma_1$ to the center of the vertical line, then moving $\gamma_2$ all the way leftward, and finally moving $\gamma_1$ up and to the right.   Similar ideas allow one to exchange the Majoranas $\gamma_3$ and $\gamma_4$ belonging to different topological segments, and can also be adapted to the corner junction of Fig.\ \ref{Braiding_fig}(c).  Interestingly, novel methods of effectively implementing exchanges without physically transporting Majoranas (as in measurement-only topological quantum computation\cite{MeasurementOnlyTQC}) have also been proposed recently.\cite{Flensberg,BraidingWithoutTransport}

Although the exchange of Majorana zero-modes becomes well-defined in these networks, a second, much subtler problem appears in this context.  Namely, the vortices that are crucial for establishing non-Abelian statistics in 2D $p+ip$ superconductors are entirely absent here---so does non-Abelian statistics still emerge?  Fortunately several studies have shown that it does, and that the Majorana zero-modes in fact transform under exchange exactly as in the $p+ip$ case despite the lack of vortices.\cite{AliceaBraiding,ClarkeBraiding,SauBraiding,HalperinBraiding}  To provide a rough flavor for how this arises, consider the physical situation where the network of Fig.\ \ref{Braiding_fig}(b) arises from spin-orbit-coupled wires adjacent to an $s$-wave superconductor with uniform phase $\phi$.  Suppose that we exchange $\gamma_{1,2}$ as described above.  Although the $s$-wave superconductor exhibits a uniform phase, as the Majoranas traverse the network the phase of the \emph{effective} $p$-wave pair field that they experience does in fact vary.  This variation causes one of the Majoranas to acquire a minus sign arising from a branch cut, precisely as in Ivanov's construction.\cite{AliceaBraiding}  For more details and complementary perspectives on this interesting problem see Refs.\ \onlinecite{AliceaBraiding,ClarkeBraiding,SauBraiding,HalperinBraiding}.

By virtue of non-Abelian statistics, braiding Majorana zero-modes in 2D $p+ip$ superconductors and 1D networks allows one to perform topological quantum information processing that is in principle immune from decoherence.\cite{TQCreview}  Equations (\ref{U12}) through (\ref{U34}) illustrate a concrete example of the protected qubit processing effected in this manner.  Unfortunately, such qubit rotations are too restrictive to permit universal quantum computation; two additional processes are needed.\cite{TQCreview}  The first is a $\pi/8$ phase gate that introduces phase factors $e^{\pm i\pi/8}$ depending on the occupation number corresponding to a given pair of Majoranas.  The second is the ability to read out the eigenvalue of the product of four Majoranas, $\gamma_i\gamma_j\gamma_k\gamma_l$, without measuring that of individual pairs.  While these processes can introduce errors this does not mean that the topologically protected braiding operations are without merit.  In fact the unprotected part of the computation enjoys a dramatically higher error threshold compared to conventional quantum computing schemes.\cite{BravyiKitaev,Bravyi}  Many ingenious proposals already exist for supplement braiding with the operations required to perform universal quantum computation, both in architectures based on 1D and 2D topological phases.\cite{TQCreview,UniversalTQC,BondersonPhaseGate,ClarkePhaseGate,Flensberg,Hassler,SauWireNetwork,TopTransmon,JiangKanePreskill,TopologicalQuantumBus,Xue}  The blueprints for a Majorana-based quantum computer are therefore already in place; we simply need to begin assembling the hardware.

\section{Outlook}
\label{Outlook}

The possibility of observing Majorana fermions in condensed matter systems now appears tantalizingly close.  In Secs.\ \ref{1D_toy_model_realizations} and \ref{2D_toy_model_realizations} we saw that the number of realistic proposals that now exist is rather immense, and most involve heterostructures with garden-variety $s$-wave superconductors.  One point worth emphasizing here is that the theory for these proximity-induced topological superconductors  centers largely around non-interacting electron models---apart from, of course, attractive interactions implicitly invoked in the parent superconductors.  We view this simplicity as an enormous virtue that affords theorists a degree of predictive power for experiments that is not often encountered in the quest for exotic phases of matter.  (One can argue that the rapid progress in the field of topological insulators arose for similar reasons.)  Given this fact, along with the accessibility of the building blocks comprising the heterostructures, it is no surprise that several of these new proposals have already inspired broad experimental efforts in leading laboratories worldwide.  If these efforts continue unabated we believe the question is not whether Majorana modes will be definitively identified, but rather when, in what system, and with what measurement.  

Amongst the numerous proposals reviewed the original topological-insulator-based devices introduced by Fu and Kane\cite{FuKane,MajoranaQSHedge} (see also Ref.\ \onlinecite{CookFranz}) remain in some ways ideal.  The elegance of the theories is hard to match, and from a practical standpoint the prospect of obtaining topological phases protected by gaps that are immune to disorder\cite{Disorder4,CookFranz} and limited only by that of the parent superconductor\cite{SauProximityEffect,Disorder4} is exceedingly attractive.  We believe strongly that experiments in this direction should be ardently pursued using both 2D and 3D topological insulators even if another system `wins the race' for Majorana; future applications may rest on such exceptional properties.  Proposals employing conventional spin-orbit-coupled 1D wires, first pioneered in Refs.\ \onlinecite{1DwiresLutchyn} and \onlinecite{1DwiresOreg}, also stand out given the comparative maturity of semiconductor technology as well as the simplicity and tunability of the required architectures.  These systems are well-poised to experimentally realize the predicted magnetic-field-driven topological phase in the near future.\cite{InAsProximityEffect,InAsProximityEffect2,InSbProximityEffect,InSbCharacterization}  More broadly, we hope that experimentalists will push many more of the newly proposed directions while theorists continue to conjure up new and improved Majorana platforms.  There may just be a sleeper in the mix, or perhaps the ideal direction is even yet to be introduced.  

The first unambiguous sighting of Majorana fermions in condensed matter would provide a landmark event in physics.  We hope to have made a compelling case in Sec.\ \ref{Detection} that numerous smoking-gun detection methods are now available to make this identification definitive.  It should be stressed that this initial observation will only herald the beginning of what is likely to be a long, fruitful subfield.  The realization of exotic physics such as fractional Josephson effects and non-Abelian statistics, as well as applications from topological quantum memory to universal quantum computation are truly fascinating goals that will keep physicists occupied for many years to come.  And as always there are bound to be many surprises along the way.  

To conclude we will briefly highlight some interesting future directions and open questions.  One intriguing alternative route to engineering topological phases involves periodically driving a system that would otherwise be trivial.  Using this mechanism proposals for artificially generating topological insulators\cite{2DFloquetTI,3DFloquetTI} and 1D superfluids supporting Majorana fermions\cite{ColdAtomMajoranas} have recently been put forth.  The possibility of moving away from static systems as a means of generating non-Abelian phases opens new avenues that warrant further exploration.  There are also many other promising routes to Majorana fermions that we have not touched on here.   Among the most interesting is the potential realization of Kitaev's 2D honeycomb model\cite{ToricCode} in a certain class of magnetic insulators\cite{Jackeli1,Jackeli2,HongChen}.  Looking forward, it is worth exploring whether the connection identified by Read and Green\cite{ReadGreen} between the Moore-Read state\cite{MooreRead} and a spinless $p+ip$ superconductor can be adapted to still more exotic fractional quantum Hall states.  As noted in the introduction it is this remarkable correspondence that led to the realization that a weakly correlated superconductor can harbor non-Abelian statistics.  Establishing a similar correspondence between quantum Hall phases supporting even richer non-Abelian anyons and phases exhibited by less strongly interacting 2D systems could open entirely new directions in the pursuit of topological quantum computation.  If this can be achieved, might there exist related 1D systems supporting these richer non-Abelian anyons, in the same way that Majorana modes can appear in either 1D or 2D topological superconductors?  

\acknowledgments

I would like to first acknowledge all of my collaborators on the subject of Majorana fermions, especially Anton Akhmerov, David Clarke, Lukasz Fidkowski, Matthew Fisher, Bert Halperin, Liang Jiang, Takuya Kitagawa, Shu-Ping Lee, Netanel Lindner, Roman Lutchyn, Yuval Oreg, David Pekker, Gil Refael, Alessandro Romito, Kirill Shtengel, Oleg Starykh, Ady Stern, Miles Stoudenmire, and Felix von Oppen.  I have also benefited enormously from interactions with Doron Bergman, Parsa Bonderson, Sankar Das Sarma, Jim Eisenstein, Marcel Franz, Michael Freedman, Erik Henriksen, Leo Kouwenhoven, Patrick Lee, Charlie Marcus, Lesik Motrunich, Morgan Page, Andrew Potter, Xiaoliang Qi, Sri Raghu, Nick Read, Jay Sau, Ari Turner, and Amir Yacoby.  This work was supported by the National Science Foundation through grant DMR-1055522.  

\appendix

\section{Derivation of the effective action for 1D and 2D systems coupled to a bulk $s$-wave superconductor}
\label{EffectiveActionDerivation}

Consider a $d$-dimensional system of electrons (with $d = 1$ or 2) proximate to a bulk $s$-wave superconductor.  Let the Hamiltonian for this structure be $H = H_d + H_{SC} + H_\Gamma$, where 
\begin{equation}
  H_d = \int \frac{d^d{\bf k}}{(2\pi)^d} \psi_{\bf k}\mathcal{H}_{\bf k}\psi_{\bf k}
\end{equation}
describes the $d$-dimensional system, 
\begin{eqnarray}
  H_{SC} = \int \frac{d^3{\bf k}}{(2\pi)^3}[\epsilon_k\eta^\dagger_{\bf k}\eta_{\bf k}
  + \Delta_{sc}(\eta_{\uparrow{\bf k}} \eta_{\downarrow-{\bf k}} + H.c.)],
\end{eqnarray}
models the $s$-wave superconductor, with $\epsilon_{k} = k^2/(2m_{sc})-\mu_{sc}$ the superconductor's kinetic energy, and 
\begin{equation}
  H_\Gamma = -\Gamma \int \frac{d^3{\bf k}}{(2\pi)^3} (\psi_{{\bf k}_d}\eta_{\bf k} + H.c.)
  \label{HGammaAppendix}
\end{equation}
the term which incorporates electron tunneling between the two subsystems.  In Eq.\ (\ref{HGammaAppendix}), ${\bf k}_d = k_x$ if $d = 1$ while ${\bf k}_d = (k_x,k_y)$ if $d = 2$.  Our goal is to obtain an effective action for the $d$-dimensional system with the gapped superconductor degrees of freedom integrated out.  To achieve this it is useful to first perform a unitary transformation which diagonalizes $H_{SC}$:
\begin{eqnarray}
  \eta_{\uparrow {\bf k}} &=& -u_k \chi_{1 {\bf k}} + v_k \chi_{2-{\bf k}}^\dagger
  \nonumber \\
  \eta_{\downarrow {\bf k}} &=& v_k \chi_{1 -{\bf k}}^\dagger + u_k \chi_{2{\bf k}}
  \nonumber \\
  u_k &=& \frac{\Delta_{sc}}{\sqrt{2E_k(E_k-\epsilon_k)}},~~~ v_k = \frac{\Delta_{sc}}{\sqrt{2E_k(E_k+\epsilon_k)}},
\end{eqnarray}
where $E_k = \sqrt{\epsilon_k^2 + \Delta^2}$ are the quasiparticle energies for the superconductor.  In this new basis one obtains
\begin{eqnarray}
  H_{SC} &=& \int \frac{d^3{\bf k}}{(2\pi)^3} E_k[\chi_{1{\bf k}}^\dagger \chi_{1{\bf k}} + \chi_{2{\bf k}}^\dagger \chi_{2{\bf k}}]
  \\
  H_\Gamma &=& - \Gamma \int  \frac{d^3{\bf k}}{(2\pi)^3}[\chi_{1{\bf k}}(u_k \psi_{\uparrow {\bf k}_d}^\dagger + v_k \psi_{\downarrow -{\bf k}_d}) 
  \nonumber \\
  &+& \chi_{2{\bf k}}(v_k \psi_{\uparrow -{\bf k}_d} - u_k \psi_{\downarrow {\bf k}_d}^\dagger) + H.c.]. 
\end{eqnarray}

It is now straightforward to write down the Euclidean path integral corresponding to $H$ and then integrate out the quasiparticle operators $\chi_{1,2}$.  This yields an effective action $S_{\rm eff} = S_d + \delta S$, with
\begin{eqnarray}
  S_d &=& \int \frac{d\omega}{2\pi}\frac{d^d{\bf k}}{(2\pi)^d} \psi_{(\bf k,\omega)}\mathcal{H}_{\bf k}\psi_{(\bf k,\omega)}
  \\
  \delta S &=& \int \frac{d\omega}{2\pi}\frac{d^d{\bf k}}{(2\pi)^d}\{\Delta_{sc}\lambda({\bf k},\omega)[\psi_{\uparrow({\bf k},\omega)}\psi_{\downarrow(-{\bf k},-\omega)} + H.c.] 
  \nonumber \\
  &+& [-i \omega \lambda({\bf k},\omega)-\delta \mu({\bf k},\omega)] \psi^\dagger_{({\bf k},\omega)} \psi_{({\bf k},\omega)}\}.
\end{eqnarray}
One sees here that the superconductor renormalizes the chemical potential for the $d$-dimensional system (in a weakly frequency- and momentum-dependent fashion) through $\delta \mu({\bf k},\omega)$; this correction is unimportant, however, and will be henceforth neglected.  The essential physics associated with the hybridization is encoded in the function $\lambda({\bf k},\omega)$ appearing in $\delta S$.  Let us focus for the moment on $d = 1$ where this is given by
\begin{equation}
  \lambda(k_x,\omega) = \int_{k_y,k_z} \frac{\Gamma^2}{\omega^2 + \Delta^2 + \left[\frac{k_y^2 + k_z^2}{2m_{sc}} + \left(\frac{k_x^2}{2m_{sc}}-\mu_{sc}\right)\right]^2},
  \label{lambda}
\end{equation}
Typically we will be concerned with one-dimensional systems of rather low density so that $k_x^2/(2m_{sc}) \ll \mu_{sc}$ for the important values of $k_x$; in this case the dependence of $\lambda$ on $k_x$ can be safely ignored.  Making the further reasonable assumption that $\mu_{sc} \gg \sqrt{\omega^2 + \Delta_{sc}^2}$ over the relevant frequencies, $\lambda$ evaluates to the following simple expression:
\begin{equation}
  \lambda(\omega) = \frac{\pi \rho_{2D} \Gamma^2}{\sqrt{\omega^2 + \Delta_{sc}^2}}
  \label{lambda2}
\end{equation}
where $\rho_{2D} = m_{sc}/(2\pi)$ is the density of states (per spin) for a two-dimensional system with effective mass $m_{sc}$.  For a $d = 2$ dimensional system $\lambda(k_x,k_y,\omega)$ follows from the right side of Eq.\ (\ref{lambda}) integrated only over $k_z$.  Under similar assumptions made for $d = 1$ $\lambda$ is again approximately momentum independent and given by Eq.\ (\ref{lambda2}) with $\rho_{2D}$ replaced by the 1D density of states $\rho_{1D} = \pi^{-1}\sqrt{m_{sc}/(2\mu_{sc})}$.  Upon defining a quasiparticle weight $Z(\omega) = [1 + \lambda(\omega)]^{-1}$, the effective action can be expressed in the desired form quoted in the main text:
\begin{eqnarray}
  S_{\rm eff} &=& \int \frac{d\omega}{2\pi}\frac{d^d{\bf k}}{(2\pi)^d} Z^{-1}(\omega)\{\psi^\dagger_{({\bf k},\omega)}[-i\omega + Z(\omega)\mathcal{H}_{\bf k}]\psi_{({\bf k},\omega)}
  \nonumber \\
  &+& \Delta_{sc}[1-Z(\omega)][\psi_{\uparrow ({\bf k},\omega)} \psi_{\downarrow (-{\bf k},-\omega)} + H.c.]\}.
\end{eqnarray}


\end{document}